\input harvmac

%
\message{S-Tables Macro v1.0, ACS, TAMU (RANHELP@VENUS.TAMU.EDU)}
%
%
\newhelp\stablestylehelp{You must choose a style between 0 and 3.}%
\newhelp\stablelinehelp{You should not use special hrules when
stretching
a table.}%
\newhelp\stablesmultiplehelp{You have tried to place an S-Table
inside another S-Table.  I would recommend not going on.}%
%
%
\newdimen\stablesthinline
\stablesthinline=0.4pt
\newdimen\stablesthickline
\stablesthickline=1pt
%
%
\newif\ifstablesborderthin
\stablesborderthinfalse
\newif\ifstablesinternalthin
\stablesinternalthintrue
\newif\ifstablesomit
\newif\ifstablemode
\newif\ifstablesright
\stablesrightfalse
%
%
\newdimen\stablesbaselineskip
\newdimen\stableslineskip
\newdimen\stableslineskiplimit
%
%
\newcount\stablesmode
\newcount\stableslines
\newcount\stablestemp
\stablestemp=3
\newcount\stablescount
\stablescount=0
\newcount\stableslinet
\stableslinet=0
%
%
%
\newcount\stablestyle
\stablestyle=0
%
%
\def\stablesleft{\quad\hfil}%
\def\stablesright{\hfil\quad}%
%
%
\catcode`\|=\active%
%
%
\newcount\stablestrutsize
\newbox\stablestrutbox
\setbox\stablestrutbox=\hbox{\vrule height10pt depth5pt width0pt}
\def\stablestrut{\relax\ifmmode%
                         \copy\stablestrutbox%
                       \else%
                         \unhcopy\stablestrutbox%
                       \fi}%
%
%
\newdimen\stablesborderwidth
\newdimen\stablesinternalwidth
\newdimen\stablesdummy
\newcount\stablesdummyc
\newif\ifstablesin
\stablesinfalse
%
%
%
%
%
\def\stablesadj{%
  \ifcase\stablestyle%
    \hbox to \hsize\bgroup\hss\vbox\bgroup%
  \or%
    \hbox to \hsize\bgroup\vbox\bgroup%
  \or%
    \hbox to \hsize\bgroup\hss\vbox\bgroup%
  \or%
    \hbox\bgroup\vbox\bgroup%
  \else%
    \errhelp=\stablestylehelp%
    \errmessage{Invalid style selected, using default}%
    \hbox to \hsize\bgroup\hss\vbox\bgroup%
  \fi}%
\def\stablesend{\egroup%
  \ifcase\stablestyle%
    \hss\egroup%
  \or%
    \hss\egroup%
  \or%
    \egroup%
  \or%
    \egroup%
  \else%
    \hss\egroup%
  \fi}%
\def\stablestart{%
  \ifstablesin%
    \errhelp=\stablesmultiplehelp%
    \errmessage{An S-Table cannot be placed within an S-Table!}%
  \fi
  \global\stablesintrue%
  \global\advance\stablescount by 1%
  \message{<S-Tables Generating Table \number\stablescount}%
  \begingroup%
  \stablestrutsize=\ht\stablestrutbox%
  \advance\stablestrutsize by \dp\stablestrutbox%
  \ifstablesborderthin%
    \stablesborderwidth=\stablesthinline%
  \else%
    \stablesborderwidth=\stablesthickline%
  \fi%
  \ifstablesinternalthin%
    \stablesinternalwidth=\stablesthinline%
  \else%
    \stablesinternalwidth=\stablesthickline%
  \fi%
  \tabskip=0pt%
  \stablesbaselineskip=\baselineskip%
  \stableslineskip=\lineskip%
  \stableslineskiplimit=\lineskiplimit%
  \offinterlineskip%
  \def\borderrule{\vrule width \stablesborderwidth}%
  \def\internalrule{\vrule width \stablesinternalwidth}%
  \def\thinline{\noalign{\hrule height \stablesthinline}}%
  \def\thickline{\noalign{\hrule height \stablesthickline}}%
  \def\trule{\omit\leaders\hrule height \stablesthinline\hfill}%
  \def\ttrule{\omit\leaders\hrule height \stablesthickline\hfill}%
  \def\tttrule##1{\omit\leaders\hrule height ##1\hfill}%
  \def\stablesel{&\omit\global\stablesmode=0%
    \global\advance\stableslines by 1\borderrule\hfil\cr}%
  \def\el{\stablesel&}%
  \def\elt{\stablesel\thinline&}%
  \def\eltt{\stablesel\thickline&}%
  \def\elttt##1{\stablesel\noalign{\hrule height ##1}&}%
  \def\elspec{&\omit\hfil\borderrule\cr\omit\borderrule&%
              \ifstablemode%
              \else%
                \errhelp=\stablelinehelp%
                \errmessage{Special ruling will not display properly}%
              \fi}%
  \def\stmultispan##1{\mscount=##1 \loop\ifnum\mscount>3
\stspan\repeat}%
  \def\stspan{\span\omit \advance\mscount by -1}%
  \def\multicolumn##1{\omit\multiply\stablestemp by ##1%
     \stmultispan{\stablestemp}%
     \advance\stablesmode by ##1%
     \advance\stablesmode by -1%
     \stablestemp=3}%
  \def\multirow##1{\stablesdummyc=##1\parindent=0pt\setbox0\hbox\bgroup%
    \aftergroup\emultirow\let\temp=}
  \def\emultirow{\setbox1\vbox to\stablesdummyc\stablestrutsize%
    {\hsize\wd0\vfil\box0\vfil}%
    \ht1=\ht\stablestrutbox%
    \dp1=\dp\stablestrutbox%
    \box1}%
%
  \def\stpar##1{\vtop\bgroup\hsize ##1%
     \baselineskip=\stablesbaselineskip%
     \lineskip=\stableslineskip%

\lineskiplimit=\stableslineskiplimit\bgroup\aftergroup\estpar\let\temp=}%
  \def\estpar{\vskip 6pt\egroup}%
  \def\stparrow##1##2{\stablesdummy=##2%
     \setbox0=\vtop to ##1\stablestrutsize\bgroup%
     \hsize\stablesdummy%
     \baselineskip=\stablesbaselineskip%
     \lineskip=\stableslineskip%
     \lineskiplimit=\stableslineskiplimit%
     \bgroup\vfil\aftergroup\estparrow%
     \let\temp=}%
  \def\estparrow{\vfil\egroup%
     \ht0=\ht\stablestrutbox%
     \dp0=\dp\stablestrutbox%
     \wd0=\stablesdummy%
     \box0}%
  \def|{\global\advance\stablesmode by 1&&&}%
  \def\|{\global\advance\stablesmode by 1&\omit\vrule width 0pt%
         \hfil&&}%
\def\vt{\global\advance\stablesmode
by 1&\omit\vrule width \stablesthinline%
          \hfil&&}%
  \def\vtt{\global\advance\stablesmode by 1&\omit\vrule width
\stablesthickline%
          \hfil&&}%
  \def\vttt##1{\global\advance\stablesmode by 1&\omit\vrule width ##1%
          \hfil&&}%
  \def\vtr{\global\advance\stablesmode by 1&\omit\hfil\vrule width%
           \stablesthinline&&}%
  \def\vttr{\global\advance\stablesmode by 1&\omit\hfil\vrule width%
            \stablesthickline&&}%
\def\vtttr##1{\global\advance\stablesmode
 by 1&\omit\hfil\vrule width ##1&&}%
  \stableslines=0%
  \stablesomitfalse}
\def\stablesdef{\bgroup\stablestrut\borderrule##\tabskip=0pt plus 1fil%
  &\stablesleft##\stablesright%
  &##\ifstablesright\hfill\fi\internalrule\ifstablesright\else\hfill\fi%
  \tabskip 0pt&&##\hfil\tabskip=0pt plus 1fil%
  &\stablesleft##\stablesright%
  &##\ifstablesright\hfill\fi\internalrule\ifstablesright\else\hfill\fi%
  \tabskip=0pt\cr%
  \ifstablesborderthin%
    \thinline%
  \else%
    \thickline%
  \fi&%
}%
\def\endtable{\advance\stableslines by 1\advance\stablesmode by 1%
   \message{- Rows: \number\stableslines, Columns:
\number\stablesmode>}%
   \stablesel%
   \ifstablesborderthin%
     \thinline%
   \else%
     \thickline%
   \fi%
   \egroup\stablesend%
\endgroup%
\global\stablesinfalse}
%


\overfullrule=0pt \abovedisplayskip=10pt plus 1pt minus 1pt
\belowdisplayskip=10pt plus 1pt minus 1pt

\noblackbox
\input epsf
\newcount\figno
\figno=0
\def\fig#1#2#3{
\par\begingroup\parindent=0pt\leftskip=1cm\rightskip=1cm\parindent=0pt
\baselineskip=11pt \global\advance\figno by 1 \midinsert
\epsfxsize=#3 \centerline{\epsfbox{#2}} \vskip 12pt
\centerline{{\bf Figure \the\figno:} #1}\par
\endinsert\endgroup\par}
\def\figlabel#1{\xdef#1{\the\figno}}

\def\IR{\relax{\rm I\kern-.18em R}}


\font\cmss=cmss10 \font\cmsss=cmss10 at 7pt
\def\rlx{\relax\leavevmode}
\def\inbar{\vrule height1.5ex width.4pt depth0pt}
\def\IC{\relax\,\hbox{$\inbar\kern-.3em{\rm C}$}}
\def\IN{\relax{\rm I\kern-.18em N}}
\def\IP{\relax{\rm I\kern-.18em P}}
\def\ZZ{\rlx\leavevmode\ifmmode\mathchoice{\hbox{\cmss Z\kern-.4em Z}}
 {\hbox{\cmss Z\kern-.4em Z}}{\lower.9pt\hbox{\cmsss Z\kern-.36em Z}}
 {\lower1.2pt\hbox{\cmsss Z\kern-.36em Z}}\else{\cmss Z\kern-.4em
 Z}\fi}
\def\IZ{\relax\ifmmode\mathchoice
{\hbox{\cmss Z\kern-.4em Z}}{\hbox{\cmss Z\kern-.4em Z}}
{\lower.9pt\hbox{\cmsss Z\kern-.4em Z}} {\lower1.2pt\hbox{\cmsss
Z\kern-.4em Z}}\else{\cmss Z\kern-.4em Z}\fi}

\def\narrowplus{\kern -.04truein + \kern -.03truein}
\def\narrowminus{- \kern -.04truein}
\def\narrowminussub{\kern -.02truein - \kern -.01truein}

\def\frac#1#2{{#1\over #2}}

\def\IZ{\relax\ifmmode\mathchoice
{\hbox{\cmss Z\kern-.4em Z}}{\hbox{\cmss Z\kern-.4em Z}}
{\lower.9pt\hbox{\cmsss Z\kern-.4em Z}} {\lower1.2pt\hbox{\cmsss
Z\kern-.4em Z}}\else{\cmss Z\kern-.4em Z}\fi}
\def\IB{\relax{\rm I\kern-.18em B}}
\def\IC{{\relax\hbox{$\inbar\kern-.3em{\rm C}$}}}
\def\ID{\relax{\rm I\kern-.18em D}}
\def\IE{\relax{\rm I\kern-.18em E}}
\def\IF{\relax{\rm I\kern-.18em F}}
\def\IG{\relax\hbox{$\inbar\kern-.3em{\rm G}$}}
\def\IGa{\relax\hbox{${\rm I}\kern-.18em\Gamma$}}
\def\IH{\relax{\rm I\kern-.18em H}}
\def\II{\relax{\rm I\kern-.18em I}}
\def\IK{\relax{\rm I\kern-.18em K}}
\def\IP{\relax{\rm I\kern-.18em P}}

\font\cmss=cmss10 \font\cmsss=cmss10 at 7pt
\def\IR{\relax{\rm I\kern-.18em R}}

\def\1{{\bf 1}}
\def\3{{\bf 3}}
\def\7{{\bf 7}}
\def\2{{\bf 2}}
\def\8{{\bf 8}}

\def\quabla{{\sqcap}\!\!\!\!{\sqcup}}

\def\o{\over}
%

%
%
\def\eqnn#1{\xdef #1{(\secsym\the\meqno)}\writedef{#1\leftbracket#1}%
\global\advance\meqno by1\wrlabeL#1}
\def\eqna#1{\xdef #1##1{\hbox{$(\secsym\the\meqno##1)$}}
\writedef{#1\numbersign1\leftbracket#1{\numbersign1}}%
\global\advance\meqno by1\wrlabeL{#1$\{\}$}}
\def\eqn#1#2{\xdef #1{(\secsym\the\meqno)}\writedef{#1\leftbracket#1}%
\global\advance\meqno by1$$#2\eqno#1\eqlabeL#1$$}



\lref\vafai{C.~Vafa,
``Superstrings and topological strings at large N,''
J.\ Math.\ Phys.\  {\bf 42}, 2798 (2001), hep-th/0008142.}

\lref\civ{F.~Cachazo, K.~A.~Intriligator and C.~Vafa,
``A large N duality via a geometric transition,''
Nucl.\ Phys.\ B {\bf 603}, 3 (2001), hep-th/0103067.}

\lref\cveticone{M.~Cvetic, G.~W.~Gibbons, H.~Lu and C.~N.~Pope,
 ``Cohomogeneity one manifolds of Spin(7) and G(2) holonomy,''
 Phys.\ Rev.\ D {\bf 65}, 106004 (2002), hep-th/0108245;
``M-theory conifolds,''
 Phys.\ Rev.\ Lett.\  {\bf 88}, 121602 (2002), hep-th/0112098;
``A G(2) unification of the deformed and resolved conifolds,''
Phys.\ Lett.\ B {\bf 534}, 172 (2002), hep-th/0112138.}

\lref\cvetictwo{M.~Cvetic, G.~W.~Gibbons, H.~Lu and C.~N.~Pope,
``New complete non-compact Spin(7) manifolds,''
 Nucl.\ Phys.\ B {\bf 620}, 29 (2002), hep-th/0103155.}

\lref\brand{A.~Brandhuber, J.~Gomis, S.~S.~Gubser and S.~Gukov,
``Gauge theory at large N and new G(2) holonomy metrics,''
Nucl.\ Phys.\ B {\bf 611}, 179 (2001), hep-th/0106034.}

\lref\fawad{S.~F.~Hassan,
``T-duality, space-time spinors and R-R fields in curved backgrounds,''
Nucl.\ Phys.\ B {\bf 568}, 145 (2000), hep-th/9907152.}

\lref\brandtwo{A.~Brandhuber,
``G(2) holonomy spaces from invariant three-forms,''
Nucl.\ Phys.\ B {\bf 629}, 393 (2002), hep-th/0112113.}

\lref\salamontwo{R. ~Bryant, S.~Salamon,
``On the construction of some complete metrics with exceptional
holonomy,'' Duke Math. J. {\bf 58} (1989) 829;
G.~W.~Gibbons, D.~N.~Page and C.~N.~Pope,
``Einstein Metrics On S**3 R**3 And R**4 Bundles,''
Commun.\ Math.\ Phys.\  {\bf 127}, 529 (1990).}

\lref\kovalev{A. Kovalev,
``Twisted connected sums and special Riemannian holonomy,''
math-DG/0012189.}

\lref\joyce{D. ~Joyce,
``Compact Riemannian 7 manifolds with holonomy $G_2$ I, J. Diff. Geom.
{\bf 43} (1996) 291;
II: J. Diff. Geom.{\bf 43} (1996) 329.}

\lref\grayone{A. Gray, L. Hervella,
``The sixteen classes of almost Hermitian manifolds and their linear
invariants,'' Ann. Mat. Pura Appl.(4) {\bf 123} (1980) 35.}

\lref\salamon{ S. Chiossi, S. Salamon,
``The intrinsic torsion of $SU(3)$ and $G_2$ structures,''
 Proc. conf. Differential Geometry Valencia 2001.}

\lref\gauntlett{J.~P.~Gauntlett, D.~Martelli and D.~Waldram,
``Superstrings with intrinsic torsion,'' hep-th/0302158.}

\lref\guth{A.~H.~Guth,
``The Inflationary Universe: A Possible Solution To The Horizon And Flatness
Problems,''
Phys.\ Rev.\ D {\bf 23}, 347 (1981).}

\lref\lindeN{A.~D.~Linde,
``A New Inflationary Universe Scenario: A Possible Solution Of The Horizon,
Flatness, Homogeneity, Isotropy And Primordial Monopole Problems,''
Phys.\ Lett.\ B {\bf 108}, 389 (1982).}

\lref\albre{A.~Albrecht and P.~J.~Steinhardt,
``Cosmology For Grand Unified Theories With Radiatively Induced Symmetry
Breaking,''
Phys.\ Rev.\ Lett.\  {\bf 48}, 1220 (1982).}

\lref\barrowlid{A.~R.~Liddle, P.~Parsons and J.~D.~Barrow,
``Formalizing the slow roll approximation in inflation,''
Phys.\ Rev.\ D {\bf 50}, 7222 (1994), astro-ph/9408015.}

\lref\lidlit{A.~R.~Liddle and D.~H.~Lyth,
``COBE, gravitational waves, inflation and extended inflation,''
Phys.\ Lett.\ B {\bf 291}, 391 (1992), astro-ph/9208007.}

\lref\lindeC{A.~D.~Linde,
``Chaotic Inflation,''
Phys.\ Lett.\ B {\bf 129}, 177 (1983).}

\lref\gvw{S.~Gukov, C.~Vafa and E.~Witten,
``CFT's from Calabi-Yau four-folds,''
Nucl.\ Phys.\ B {\bf 584}, 69 (2000)
[Erratum-ibid.\ B {\bf 608}, 477 (2001)], hep-th/9906070.}

\lref\gukov{S.~Gukov,
``Solitons, superpotentials and calibrations,''
Nucl.\ Phys.\ B {\bf 574}, 169 (2000), hep-th/9911011.}

\lref\ach{B.~S.~Acharya and B.~Spence,
``Flux, supersymmetry and M theory on 7-manifolds,''
arXiv:hep-th/0007213.}

\lref\bw{C.~Beasley and E.~Witten,
 ``A note on fluxes and superpotentials in M-theory compactifications on
manifolds of G(2) holonomy,'' JHEP {\bf 0207}, 046 (2002),
hep-th/0203061.}

\lref\behr{K.~Behrndt and C.~Jeschek,
``Fluxes in M-theory on 7-manifolds and G structures,''
JHEP {\bf 0304}, 002 (2003), hep-th/0302047;
``Fluxes in M-theory on 7-manifolds: G-structures and
superpotential,'' hep-th/0311119.}

\lref\bertwo{K.~Behrndt and C.~Jeschek,
``Superpotentials from flux compactifications of M-theory,'', hep-th/0401019.}

\lref\gray{M. Fernandez and A. Gray, ``Riemannian manifolds with
structure group $G_2$,'' Ann. Mat. Pura. Appl. {\bf 32} (1982),
19-45.}

\lref\graytwo{M. Fernandez and L. Ugarte,
``Dolbeault cohomology for $G_2$ manifolds,''
Geom. Dedicata, {\bf 70} (1998) 57.}

\lref\ivan{T.~Friedrich and S.~Ivanov,
 ``Parallel spinors and connections with skew-symmetric torsion
  in string theory,'' math.dg/0102142;
T.~Friedrich and S.~Ivanov,
``Killing spinor equations in dimension 7 and geometry of integrable
 $G_2$-manifolds,'' math.dg/0112201.
P.~Ivanov and S.~Ivanov,
``SU(3)-instantons and $G_2$, Spin(7)-heterotic string solitons,'' math.dg/0312094.}

\lref\tp{G.~Papadopoulos and A.~A.~Tseytlin, ``Complex geometry of conifolds
and 5-brane wrapped on 2-sphere,''
Class.\ Quant.\ Grav.\  {\bf 18}, 1333 (2001).hep-th/0012034.}

\lref\lust{G.~L.~Cardoso, G.~Curio, G.~Dall'Agata, D.~Lust, P.~Manousselis and G.~Zoupanos,
``Non-Kaehler string backgrounds and their five torsion classes,''
Nucl.\ Phys.\ B {\bf 652}, 5 (2003), hep-th/0211118.}

\lref\louis{S.~Gurrieri, J.~Louis, A.~Micu and D.~Waldram,
``Mirror symmetry in generalized Calabi-Yau compactifications,''
Nucl.\ Phys.\ B {\bf 654}, 61 (2003), hep-th/0211102.}

\lref\rstrom{A.~Strominger, ``Superstrings with torsion,'' Nucl.\
Phys.\ B {\bf 274}, 253 (1986).}

\lref\mal{J.~M.~Maldacena,
``The large N limit of superconformal field theories and supergravity,''
Adv.\ Theor.\ Math.\ Phys.\  {\bf 2}, 231 (1998)
[Int.\ J.\ Theor.\ Phys.\  {\bf 38}, 1113 (1999), hep-th/9711200.}

\lref\ks{I.~R.~Klebanov and M.~J.~Strassler,
``Supergravity and a confining gauge theory: Duality cascades and
chiSB-resolution of naked singularities,'' JHEP {\bf 0008}, 052 (2000), hep-th/0007191.}

\lref\paddy{T.~Padmanabhan,
``Cosmological constant: The weight of the vacuum,''
Phys.\ Rept.\  {\bf 380}, 235 (2003), hep-th/0212290.}

\lref\carroll{S.~Carroll, M.~Trodden, ``TASI lectures: introduction to 
cosmology'', astro-ph/0401547.}

\lref\vijay{V.~Balasubramanian,
``Accelerating universes and string theory,''
Class.\ Quant.\ Grav.\  {\bf 21}, S1337 (2004), hep-th/0404075.}

\lref\banks{T.~Banks, W.~Fischler, ``M-theory observables for cosmological 
spacetimes'', hep-th/0102077 (see also ``Holographic cosmology'', 
hep-th/0405200 and references therein)}

\lref\mn{J.~M.~Maldacena and C.~Nunez,
``Towards the large N limit of pure N = 1 super Yang Mills,''
Phys.\ Rev.\ Lett.\  {\bf 86}, 588 (2001), hep-th/0008001.}

\lref\bdkt{M.~Becker, K.~Dasgupta, A.~Knauf and R.~Tatar,
``Geometric transitions, flops and non-Kaehler manifolds. I,''
Nucl.\ Phys.\ B {\bf 702}, 207 (2004), hep-th/0403288.}

\lref\work{P.~Chen, K.~Dasgupta, K.~Narayan, M.~Shmakova and M.~Zagermann,
``Work in progress.''}

\lref\motu{T.~Mohaupt and F.~Saueressig,
``Dynamical conifold transitions and moduli trapping in M-theory cosmology,''
hep-th/0410273;
A.~Lukas, E.~Palti and P.~M.~Saffin,
``Type IIB conifold transitions in cosmology,'' hep-th/0411033.}

\lref\senlater{A.~Sen,
``String network,''
JHEP {\bf 9803}, 005 (1998), hep-th/9711130.}

\lref\nakahara{M.~Nakahara,
Geometry, Topology and Physics. IOP Publication.}

\lref\schkol{J.~H.~Schwarz,
``Lectures on superstring and M theory dualities,''
Nucl.\ Phys.\ Proc.\ Suppl.\  {\bf 55B}, 1 (1997), hep-th/9607201;
O.~Aharony, A.~Hanany and B.~Kol,
``Webs of (p,q) 5-branes, five dimensional field theories and grid
diagrams,''
JHEP {\bf 9801}, 002 (1998), hep-th/9710116.}

\lref\vafai{C.~Vafa,
``Superstrings and topological strings at large N,''
J.\ Math.\ Phys.\  {\bf 42}, 2798 (2001), hep-th/0008142.}

\lref\civ{F.~Cachazo, K.~A.~Intriligator and C.~Vafa,
``A large N duality via a geometric transition,''
Nucl.\ Phys.\ B {\bf 603}, 3 (2001), hep-th/0103067.}

\lref\syz{A.~Strominger, S.~T.~Yau and E.~Zaslow,
``Mirror symmetry is T-duality,''
Nucl.\ Phys.\ B {\bf 479}, 243 (1996), hep-th/9606040.}

\lref\tduality{E.~Bergshoeff, C.~M.~Hull and T.~Ortin,
``Duality in the type II superstring effective action,''
Nucl.\ Phys.\ B {\bf 451}, 547 (1995), hep-th/9504081;
P.~Meessen and T.~Ortin,
``An Sl(2,Z) multiplet of nine-dimensional type II supergravity theories,''
Nucl.\ Phys.\ B {\bf 541}, 195 (1999), hep-th/9806120.}

\lref\eot{J.~D.~Edelstein, K.~Oh and R.~Tatar,
``Orientifold,
 geometric transition and large N duality for SO/Sp gauge  theories,''
JHEP {\bf 0105}, 009 (2001), hep-th/0104037.}

\lref\dotu{K.~Dasgupta, K.~Oh and R.~Tatar, {``Geometric
transition, large N dualities and MQCD dynamics,''} Nucl.\ Phys.\
B {\bf 610}, 331 (2001), hep-th/0105066; {``Open/closed string
dualities and Seiberg duality from geometric transitions in
M-theory,''} JHEP {\bf 0208}, 026 (2002), hep-th/0106040.}

\lref\dotd{K.~Dasgupta, K.~h.~Oh, J.~Park and R.~Tatar, ``Geometric
transition versus cascading solution,'' JHEP {\bf 0201}, 031
(2002), hep-th/0110050;
K.~Dasgupta, K.~Oh and R.~Tatar, {``Geometric
transition, large N dualities and MQCD dynamics,''} Nucl.\ Phys.\
B {\bf 610}, 331 (2001), hep-th/0105066; {``Open/closed string
dualities and Seiberg duality from geometric transitions in
M-theory,''} JHEP {\bf 0208}, 026 (2002), hep-th/0106040.}

\lref\ohta{K.~Ohta and T.~Yokono,
``Deformation of conifold and intersecting branes,''
JHEP {\bf 0002}, 023 (2000), hep-th/9912266.}

\lref\dott{K.~h.~Oh and R.~Tatar,
``Duality and confinement
in N = 1 supersymmetric theories from geometric  transitions,''
Adv.\ Theor.\ Math.\ Phys.\  {\bf 6}, 141 (2003), hep-th/0112040.}

\lref\edelstein{J.~D.~Edelstein and C.~Nunez,
``D6 branes and M-theory geometrical transitions from gauged  supergravity,''
JHEP {\bf 0104}, 028 (2001), hep-th/0103167.}

\lref\candelas{P.~Candelas and X.~C.~de la Ossa, ``Comments on
conifolds,'' Nucl.\ Phys.\ B {\bf 342}, 246 (1990).}

\lref\minasianone{R.~Minasian and D.~Tsimpis,
``Hopf reductions, fluxes and branes,''
Nucl.\ Phys.\ B {\bf 613}, 127 (2001), hep-th/0106266.}

\lref\imamura{Y.~Imamura,
``Born-Infeld action and Chern-Simons term from Kaluza-Klein monopole in
M-theory,''
Phys.\ Lett.\ B {\bf 414}, 242 (1997), hep-th/9706144;
A.~Sen,
``Dynamics of multiple Kaluza-Klein monopoles in M and string theory,''
Adv.\ Theor.\ Math.\ Phys.\  {\bf 1}, 115 (1998), hep-th/9707042;
``A note on enhanced gauge symmetries in M- and string theory,''
JHEP {\bf 9709}, 001 (1997), hep-th/9707123.}

\lref\robbins{K.~Dasgupta, G.~Rajesh, D.~Robbins and S.~Sethi,
``Time-dependent warping, fluxes, and NCYM,''
JHEP {\bf 0303}, 041 (2003), hep-th/0302049;
K.~Dasgupta and M.~Shmakova,
``On branes and oriented B-fields,''
Nucl.\ Phys.\ B {\bf 675}, 205 (2003), hep-th/0306030.}

\lref\marinaCY{M.~Shmakova,
``Calabi-Yau black holes,''
Phys.\ Rev.\ D {\bf 56}, 540 (1997), hep-th/9612076.}

\lref\senBH{A.~Sen,
``Extremal black holes and elementary string states,''
Mod.\ Phys.\ Lett.\ A {\bf 10}, 2081 (1995), hep-th/9504147.}

\lref\Atish{A.~Dabholkar, ``Exact counting of black hole microstates'',
hep-th/0409148.}

\lref\maloney{A.~Dabholkar, R.~Kallosh, A.~Maloney, ``A stringy cloak
for a classical singularity,
JHEP {\bf 0412}, 059 (2004),
 hep-th/0410076; V.~Hubeny, A.~Maloney,
M.~Rangamani, ``String-corrected black holes'', hep-th/0411272.}

\lref\dewit{G.~Lopes Cardoso, B.~de Wit, T.~Mohaupt,
``Corrections to macroscopic supersymmetric black-hole entropy,''
Phys.\ Lett.\ B {\bf 451}, 309 (1999), hep-th/9812082;
``Deviations from the area law for supersymmetric black holes,''
Fortsch.\ Phys.\  {\bf 48}, 49 (2000), hep-th/9904005;
``Area law corrections from state counting and supergravity,''
Class.\ Quant.\ Grav.\  {\bf 17}, 1007 (2000), hep-th/9910179;
G.~Lopes Cardoso, B.~de Wit, J.~Kappeli,
T.~Mohaupt,
``Stationary BPS solutions in N = 2 supergravity with R**2 interactions,''
JHEP {\bf 0012}, 019 (2000), hep-th/0009234;
G.~L.~Cardoso, B.~de Wit, J.~Kappeli and T.~Mohaupt,
``Asymptotic degeneracy of dyonic N = 4 string states and black hole
entropy,'' 
JHEP {\bf 0412}, 075 (2004),
hep-th/0412287.}

\lref\senhetbh{A.~Sen, ``How does a fundamental string stretch its
horizon ?'', 
JHEP {\bf 0505}, 059 (2005),
hep-th/0411255.}

\lref\svw{S.~Sethi, C.~Vafa and E.~Witten,
``Constraints on low-dimensional string compactifications,''
Nucl.\ Phys.\ B {\bf 480}, 213 (1996), hep-th/9606122;
K.~Dasgupta and S.~Mukhi,
``A note on low-dimensional string compactifications,''
Phys.\ Lett.\ B {\bf 398}, 285 (1997), hep-th/9612188.}

\lref\tliz{N.~Iizuka and S.~P.~Trivedi,
``An inflationary model in string theory,''
Phys.\ Rev.\ D {\bf 70}, 043519 (2004), hep-th/0403203.}

\lref\ano{A.~A.~Abrikosov,
``On The Magnetic Properties Of Superconductors Of The Second Group,''
Sov.\ Phys.\ JETP {\bf 5}, 1174 (1957)
[Zh.\ Eksp.\ Teor.\ Fiz.\  {\bf 32}, 1442 (1957)];
H.~B.~Nielsen and P.~Olesen,
``Vortex-Line Models For Dual Strings,''
Nucl.\ Phys.\ B {\bf 61}, 45 (1973).}

\lref\gunah{
M.~Gunaydin, L.~J.~Romans and N.~P.~Warner,
``Compact And Noncompact Gauged Supergravity Theories In
Five-Dimensions,''
Nucl.\ Phys.\ B {\bf 272}, 598 (1986).}

\lref\pernici{M.~Pernici, K.~Pilch and P.~van Nieuwenhuizen,
``Gauged N=8 D = 5 Supergravity,''
Nucl.\ Phys.\ B {\bf 259}, 460 (1985).}

\lref\ctheorem{D.~Z.~Freedman, S.~S.~Gubser, K.~Pilch and N.~P.~Warner,
``Renormalization group flows from holography supersymmetry and a
c-theorem,''
Adv.\ Theor.\ Math.\ Phys.\  {\bf 3}, 363 (1999), hep-th/9904017.}

\lref\bergg{M.~Berg, M.~Haack and B.~Kors,
``Loop corrections to volume moduli and inflation in string theory,''
Phys.\ Rev.\ D {\bf 71}, 026005 (2005),
hep-th/0404087;
``On the moduli dependence of nonperturbative superpotentials in brane
inflation,'' hep-th/0409282.}

\lref\mcall{L.~McAllister,
``An inflaton mass problem in string inflation from threshold corrections to volume stabilization,''
 hep-th/0502001.}

\lref\bankcrit{T.~Banks,
``Landskepticism or why effective potentials don't count string models,''
hep-th/0412129.}

\lref\Kawa{
M.~Kawasaki, M.~Yamaguchi and T.~Yanagida,
``Natural chaotic inflation in supergravity,''
Phys.\ Rev.\ Lett.\  {\bf 85}, 3572 (2000), hep-ph/0004243.}

\lref\hsushift{J.~P.~Hsu, R.~Kallosh and S.~Prokushkin,
``On brane inflation with volume stabilization,''
JCAP {\bf 0312}, 009 (2003), hep-th/0311077;
J.~P.~Hsu and R.~Kallosh,
``Volume stabilization and the origin of the inflaton shift symmetry in string
theory,''
JHEP {\bf 0404}, 042 (2004), hep-th/0402047.}

\lref\senreview{A.~Sen, "Tachyon dynamics in open string theory",
hep-th/0410103.}

\lref\senrolltach{A.~Sen,
``Rolling tachyon,''
JHEP {\bf 0204}, 048 (2002), hep-th/0203211.}

\lref\llm{N.~Lambert, H.~Liu, J.~Maldacena, "Closed strings
from decaying D-branes", hep-th/0303139.}

\lref\BraxMandalOz{P.~Brax, G.~Mandal and Y.~Oz,
``Supergravity description of non-BPS branes,''
Phys.\ Rev.\ D {\bf 63}, 064008 (2001), hep-th/0005242.}

\lref\wmap{H.~V.~Peiris {\it et al.},
``First year Wilkinson Microwave Anisotropy Probe (WMAP) observations:
Implications for inflation,''
Astrophys.\ J.\ Suppl.\  {\bf 148}, 213 (2003), astro-ph/0302225;
M.~Tegmark {\it et al.}  [SDSS Collaboration],
``Cosmological parameters from SDSS and WMAP,''
Phys.\ Rev.\ D {\bf 69}, 103501 (2004), astro-ph/0310723.}

\lref\landscape{M.~R.~Douglas,
``The statistics of string / M theory vacua,''
JHEP {\bf 0305}, 046 (2003), hep-th/0303194;
S.~Ashok and M.~R.~Douglas,
``Counting flux vacua,''
JHEP {\bf 0401}, 060 (2004), hep-th/0307049;
L.~Susskind,
``The anthropic landscape of string theory,''hep-th/0302219.}

\lref\kklmmt{S.~Kachru, R.~Kallosh, A.~Linde,
J.~Maldacena, L.~McAllister and S.~P.~Trivedi,
``Towards inflation in string theory,''
JCAP {\bf 0310}, 013 (2003), hep-th/0308055.}

\lref\kachruone{S.~Kachru, M.~B.~Schulz, P.~K.~Tripathy and S.~P.~Trivedi,
``New supersymmetric string compactifications,''
JHEP {\bf 0303}, 061 (2003), hep-th/0211182;
S.~Kachru, M.~B.~Schulz and S.~Trivedi,
``Moduli stabilization from fluxes in a simple IIB orientifold,''
JHEP {\bf 0310}, 007 (2003), hep-th/0201028.}

\lref\denef{F.~Denef, M.~R.~Douglas and B.~Florea,
``Building a better racetrack,''
JHEP {\bf 0406}, 034 (2004), hep-th/0404257.}

\lref\hash{A.~Hashimoto and S.~Sethi,
``Holography and string dynamics in time-dependent backgrounds,''
Phys.\ Rev.\ Lett.\  {\bf 89}, 261601 (2002), hep-th/0208126;
R.~G.~Cai, J.~X.~Lu and N.~Ohta,
  ``NCOS and D-branes in time-dependent backgrounds,''
  Phys.\ Lett.\ B {\bf 551}, 178 (2003), hep-th/0210206.}

\lref\savrobbins{D.~Robbins and S.~Sethi,
``A barren landscape,'' 
Phys.\ Rev.\ D {\bf 71}, 046008 (2005),
hep-th/0405011.}

\lref\berghaack{M.~Haack and J.~Louis,
``M-theory compactified on Calabi-Yau fourfolds with background flux,''
Phys.\ Lett.\ B {\bf 507} (2001) 296, hep-th/0103068;
M.~Berg, M.~Haack and H.~Samtleben,
``Calabi-Yau fourfolds with flux and supersymmetry breaking,''
JHEP {\bf 0304} (2003) 046, hep-th/0212255.}

\lref\burt{E.~I.~Buchbinder and B.~A.~Ovrut,
``Vacuum stability in heterotic M-theory,''
Phys.\ Rev.\ D {\bf 69}, 086010 (2004), hep-th/0310112;
M.~Becker, G.~Curio and A.~Krause,
``De Sitter vacua from heterotic M-theory,''
Nucl.\ Phys.\ B {\bf 693}, 223 (2004), hep-th/0403027.}

\lref\gkp{S.~B.~Giddings, S.~Kachru and J.~Polchinski,
``Hierarchies from fluxes in string compactifications,''
Phys.\ Rev.\ D {\bf 66}, 106006 (2002), hep-th/0105097.}

\lref\kklt{S.~Kachru, R.~Kallosh, A.~Linde and S.~P.~Trivedi,
``De Sitter vacua in string theory,''
Phys.\ Rev.\ D {\bf 68}, 046005 (2003), hep-th/0301240.}

\lref\hitchin{N. ~Hitchin,
``Stable forms and special metrics'',
Contemp. Math., {\bf 288}, Amer. Math. Soc. (2000).}

\lref\giveon{S.~S.~Gubser,
 ``Supersymmetry and F-theory realization of the deformed conifold with
three-form flux,'' hep-th/0010010;
A.~Giveon, A.~Kehagias and H.~Partouche,
``Geometric transitions, brane dynamics and gauge theories,''
JHEP {\bf 0112}, 021 (2001), hep-th/0110115.}

\lref\bonan{E. Bonan,
``Sur le varietes remanniennes a groupe d'holonomie $G_2$ ou Spin(7),''
C. R. Acad. Sci. paris {\bf 262} (1966) 127.}



\lref\amv{M.~Atiyah, J.~M.~Maldacena and C.~Vafa,
``An M-theory flop as a large N duality,''
J.\ Math.\ Phys.\  {\bf 42}, 3209 (2001), hep-th/0011256.}

\lref\syz{A.~Strominger, S.~T.~Yau and E.~Zaslow,
``Mirror symmetry is T-duality,''
Nucl.\ Phys.\ B {\bf 479}, 243 (1996), hep-th/9606040.}

\lref\dvhal{P.~Binetruy and G.~R.~Dvali,
``D-term inflation,''
Phys.\ Lett.\ B {\bf 388}, 241 (1996), hep-ph/9606342;
E.~Halyo,
``Hybrid inflation from supergravity D-terms,''
Phys.\ Lett.\ B {\bf 387}, 43 (1996), hep-ph/9606423.}

\lref\tduality{E.~Bergshoeff, C.~M.~Hull and T.~Ortin,
``Duality in the type II superstring effective action,''
Nucl.\ Phys.\ B {\bf 451}, 547 (1995), hep-th/9504081;
P.~Meessen and T.~Ortin,
``An Sl(2,Z) multiplet of nine-dimensional type II supergravity theories,''
Nucl.\ Phys.\ B {\bf 541}, 195 (1999), hep-th/9806120.}

\lref\adoptone{J.~D.~Edelstein, K.~Oh and R.~Tatar,
``Orientifold, geometric transition and large N duality for SO/Sp
gauge theories,'' JHEP {\bf 0105}, 009 (2001), hep-th/0104037.}

\lref\ohta{K.~Ohta and T.~Yokono,
``Deformation of conifold and intersecting branes,''
JHEP {\bf 0002}, 023 (2000), hep-th/9912266.}

\lref\adoptfo{K.~h.~Oh and R.~Tatar, ``Duality and confinement in
N = 1 supersymmetric theories from geometric transitions,'' Adv.\
Theor.\ Math.\ Phys.\  {\bf 6}, 141 (2003), hep-th/0112040.}

\lref\edelstein{J.~D.~Edelstein and C.~Nunez, ``D6 branes and
M-theory geometrical transitions from gauged supergravity,'' JHEP
{\bf 0104}, 028 (2001), hep-th/0103167.}

\lref\candelas{P.~Candelas and X.~C.~de la Ossa,
``Comments On Conifolds,''
Nucl.\ Phys.\ B {\bf 342}, 246 (1990).}

\lref\minasianone{R.~Minasian and D.~Tsimpis,
``Hopf reductions, fluxes and branes,''
Nucl.\ Phys.\ B {\bf 613}, 127 (2001), hep-th/0106266.}

\lref\pandoz{L.~A.~Pando Zayas and A.~A.~Tseytlin,
``3-branes on resolved conifold,''
JHEP {\bf 0011}, 028 (2000), hep-th/0010088.}

\lref\rBB{K.~Becker and M.~Becker, ``M-Theory on
eight-manifolds,'' Nucl.\ Phys.\ B {\bf 477}, 155 (1996),
hep-th/9605053.}

\lref\bbdgs{K.~Becker, M.~Becker, P.~S.~Green, K.~Dasgupta and
E.~Sharpe, ``Compactifications of heterotic strings on
non-K\"ahler complex manifolds. II,'' Nucl.\ Phys.\ B {\bf 678},
19 (2004), hep-th/0310058.}

\lref\bbdg{K.~Becker, M.~Becker, K.~Dasgupta and P.~S.~Green,
``Compactifications of heterotic theory on non-K\"ahler complex
manifolds. I,'' JHEP {\bf 0304}, 007 (2003), hep-th/0301161.}

\lref\GP{E.~Goldstein and S.~Prokushkin, ``Geometric model for
complex non-K\"ahler manifolds with SU(3) structure,''
hep-th/0212307.}

\lref\bbdp{K.~Becker, M.~Becker, K.~Dasgupta and S.~Prokushkin,
``Properties of heterotic vacua from superpotentials,''
Nucl.\ Phys.\ B {\bf 666}, 144 (2003), hep-th/0304001.}

\lref\townsend{P.~K.~Townsend,
``D-branes from M-branes,''
Phys.\ Lett.\ B {\bf 373}, 68 (1996), hep-th/9512062.}

\lref\chodos{A.~Chodos and S.~Detweiler,
``Where Has The Fifth-Dimension Gone?,''
Phys.\ Rev.\ D {\bf 21}, 2167 (1980);
A.~Chodos and S.~Detweiler,
``Spherically Symmetric Solutions In Five-Dimensional General Relativity,''
Gen.\ Rel.\ Grav.\  {\bf 14}, 879 (1982);
P.~G.~O.~Freund,
``Kaluza-Klein Cosmologies,''
Nucl.\ Phys.\ B {\bf 209}, 146 (1982);
P.~G.~O.~Freund and M.~A.~Rubin,
``Dynamics Of Dimensional Reduction,''
Phys.\ Lett.\ B {\bf 97}, 233 (1980);
T.~Appelquist and A.~Chodos,
``The Quantum Dynamics Of Kaluza-Klein Theories,''
Phys.\ Rev.\ D {\bf 28}, 772 (1983);
S.~Randjbar-Daemi, A.~Salam and J.~Strathdee,
``On Kaluza-Klein Cosmology,''
Phys.\ Lett.\ B {\bf 135}, 388 (1984);
W.~J.~Marciano,
``Time Variation Of The Fundamental 'Constants' And Kaluza-Klein Theories,''
Phys.\ Rev.\ Lett.\  {\bf 52}, 489 (1984);
R.~B.~Abbott, S.~M.~Barr and S.~D.~Ellis,
``Kaluza-Klein Cosmologies And Inflation,''
Phys.\ Rev.\ D {\bf 30}, 720 (1984);
E.~W.~Kolb, M.~J.~Perry and T.~P.~Walker,
``Time Variation Of Fundamental Constants, Primordial Nucleosynthesis And The
Size Of Extra Dimensions,''
Phys.\ Rev.\ D {\bf 33}, 869 (1986);
D.~La and P.~J.~Steinhardt,
``Extended Inflationary Cosmology,''
Phys.\ Rev.\ Lett.\  {\bf 62}, 376 (1989)
[Erratum-ibid.\  {\bf 62}, 1066 (1989)];
V.~A.~Kostelecky and S.~Samuel,
``Gravitational Phenomenology In Higher Dimensional Theories And Strings,''
Phys.\ Rev.\ D {\bf 40}, 1886 (1989).}

\lref\buwal{K.~Behrndt and S.~Forste,
``String Kaluza-Klein cosmology,''
Nucl.\ Phys.\ B {\bf 430}, 441 (1994), hep-th/9403179;
A.~Lukas, B.~A.~Ovrut and D.~Waldram,
``Cosmological solutions of type II string theory,''
Phys.\ Lett.\ B {\bf 393}, 65 (1997), hep-th/9608195;
R.~Poppe and S.~Schwager,
``String Kaluza-Klein cosmologies with RR-fields,''
Phys.\ Lett.\ B {\bf 393}, 51 (1997), hep-th/9610166;
``String and M-theory cosmological solutions with Ramond forms,''
Nucl.\ Phys.\ B {\bf 495}, 365 (1997), hep-th/9610238;
F.~Larsen and F.~Wilczek,
``Resolution of cosmological singularities,''
Phys.\ Rev.\ D {\bf 55}, 4591 (1997), hep-th/9610252;
J.~E.~Lidsey, D.~Wands and E.~J.~Copeland,
``Superstring cosmology,''
Phys.\ Rept.\  {\bf 337}, 343 (2000), hep-th/9909061.}

\lref\kasner{E.~Kasner, ``Geometrical theorems on Einstein's cosmological equation"
Am. J. Math. 43 (1921) 217;
P.~Jordan,
``The Present State Of Dirac's Cosmological Hypothesis,''
Z.\ Phys.\  {\bf 157}, 112 (1959);
C.~Brans and .~H.~Dicke,
``Mach's Principle And A Relativistic Theory Of Gravitation,''
Phys.\ Rev.\  {\bf 124}, 925 (1961).}

\lref\nogo{G.~W.~Gibbons,
``Aspects Of Supergravity Theories,''in {\it Supersymmetry, Supergravity and Related Topics}, eds. F. del Aguila, J. A. de Azcarraga and
L. E. Ibanez (World Scientific 1985) 344;
B.~de Wit, D.~J.~Smit and N.~D.~Hari Dass,
  ``Residual Supersymmetry Of Compactified D = 10 Supergravity,''
  Nucl.\ Phys.\ B {\bf 283}, 165 (1987);
J.~M.~Maldacena and C.~Nunez,
``Supergravity description of field theories on curved manifolds and a no  go
theorem,''
Int.\ J.\ Mod.\ Phys.\ A {\bf 16}, 822 (2001), hep-th/0007018.}

\lref\wittenflux{E.~Witten,
``On flux quantization in M-theory and the effective action,''
J.\ Geom.\ Phys.\  {\bf 22}, 1 (1997), hep-th/9609122.}

\lref\sav{K.~Dasgupta, G.~Rajesh and S.~Sethi,
``M theory, orientifolds and G-flux,''
JHEP {\bf 9908}, 023 (1999), hep-th/9908088.}

\lref\beckerD{K.~Becker and K.~Dasgupta,
``Heterotic strings with torsion,''
JHEP {\bf 0211}, 006 (2002), hep-th/0209077.}

\lref\ks{I.~R.~Klebanov and M.~J.~Strassler,
 ``Supergravity and a confining gauge theory: Duality cascades
  and $\chi_{SB}$-resolution of naked singularities,''
JHEP {\bf 0008}, 052 (2000), hep-th/0007191.}

\lref\gv{R.~Gopakumar and C.~Vafa,
``On the gauge theory/geometry correspondence,''
Adv.\ Theor.\ Math.\ Phys.\  {\bf 3}, 1415 (1999),hep-th/9811131.}

\lref\dvu{R.~Dijkgraaf and C.~Vafa,
``Matrix models, topological strings, and supersymmetric gauge theories,''
Nucl.\ Phys.\ B {\bf 644}, 3 (2002), hep-th/0206255.}

\lref\ps{J.~Polchinski and M.~J.~Strassler, ``The String Dual of a
Confining Four-Dimensional Gauge Theory ,'' hep-th/0003136.}

\lref\matsuda{T.~Matsuda,
  ``Brane necklaces and brane coils,'' 
JHEP {\bf 0505}, 015 (2005),
hep-ph/0412290;
``Formation of monopoles and domain walls after brane inflation,''
  JHEP {\bf 0410}, 042 (2004), hep-ph/0406064;
``String production after angled brane inflation,''
  Phys.\ Rev.\ D {\bf 70}, 023502 (2004), hep-ph/0403092;
``Formation of cosmological brane defects,''
  JHEP {\bf 0411}, 039 (2004), hep-ph/0402232;
``Brane Q-ball, branonium and brane Q-ball inflation,''
  JCAP {\bf 0410}, 014 (2004), hep-ph/0402223.}

\lref\drmknmrp{David R.~Morrison, K.~Narayan, M.~Ronen Plesser,
``Localized tachyons in $C^3/Z_N$,''
JHEP {\bf 0408}, 047 (2004),
 hep-th/0406039.}

\lref\drmkn{David R.~Morrison, K.~Narayan, ``On tachyons, gauged linear 
sigma models, and flip transitions'', 
JHEP {\bf 0502}, 062 (2005),
hep-th/0412337.}

\lref\aps{A.~Adams, J.~Polchinski, E.~Silverstein,
``Don't panic! Closed string tachyons in ALE spacetimes,'' 
JHEP {\bf 0110}, 029 (2001),
hep-th/0108075.}

\lref\hkmm{J.~Harvey, D.~Kutasov, E.~Martinec, G.~Moore,
``Localized tachyons and RG flows,'' hep-th/0111154.}

\lref\slanskie{R.~Slansky,
  ``Group Theory For Unified Model Building,''
  Phys.\ Rept.\  {\bf 79}, 1 (1981).}

\lref\vafatach{C.~Vafa,
``Mirror symmetry and closed string tachyon condensation,'' hep-th/0111051.}

\lref\evatach{E.~Silverstein,
``Counter-intuition and scalar masses,'' hep-th/0407202.}

\lref\dvd{R.~Dijkgraaf and C.~Vafa,
``A perturbative window into non-perturbative physics,'' hep-th/0208048.}

\lref\civu{F.~Cachazo, S.~Katz and C.~Vafa,
``Geometric transitions and N = 1 quiver theories,'' hep-th/0108120.}

\lref\civd{F.~Cachazo, B.~Fiol, K.~A.~Intriligator, S.~Katz and C.~Vafa,
``A geometric unification of dualities,'' Nucl.\ Phys.\ B {\bf 628}, 3 (2002),
hep-th/0110028.}

\lref\radu{R.~Roiban, R.~Tatar and J.~Walcher,
``Massless flavor in geometry and matrix models,''
Nucl.\ Phys.\ B {\bf 665}, 211 (2003), hep-th/0301217.}

\lref\radd{K.~Landsteiner, C.~I.~Lazaroiu and R.~Tatar,
``(Anti)symmetric matter and superpotentials from IIB orientifolds,''
JHEP {\bf 0311}, 044 (2003), hep-th/0306236.}

\lref\radt{K.~Landsteiner, C.~I.~Lazaroiu and R.~Tatar,
``Chiral field theories from conifolds,''
JHEP {\bf 0311}, 057 (2003), hep-th/0310052.}

\lref\radp{K.~Landsteiner, C.~I.~Lazaroiu and R.~Tatar,
``Puzzles for matrix models of chiral field theories,'' hep-th/0311103.}

\lref\ber{M.~Bershadsky, S.~Cecotti, H.~Ooguri and C.~Vafa,
 ``Kodaira-Spencer theory of gravity and exact results for quantum string amplitudes,''
Commun.\ Math.\ Phys.\  {\bf 165}, 311 (1994), hep-th/9309140.}

\lref\bismut{J. M. Bismut,
``A local index theorem for non-K\"ahler manifolds,''
Math. Ann. {\bf 284} (1989) 681.}

\lref\monar{F. Cabrera, M. Monar and A. Swann,
``Classification of $G_2$ structures,''
J. London Math. Soc. {\bf 53} (1996) 98;
F. Cabrera,
``On Riemannian manifolds with $G_2$ structure,''
Bolletino UMI A {\bf 10} (1996) 98.}

\lref\kath{Th. Friedrich and I. Kath,
``7-dimensional compact Riemannian manifolds with killing spinors,''
Comm. Math. Phys. {\bf 133} (1990) 543;
Th. Friedrich, I. Kath, A. Moroianu and U. Semmelmann,
``On nearly parallel $G_2$ structures,'' J. geom. Phys. {\bf 23} (1997) 259;
S. Salamon,
``Riemannian geometry and holonomy groups,''
Pitman Res. Notes Math. Ser., {\bf 201} (1989);
V. Apostolov and S. Salamon,
``K\"ahler reduction of metrics with holonomy $G_2$,''
math-DG/0303197.}

\lref\ooguri{H.~Ooguri and C.~Vafa,
``The C-deformation of gluino and non-planar diagrams,''
Adv.\ Theor.\ Math.\ Phys.\  {\bf 7}, 53 (2003), hep-th/0302109;
``Gravity induced C-deformation,''
Adv.\ Theor.\ Math.\ Phys.\  {\bf 7}, 405 (2004), hep-th/0303063.}

\lref\tv{T.~R.~Taylor and C.~Vafa,
``RR flux on Calabi-Yau and partial supersymmetry breaking,''
Phys.\ Lett.\ B {\bf 474}, 130 (2000), hep-th/9912152.}

\lref\nekra{N.~Nekrasov,~H.~Ooguri and C.~Vafa, `S-duality and
Topological Strings,'' hep-th/0403167.}

\lref\lustu{G.~L.~Cardoso, G.~Curio, G.~Dall'Agata and D.~Lust,
``BPS action and superpotential for heterotic string compactifications  with
fluxes,'' JHEP {\bf 0310}, 004 (2003),hep-th/0306088.}

\lref\lustd{G.~L.~Cardoso, G.~Curio, G.~Dall'Agata and D.~Lust,
``Heterotic string theory on non-K\"ahler manifolds with H-flux
and gaugino condensate,'' hep-th/0310021.}

\lref\bd{M.~Becker and K.~Dasgupta, ``K\"ahler versus non-K\"ahler
compactifications,'' hep-th/0312221.}

\lref\micu{S.~Gurrieri and A.~Micu,
``Type IIB theory on half-flat manifolds,''
Class.\ Quant.\ Grav.\  {\bf 20}, 2181 (2003), hep-th/0212278.}

\lref\dal{G.~Dall'Agata and N.~Prezas,
``N = 1 geometries for M-theory and type IIA strings with fluxes,''
hep-th/0311146.}

\lref\douo{M.~R.~Douglas,``The statistics of string / M theory vacua,''
JHEP {\bf 0305}, 046 (2003), hep-th/0303194.}

\lref\wittenchern{E.~Witten,``Chern-Simons gauge theory as a
string theory,'' hep-th/9207094.}

\lref\ms{D.~Martelli and J.~Sparks,``G Structures, fluxes and
calibrations in M Theory,'' hep-th/0306225.}

\lref\fg{A.~F.~Frey and A.~Grana,``Type IIB solutions with
interpolating supersymetries ,'' hep-th/0307142.}

\lref\bbs{K.~Becker, M.~Becker and R.~Sriharsha,``PP-waves,
M-theory and fluxes,'' hep-th/0308014.}

\lref\minu{P. Kaste,~R. Minasian,~A. Tomasiello,''Supersymmetric M theory
Compactifications with
Fluxes on Seven-Manifolds and G Structures'', JHEP {\bf 0307} 004, 2003,
hep-th/0303127.}

 \lref\mind{S. Fidanza,~R. Minasian,~A. Tomasiello,''Mirror Symmetric
SU(3) Structure Manifolds with NS fluxes'',
hep-th/0311122.}

\lref\beru{K.~Behrndt and M.~Cvetic, ``General N=1 Supersymmetric Flux
Vacua of
(Massive) Type IIA String Theory'', hep-th/0403049.}

\lref\dalu{G.~Dall'Agata, ``On Supersymmetric Solutions of Type IIB
Supergravity with General Fluxes'', hep-th/0403220.}

\lref\kleb{I.~R.~Klebanov and M.~J.~Strassler,
``Supergravity and a confining gauge theory: Duality cascades and
$\chi$ - SB-resolution of naked singularities,''
JHEP {\bf 0008}, 052 (2000), hep-th/0007191.}

\lref\realm{S.~Alexander, K.~Becker, M.~Becker, K.~Dasgupta, A.~Knauf, R.~Tatar,
``In the realm of the geometric transitions,''
Nucl.\ Phys.\ B {\bf 704}, 212 (2005), hep-th/0408192.}

\lref\randall{L.~Randall, M.~Soljacic and A.~Guth,
``Supernatural Inflation: Inflation from Supersymmetry with No (Very)
Small Parameters,''
Nucl.\ Phys.\ B {\bf 472}, 377 (1996), hep-ph/9512439, hep-ph/9601296.}

\lref\panang{G.~Lazarides and C.~Panangiotakopoulos,
``Smooth hybrid inflation,''
Phys.\ Rev.\ D {\bf 52}, 559 (1995), hep-ph/9506325;
G.~Lazarides and Q.~Shafi,
``Topological defects and inflation,''
Phys.\ Lett.\ B {\bf 372}, 20 (1996), hep-ph/9510275.}

\lref\lindehy{A.~D.~Linde,
``Hybrid inflation,''
Phys.\ Rev.\ D {\bf 49}, 748 (1994), astro-ph/9307002.}

\lref\ori{S.~Chakravarty, K.~Dasgupta, O.~J.~Ganor and G.~Rajesh,
``Pinned branes and new non Lorentz invariant theories,''
Nucl.\ Phys.\ B {\bf 587}, 228 (2000), hep-th/0002175.}

\lref\marina{K.~Dasgupta and M.~Shmakova,
``On branes and oriented B-fields,''
Nucl.\ Phys.\ B {\bf 675}, 205 (2003), hep-th/0306030.}

\lref\gross{H.~Kodama, M.~Sasaki and K.~Sato,
``Abundance Of Primordial Holes Produced By Cosmological First Order Phase
Transition,''
Prog.\ Theor.\ Phys.\  {\bf 68}, 1979 (1982);
D.~J.~Gross, M.~J.~Perry and L.~G.~Yaffe,
``Instability Of Flat Space At Finite Temperature,''
Phys.\ Rev.\ D {\bf 25}, 330 (1982).}

\lref\liddle{J.~D.~Barrow, E.~J.~Copeland, E.~W.~Kolb and A.~R.~Liddle,
``Baryogenesis In Extended Inflation. 2. Baryogenesis Via Primordial Black
Holes,''
Phys.\ Rev.\ D {\bf 43}, 984 (1991).}

\lref\greenliddle{A.~M.~Green and A.~R.~Liddle,
``Constraints on the density perturbation spectrum from primordial black
holes,''
Phys.\ Rev.\ D {\bf 56}, 6166 (1997), astro-ph/9704251.}

\lref\glid{A.~M.~Green, A.~R.~Liddle and A.~Riotto,
``Primordial black hole constraints in cosmologies with early matter
domination,''
Phys.\ Rev.\ D {\bf 56}, 7559 (1997), astro-ph/9705166;
A.~R.~Liddle and A.~M.~Green,
``Cosmological constraints from primordial black holes,''
Phys.\ Rept.\  {\bf 307}, 125 (1998), gr-qc/9804034;
``Primordial black holes and early cosmology,'' astro-ph/9710235.}

\lref\liddletwo{
J.~D.~Barrow, E.~J.~Copeland and A.~R.~Liddle,
``The Cosmology of black hole relics,''
Phys.\ Rev.\ D {\bf 46}, 645 (1992).}

\lref\carrtwo{B.~J.~Carr and J.~H.~MacGibbon,
``Cosmic rays from primordial black holes and constraints on the early
universe,''
Phys.\ Rept.\  {\bf 307}, 141 (1998).}

\lref\crawford{M.~Crawford and D.~N.~Schramm,
``Spontaneous Generation Of Density Perturbations In The Early Universe,''
Nature {\bf 298}, 538 (1982).}

\lref\lyth{D.~H.~Lyth and E.~D.~Stewart,
``Cosmology with a TeV mass GUT Higgs,''
Phys.\ Rev.\ Lett.\  {\bf 75}, 201 (1995), hep-ph/9502417;
``Thermal inflation and the moduli problem,''
Phys.\ Rev.\ D {\bf 53}, 1784 (1996), hep-ph/9510204.}

\lref\moduli{G.~D.~Coughlan, W.~Fischler, E.~W.~Kolb, S.~Raby and G.~G.~Ross,
``Cosmological Problems For The Polonyi Potential,''
Phys.\ Lett.\ B {\bf 131}, 59 (1983);
J.~R.~Ellis, D.~V.~Nanopoulos and M.~Quiros,
``On The Axion, Dilaton, Polonyi, Gravitino And Shadow Matter Problems In
Supergravity And Superstring Models,''
Phys.\ Lett.\ B {\bf 174}, 176 (1986);
B.~de Carlos, J.~A.~Casas, F.~Quevedo and E.~Roulet,
``Model independent properties and cosmological implications of the dilaton
and moduli sectors of 4-d strings,''
Phys.\ Lett.\ B {\bf 318}, 447 (1993), hep-ph/9308325.}

\lref\lisa{T.~Banks, D.~B.~Kaplan and A.~E.~Nelson,
Phys.\ Rev.\ D {\bf 49}, 779 (1994), hep-ph/9308292;
L.~Randall and S.~Thomas,
``Solving the cosmological moduli problem with weak scale inflation,''
Nucl.\ Phys.\ B {\bf 449}, 229 (1995), hep-ph/9407248;
M.~Dine, L.~Randall and S.~Thomas,
``Baryogenesis from flat directions of the supersymmetric standard model,''
Nucl.\ Phys.\ B {\bf 458}, 291 (1996), hep-ph/9507453.}

\lref\dsrun{M.~Dine and N.~Seiberg,
``Couplings And Scales In Superstring Models,''
Phys.\ Rev.\ Lett.\  {\bf 55}, 366 (1985).}

\lref\carr{B.~J.~Carr,
``The Primordial Black Hole Mass Spectrum,''
Astrophys.\ J.\  {\bf 201}, 1 (1975).
B.~J.~Carr, J.~H.~Gilbert and J.~E.~Lidsey,
``Black hole relics and inflation: Limits on blue perturbation spectra,''
Phys.\ Rev.\ D {\bf 50}, 4853 (1994), astro-ph/9405027;
``Black holes from blue spectra,''
arXiv:astro-ph/9406028;
``Primordial black holes and the spectral index,''
Nucl.\ Phys.\ Proc.\ Suppl.\  {\bf 43}, 75 (1995).}

\lref\hanany{A.~Hanany and A.~Zaffaroni,
``On the realization of chiral four-dimensional gauge theories using  branes,''
JHEP {\bf 9805}, 001 (1998), hep-th/9801134.}

\lref\dtds{K.~Dasgupta, C.~Herdeiro, S.~Hirano and R.~Kallosh,
``D3/D7 inflationary model and M-theory,''
Phys.\ Rev.\ D {\bf 65}, 126002 (2002), hep-th/0203019.}

\lref\karch{M.~Aganagic, A.~Karch, D.~Lust and A.~Miemiec,
``Mirror symmetries for brane configurations and branes at singularities,''
Nucl.\ Phys.\ B {\bf 569}, 277 (2000), hep-th/9903093.}

\lref\linrio{A.~D.~Linde and A.~Riotto,
``Hybrid inflation in supergravity,''
Phys.\ Rev.\ D {\bf 56}, 1841 (1997), hep-ph/9703209.}

\lref\henry{G.~R.~Dvali and S.~H.~H.~Tye,
``Brane inflation,''
Phys.\ Lett.\ B {\bf 450}, 72 (1999), hep-ph/9812483.}

\lref\strinf{
S.~H.~S.~Alexander,
``Inflation from D - anti-D brane annihilation,''
Phys.\ Rev.\ D {\bf 65}, 023507 (2002), hep-th/0105032;
C.~P.~Burgess, M.~Majumdar, D.~Nolte, F.~Quevedo, G.~Rajesh and R.~J.~Zhang,
``The inflationary brane-antibrane universe,''
JHEP {\bf 0107}, 047 (2001), hep-th/0105204;
C.~P.~Burgess, P.~Martineau, F.~Quevedo, G.~Rajesh and R.~J.~Zhang,
``Brane antibrane inflation in orbifold and orientifold models,''
JHEP {\bf 0203}, 052 (2002), hep-th/0111025.}

\lref\shandera{S.~E.~Shandera,
``Slow roll in brane inflation,'' 
JCAP {\bf 0504}, 011 (2005),
hep-th/0412077.}

\lref\shafi{
B.~s.~Kyae and Q.~Shafi,
``Branes and inflationary cosmology,''
Phys.\ Lett.\ B {\bf 526}, 379 (2002), hep-ph/0111101;
J.~Garcia-Bellido, R.~Rabadan and F.~Zamora,
``Inflationary scenarios from branes at angles,''
JHEP {\bf 0201}, 036 (2002), hep-th/0112147;
R.~Blumenhagen, B.~Kors, D.~Lust and T.~Ott,
``Hybrid inflation in intersecting brane worlds,''
Nucl.\ Phys.\ B {\bf 641}, 235 (2002), hep-th/0202124.}

\lref\carlos{C.~Herdeiro, S.~Hirano and R.~Kallosh,
``String theory and hybrid inflation / acceleration,''
JHEP {\bf 0112}, 027 (2001), hep-th/0110271.}

\lref\dmukhi{K.~Dasgupta and S.~Mukhi,
``Brane constructions, fractional branes and anti-de Sitter domain walls,''
JHEP {\bf 9907}, 008 (1999), hep-th/9904131;
``Brane constructions, conifolds and M-theory,''
Nucl.\ Phys.\ B {\bf 551}, 204 (1999), hep-th/9811139.}

\lref\highd{A.~A.~Tseytlin,
``$R^4$ terms in 11 dimensions and conformal anomaly of (2,0) theory,''
Nucl.\ Phys.\ B {\bf 584}, 233 (2000), hep-th/0005072;
A.~Strominger,
``Loop corrections to the universal hypermultiplet,''
Phys.\ Lett.\ B {\bf 421}, 139 (1998), hep-th/9706195;
S.~Deser and D.~Seminara,
`Counterterms/M-theory corrections to D = 11 supergravity,''
Phys.\ Rev.\ Lett.\  {\bf 82}, 2435 (1999), hep-th/9812136.}

\lref\swmap{N.~Seiberg and E.~Witten,
``String theory and noncommutative geometry,''
JHEP {\bf 9909}, 032 (1999), hep-th/9908142.}

\lref\pisin{P.~Chen and R.~J.~Adler,
``Black hole remnants and dark matter,''
Nucl.\ Phys.\ Proc.\ Suppl.\  {\bf 124}, 103 (2003), gr-qc/0205106;
P.~Chen,
``Might Dark Matter be Actually Black?,'' astro-ph/0303349;
``Generalized uncertainty principle and dark matter,'' astro-ph/0305025;
``Inflation induced Planck-size black hole remnants as dark matter,'' 
New Astron.\ Rev.\  {\bf 49}, 233 (2005),
astro-ph/0406514.}

\lref\maeda{K.~i.~Maeda and N.~Ohta,
``Inflation from M-theory with fourth-order corrections and large extra
dimensions,''
Phys.\ Lett.\ B {\bf 597}, 400 (2004), hep-th/0405205;
``Inflation from superstring / M theory compactification with higher order
corrections. I,'' 
Phys.\ Rev.\ D {\bf 71}, 063520 (2005),
hep-th/0411093.}

\lref\gutper{M.~Gutperle and A.~Strominger,
``Spacelike branes,''
JHEP {\bf 0204}, 018 (2002), hep-th/0202210;
C.~M.~Chen, D.~V.~Gal'tsov and M.~Gutperle,
``S-brane solutions in supergravity theories,''
Phys.\ Rev.\ D {\bf 66}, 024043 (2002), hep-th/0204071;
N.~Ohta,
 ``Intersection rules for S-branes,''
  Phys.\ Lett.\ B {\bf 558}, 213 (2003), hep-th/0301095.}

\lref\gkal{M.~Gutperle, R.~Kallosh and A.~Linde,
``M / string theory, S-branes and accelerating universe,''
JCAP {\bf 0307}, 001 (2003), hep-th/0304225.}

\lref\coupling{S.~Sarangi, S.~H.~H.~Tye, ``Cosmic string production
towards the end of brane inflation'',
Phys.\ Lett.\ B {\bf 536}, 185 (2002), hep-th/0204074;
N.~T.~Jones, H.~Stoica and S.~H.~H.~Tye,
``The production, spectrum and evolution of cosmic strings in brane
inflation,''
Phys.\ Lett.\ B {\bf 563}, 6 (2003), hep-th/0303269;
G.~Dvali and A.~Vilenkin,
``Formation and evolution of cosmic D-strings,''
JCAP {\bf 0403}, 010 (2004), jhep-th/0312007;
G.~Dvali, R.~Kallosh and A.~Van Proeyen,
``D-term strings,''
JHEP {\bf 0401}, 035 (2004), hep-th/0312005;
E.~Halyo,
``Cosmic D-term strings as wrapped D3 branes,''
JHEP {\bf 0403}, 047 (2004), hep-th/0312268;
E.~J.~Copeland, R.~C.~Myers and J.~Polchinski,
``Cosmic F- and D-strings,''
JHEP {\bf 0406}, 013 (2004), hep-th/0312067;
M.~G.~Jackson, N.~T.~Jones and J.~Polchinski,
``Collisions of cosmic F- and D-strings,'' hep-th/0405229;
J.~Polchinski,
``Cosmic superstrings revisited,'' 
AIP Conf.\ Proc.\  {\bf 743}, 331 (2005),
hep-th/0410082;
``Introduction to cosmic F- and D-strings,'' hep-th/0412244;
N.~Barnaby and J.~M.~Cline,
``Tachyon defect formation and reheating in brane-antibrane inflation,''
Int.\ J.\ Mod.\ Phys.\ A {\bf 19}, 5455 (2004),
hep-th/0410030;
N.~Barnaby, A.~Berndsen, J.~M.~Cline and H.~Stoica,
``Overproduction of cosmic superstrings,'' 
JHEP {\bf 0506}, 075 (2005),
hep-th/0412095.}

\lref\gutparuli{N.~Ohta,
``Accelerating cosmologies from S-branes,''
Phys.\ Rev.\ Lett.\  {\bf 91}, 061303 (2003), hep-th/0303238;
``A study of accelerating cosmologies from superstring / M theories,''
Prog.\ Theor.\ Phys.\  {\bf 110}, 269 (2003), hep-th/0304172;
``Accelerating cosmologies and inflation from M / superstring theories,''
hep-th/0411230.}

\lref\wohl{M.~N.~R.~Wohlfarth,
``Accelerating cosmologies and a phase transition in M-theory,''
Phys.\ Lett.\ B {\bf 563}, 1 (2003), hep-th/0304089;
``Inflationary cosmologies from compactification?,''
Phys.\ Rev.\ D {\bf 69}, 066002 (2004), hep-th/0307179.}

\lref\teo{E.~Teo,
  ``A no-go theorem for accelerating cosmologies from M-theory
  compactifications,'' 
Phys.\ Lett.\ B {\bf 609}, 181 (2005),
hep-th/0412164.}

\lref\lowaction{D.~Lust, P.~Mayr, R.~Richter and S.~Stieberger,
  ``Scattering of gauge, matter, and moduli fields from intersecting branes,''
  Nucl.\ Phys.\ B {\bf 696}, 205 (2004), hep-th/0404134;
D.~Lust, S.~Reffert and S.~Stieberger,
  ``Flux-induced soft supersymmetry breaking in chiral type IIb orientifolds
  with D3/D7-branes,''
  Nucl.\ Phys.\ B {\bf 706}, 3 (2005), hep-th/0406092;
H.~Jockers and J.~Louis,
  ``The effective action of D7-branes in N = 1 Calabi-Yau orientifolds,''
  Nucl.\ Phys.\ B {\bf 705}, 167 (2005), hep-th/0409098;
D.~Lust, S.~Reffert and S.~Stieberger,
  ``MSSM with soft SUSY breaking terms from D7-branes with fluxes,'' hep-th/0410074;
P.~G.~Camara, L.~E.~Ibanez and A.~M.~Uranga,
  ``Flux-induced SUSY-breaking soft terms,''
  Nucl.\ Phys.\ B {\bf 689}, 195 (2004), hep-th/0311241;
A.~Font and L.~E.~Ibanez,
  ``SUSY-breaking soft terms in a MSSM magnetized D7-brane model,'' hep-th/0412150;
D.~Lust, P.~Mayr, S.~Reffert and S.~Stieberger,
  ``F-theory flux, destabilization of orientifolds and soft terms on
  D7-branes,'' hep-th/0501139;
H.~Jockers and J.~Louis,
  ``D-terms and F-terms from D7-brane fluxes,'' hep-th/0502059.}

\lref\ceng{C.~M.~Chen, P.~M.~Ho, I.~P.~Neupane and J.~E.~Wang,
``A note on acceleration from product space compactification,''
JHEP {\bf 0307}, 017 (2003), hep-th/0304177;
C.~M.~Chen, P.~M.~Ho, I.~P.~Neupane, N.~Ohta and J.~E.~Wang,
``Hyperbolic space cosmologies,''
JHEP {\bf 0310}, 058 (2003), hep-th/0306291.}

\lref\roy{S.~Roy,
``Accelerating cosmologies from M/string theory compactifications,''
Phys.\ Lett.\ B {\bf 567}, 322 (2003), hep-th/0304084.}

\lref\cornalba{L.~Cornalba and M.~S.~Costa,
``A new cosmological scenario in string theory,''
Phys.\ Rev.\ D {\bf 66}, 066001 (2002), hep-th/0203031;
L.~Cornalba, M.~S.~Costa and C.~Kounnas,
``A resolution of the cosmological singularity with orientifolds,''
Nucl.\ Phys.\ B {\bf 637}, 378 (2002), hep-th/0204261.}

\lref\emparan{R.~Emparan and J.~Garriga,
``A note on accelerating cosmologies from compactifications and S-branes,''
JHEP {\bf 0305}, 028 (2003), hep-th/0304124.}

\lref\teo{E.~Teo,
``A no-go theorem for accelerating cosmologies from M-theory
compactifications,'' hep-th/0412164.}

\lref\neupane{I.~P.~Neupane,
 ``Accelerating cosmologies from exponential potentials,''
  Class.\ Quant.\ Grav.\  {\bf 21}, 4383 (2004), hep-th/0311071;
I.~P.~Neupane and D.~L.~Wiltshire,
  ``Accelerating cosmologies from compactification with a twist,'' hep-th/0502003.}

\lref\town{P.~K.~Townsend and M.~N.~R.~Wohlfarth,
``Accelerating cosmologies from compactification,''
Phys.\ Rev.\ Lett.\  {\bf 91}, 061302 (2003), hep-th/0303097.}

\lref\jarv{L.~Jarv, T.~Mohaupt and F.~Saueressig,
``M-theory cosmologies from singular Calabi-Yau compactifications,''
JCAP {\bf 0402}, 012 (2004), hep-th/0310174;
``Quintessence cosmologies with a double exponential potential,''
JCAP {\bf 0408}, 016 (2004), hep-th/0403063.}

\lref\renand{J.~J.~Blanco-Pillado,  C.P. Burgess, J.M. Cline, C. Escoda, M. Gomez-Reino, R. Kallosh, A. Linde, F. Quevedo, 
``Racetrack inflation,''
JHEP {\bf 0411}, 063 (2004), hep-th/0406230.}

\lref\gins{P.~H.~Ginsparg and M.~J.~Perry,
``Semiclassical Perdurance Of De Sitter Space,''
Nucl.\ Phys.\ B {\bf 222}, 245 (1983);
G.~W.~Gibbons, S.~W.~Hawking and M.~J.~Perry,
``Path Integrals And The Indefiniteness Of The Gravitational Action,''
Nucl.\ Phys.\ B {\bf 138}, 141 (1978);
G.~W.~Gibbons and M.~J.~Perry,
``Quantizing Gravitational Instantons,''
Nucl.\ Phys.\ B {\bf 146}, 90 (1978);
S.~M.~Christensen and M.~J.~Duff,
``Flat Space As A Gravitational Instanton,''
Nucl.\ Phys.\ B {\bf 146}, 11 (1978).
M.~S.~Volkov and A.~Wipf,
``Black hole pair creation in de Sitter space: A complete one-loop
analysis,''
Nucl.\ Phys.\ B {\bf 582}, 313 (2000), hep-th/0003081.}

\lref\qued{F.~Quevedo,
``Lectures on string / brane cosmology,''
Class.\ Quant.\ Grav.\  {\bf 19}, 5721 (2002), hep-th/0210292.}

\lref\kalth{R.~Kallosh,
``Deformation, non-commutativity and the cosmological constant problem,'' hep-th/0405246.}

\lref\ren{C.~P.~Burgess, R.~Kallosh and F.~Quevedo,
``de Sitter string vacua from supersymmetric D-terms,''
JHEP {\bf 0310}, 056 (2003), hep-th/0309187}

\lref\bsv{M.~Bershadsky, C.~Vafa and V.~Sadov,
``D-Strings on D-Manifolds,''
Nucl.\ Phys.\ B {\bf 463}, 398 (1996), hep-th/9510225.}

\lref\hura{A.~Hanany and A.~M.~Uranga,
``Brane boxes and branes on singularities,''
JHEP {\bf 9805}, 013 (1998), hep-th/9805139.}

\lref\ohta{K.~Ohta and T.~Yokono,
``Deformation of conifold and intersecting branes,''
JHEP {\bf 0002}, 023 (2000), hep-th/9912266.}

\lref\andyS{A.~Strominger,
``Massless black holes and conifolds in string theory,''
Nucl.\ Phys.\ B {\bf 451}, 96 (1995), hep-th/9504090.}

\lref\pterm{R.~Kallosh and A.~Linde,
``P-term, D-term and F-term inflation,''
JCAP {\bf 0310}, 008 (2003), hep-th/0306058.}

\lref\semil{K.~Dasgupta, J.~P.~Hsu, R.~Kallosh, A.~Linde and M.~Zagermann,
``D3/D7 brane inflation and semilocal strings,''
JHEP {\bf 0408}, 030 (2004), hep-th/0405247.}

\lref\dm{K.~Dasgupta and S.~Mukhi,
``F-theory at constant coupling,''
Phys.\ Lett.\ B {\bf 385}, 125 (1996), hep-th/9606044.}

\lref\dowker{F.~Dowker, J.~P.~Gauntlett, S.~B.~Giddings and G.~T.~Horowitz,
``On pair creation of extremal black holes and Kaluza-Klein monopoles,''
Phys.\ Rev.\ D {\bf 50}, 2662 (1994), hep-th/9312172;
F.~Dowker, J.~P.~Gauntlett, D.~A.~Kastor and J.~H.~Traschen,
``Pair creation of dilaton black holes,''
Phys.\ Rev.\ D {\bf 49}, 2909 (1994), hep-th/9309075;
J.~P.~Gauntlett,
``Pair creation of black holes,'' hep-th/9404076,
2nd TIFR International Colloquium on Modern Quantum Field Theory, Bombay, India.}

\lref\hawrad{S.~W.~Hawking,
``Black Hole Explosions,''
Nature {\bf 248}, 30 (1974);
``Particle Creation By Black Holes,''
Commun.\ Math.\ Phys.\  {\bf 43}, 199 (1975).}


\lref\ggs{D.~Garfinkle, S.~B.~Giddings and A.~Strominger,
``Entropy in black hole pair production,''
Phys.\ Rev.\ D {\bf 49}, 958 (1994), gr-qc/9306023.}

\lref\hawking{S.~W.~Hawking, G.~T.~Horowitz and S.~F.~Ross,
``Entropy, Area, and black hole pairs,''
Phys.\ Rev.\ D {\bf 51}, 4302 (1995), gr-qc/9409013;
R.~B.~Mann and S.~F.~Ross,
``Cosmological production of charged black hole pairs,''
Phys.\ Rev.\ D {\bf 52}, 2254 (1995), gr-qc/9504015;
R.~R.~Caldwell, A.~Chamblin and G.~W.~Gibbons,
``Pair Creation of Black Holes by Domain Walls,''
Phys.\ Rev.\ D {\bf 53}, 7103 (1996), hep-th/9602126;
R.~Bousso and S.~W.~Hawking,
``Pair creation of black holes during inflation,''
Phys.\ Rev.\ D {\bf 54}, 6312 (1996), gr-qc/9606052;
``Pair creation and evolution of black holes in inflation,''
Helv.\ Phys.\ Acta {\bf 69}, 261 (1996), gr-qc/9608008.}

\lref\infrev{A.~Linde,
``Prospects of inflation,'' hep-th/0402051;
A.~H.~Guth,
``Inflation,'' astro-ph/0404546;
A.~H.~Guth,
``The Big bang and cosmic inflation,''
{\it The Oskar Klein memorial lectures, vol. 2* 27-70, and MIT Cambridge - CTP-2144};
A.~R.~Liddle,
``The early universe,'' astro-ph/9612093;
A.~D.~Linde,
``Inflation and creation of matter in the universe,''
{\it 71st Les Houches Summer School: The Primordial Universe, Les Houches, France, 28 Jun - 23 Jul 1999.}}

\lref\duff{M.~J.~Duff, R.~R.~Khuri and J.~X.~Lu,
``String solitons,''
Phys.\ Rept.\  {\bf 259}, 213 (1995), hep-th/9412184;
``String and five-brane solitons: Singular or nonsingular?,''
Nucl.\ Phys.\ B {\bf 377}, 281 (1992), hep-th/9112023.}

\lref\lindeBH{J.~Garcia-Bellido, A.~D.~Linde and D.~Wands,
``Density perturbations and black hole formation in hybrid inflation,''
Phys.\ Rev.\ D {\bf 54}, 6040 (1996), astro-ph/9605094.}

\lref\hindvac{M.~Hindmarsh, R.~Holman, T.~W.~Kephart and T.~Vachaspati,
``Generalized semilocal theories and higher Hopf maps,''
Nucl.\ Phys.\ B {\bf 404}, 794 (1993), hep-th/9209088.}

\lref\shiu{B.~R.~Greene, K.~Schalm and G.~Shiu,
``Warped compactifications in M and F theory,''
Nucl.\ Phys.\ B {\bf 584}, 480 (2000), hep-th/0004103.}

\lref\anaren{J.~Urrestilla, A.~Achucarro and A.~C.~Davis,
``D-term inflation without cosmic strings,''
Phys.\ Rev.\ Lett.\  {\bf 92}, 251302 (2004), hep-th/0402032;
P.~Binetruy, G.~Dvali, R.~Kallosh and A.~Van Proeyen,
``Fayet-Iliopoulos terms in supergravity and cosmology,''
Class.\ Quant.\ Grav.\  {\bf 21}, 3137 (2004), hep-th/0402046.}

\lref\borill{A.~Achucarro, J.~Borrill and A.~R.~Liddle,
Phys.\ Rev.\ Lett.\  {\bf 82}, 3742 (1999), hep-ph/9802306;
J.~Urrestilla, A.~Achucarro, J.~Borrill and A.~R.~Liddle,
``The evolution and persistence of dumbbells in electroweak theory,''
JHEP {\bf 0208}, 033 (2002), hep-ph/0106282.}

\lref\anaV{T.~Vachaspati and A.~Achucarro,
``Semilocal cosmic strings,''
Phys.\ Rev.\ D {\bf 44}, 3067 (1991);
``Semilocal and electroweak strings,''
Phys.\ Rept.\  {\bf 327}, 347 (2000), hep-ph/9904229.}

\lref\hindmarsh{M.~Hindmarsh,
``Existence and stability of semilocal strings,''
Phys.\ Rev.\ Lett.\  {\bf 68}, 1263 (1992);
``Semilocal topological defects,''
Nucl.\ Phys.\ B {\bf 392}, 461 (1993), hep-ph/9206229.}

\lref\preskill{J.~P.~Preskill,
``Semilocal defects,''
Phys.\ Rev.\ D {\bf 46}, 4218 (1992), hep-ph/9206216;
J.~P.~Preskill and A.~Vilenkin,
``Decay of metastable topological defects,''
Phys.\ Rev.\ D {\bf 47}, 2324 (1993), hep-ph/9209210.}

\lref\greeneyau{B.~R.~Greene, A.~D.~Shapere, C.~Vafa and S.~T.~Yau,
``Stringy Cosmic Strings And Noncompact Calabi-Yau Manifolds,''
Nucl.\ Phys.\ B {\bf 337}, 1 (1990).}

\lref\tate{J.~Tate, ``Algorithm for Determining the Type of a Singular
Fiber in an Elliptic Pencil,'' in {\it Modular Functions of One Variable
IV}, Lecture Notes in Math. vol. 476, Springer-Verlag, Berlin (1975)}

\lref\bikmsv{M.~Bershadsky, K.~A.~Intriligator, S.~Kachru, D.~R.~Morrison,
V.~Sadov and C.~Vafa,
``Geometric singularities and enhanced gauge symmetries,''
Nucl.\ Phys.\ B {\bf 481}, 215 (1996), hep-th/9605200.}

\lref\AspKal{P.~S.~Aspinwall and R.~Kallosh,
  ``Fixing all moduli for M-theory on K3 x K3,'' hep-th/0506014.}

\lref\Deneftalk{F.~Denef, 
Talk at Strings 2005, Toronto.}

\lref\gidm{S.~B.~Giddings and A.~Maharana,
  ``Dynamics of warped compactifications and the shape of the warped
  landscape,'' hep-th/0507158.}

\lref\vwc{C.~Vafa and N.~P.~Warner,
``Catastrophes And The Classification Of Conformal Theories,''
Phys.\ Lett.\ B {\bf 218}, 51 (1989).}

\lref\wolf{J.~A.~ Wolf,
``Complex homogeneous contact structures and quaternionic symmetric spaces'',
J. Math. Mech. {\bf 14} (1965) 1033.}

\lref\alex{D.~V.~ Alekseevskii,
``Classification of quaternionic spaces with transitive solvable group of
motions,''
Math. USSR $-$Izv. {\bf 9}, 297 (1975);
``Riemannian manifolds with exceptional holonomy groups,''
Funct. Anal. Appl. {\bf 2}, 97 (1968);
``Compact quaternionic spaces,''
Funct. Anal. Appl. {\bf 2}, 106 (1968).}

\lref\dWVP{
B.~de Wit and A.~Van Proeyen,
``Special geometry, cubic polynomials and homogeneous quaternionic spaces,''
Commun.\ Math.\ Phys.\  {\bf 149}, 307 (1992), hep-th/9112027.}

\lref\galicki{K.~ Galicki,
``A generalization of the momentum mapping construction for quaternionic
K\"ahler manifolds,''
Commun. Math. Phys. {\bf 108} (1987) 117;
``Quaternionic K\"ahler And Hyper-K\"ahler Nonlinear Sigma Models,''
Nucl.\ Phys.\ B {\bf 271}, 402 (1986).}

\lref\barton{A.~Johansen,
``A comment on BPS states in F-theory in 8 dimensions,''
Phys.\ Lett.\ B {\bf 395}, 36 (1997), hep-th/9608186;
M.~R.~Gaberdiel and B.~Zwiebach,
``Exceptional groups from open strings,''
Nucl.\ Phys.\ B {\bf 518}, 151 (1998), hep-th/9709013.}

\lref\bagger{J.~Bagger and E.~Witten,
``Quantization Of Newton's Constant In Certain Supergravity Theories,''
Phys.\ Lett.\ B {\bf 115}, 202 (1982);
``The Gauge Invariant Supersymmetric Nonlinear Sigma Model,''
Phys.\ Lett.\ B {\bf 118}, 103 (1982);
``Matter Couplings In ${\cal N}=2$ Supergravity ,''
Nucl.\ Phys.\ B {\bf 222}, 1 (1983).}

\lref\rocek{U.~Lindstrom and M.~Rocek,
``Scalar Tensor Duality And N=1, N=2 Nonlinear Sigma Models,''
Nucl.\ Phys.\ B {\bf 222}, 285 (1983)}

\lref\dmbps{K.~Dasgupta and S.~Mukhi,
``BPS nature of 3-string junctions,''
Phys.\ Lett.\ B {\bf 423}, 261 (1998), hep-th/9711094.}

\lref\tong{D.~Tong,
 ``The moduli space of BPS domain walls,''
 Phys.\ Rev.\ D {\bf 66}, 025013 (2002), hep-th/0202012.}

\lref\hw{A.~Hanany and E.~Witten,
``Type IIB superstrings, BPS monopoles, and three-dimensional gauge dynamics,''
Nucl.\ Phys.\ B {\bf 492}, 152 (1997), hep-th/9611230.}

\lref\seiwit{N.~Seiberg and E.~Witten,
``Electric - magnetic duality, monopole condensation, and confinement in N=2
Nucl.\ Phys.\ B {\bf 426}, 19 (1994)
[Erratum-ibid.\ B {\bf 430}, 485 (1994)], hep-th/9407087.}

\lref\ds{M.~R.~Douglas, D.~A.~Lowe and J.~H.~Schwarz,
``Probing F-theory with multiple branes,''
Phys.\ Lett.\ B {\bf 394}, 297 (1997), hep-th/9612062.}

\Title{\vbox{\hbox{hep-th/0501185} \hbox{SU-ITP-05/04,
TIFR/TH/05-02} \hbox{ILL-(TH)-05/02, SLAC-PUB-10982}}} {\vbox{
\vskip-4.5in \hbox{\centerline{Brane Inflation,}}
\hbox{\centerline{Solitons and Cosmological Solutions: I}}}}
\vskip-.2in \centerline{\bf Pisin Chen${}^1$,~~ Keshav
Dasgupta${}^2$,~~ K. Narayan${}^3$} \centerline{\bf Marina
Shmakova${}^1$~~and~~Marco Zagermann${}^4$} \vskip.1in
\centerline{\it ${}^1$Stanford Linear Accelerator Center, Stanford
University, Stanford CA 94309, USA}
\centerline{\it ${}^2$Loomis Lab, University of
Illinois at UC, Urbana IL 61801, USA.}
\centerline{\it ${}^3$Tata Institute of Fundamental Research,
Homi Bhabha Road, Mumbai 400 005, India.}
\centerline{\it ${}^4$Varian Lab, Stanford University, Stanford CA 94305, USA.}

\vskip.1in

\centerline{\tt chen,shmakova@slac.stanford.edu, ~ keshav@uiuc.edu}
\centerline{\tt narayan@theory.tifr.res.in, ~zagermann@itp.stanford.edu}

\vskip.2in

\centerline{\bf Abstract}

\noindent
In this paper we study various  cosmological solutions for a D3/D7
system directly from M-theory with fluxes and M2-branes.  In M-theory, these  
solutions
exist only if we incorporate higher derivative corrections from the
curvatures as well as G-fluxes. We take these corrections into account
and study  a number of   toy cosmologies, including one with a novel background
for the D3/D7 system whose supergravity solution can be completely
determined.
This new background preserves all the good
properties of the original model and opens up avenues to investigate
cosmological effects from wrapped branes and brane-antibrane
annihilation, to name a few.  We also discuss in some detail semilocal
defects with higher global symmetries, for example exceptional ones,
that could occur in a slightly different regime of our D3/D7 model.
We show that the D3/D7 system
does have the required ingredients to realise these configurations as
non-topological solitons of the theory. These constructions also allow us
to give a physical meaning to the existence of  certain 
  underlying homogeneous
quaternionic K\"ahler manifolds.

\Date{}

\listtoc
\writetoc

\newsec{Introduction}

One of the most intriguing  results of the  recent   spectacular
advances in observational  cosmology is the strong evidence for at
least two periods of accelerated   expansion  in the history of our
universe. The          second,        presently  ongoing, of these
two periods is commonly  attributed to a tiny positive cosmological
constant, or some similar form of {\it dark energy}. The     other,
much earlier, phase of accelerated expansion refers to some period
well before Big Bang Nucleosynthesis and        is commonly known
as {\it inflation} \guth,\lindeN,\albre. Inflation is usually modelled by a scalar field, the
inflaton, which is slowly rolling down a sufficiently flat potential
that
  dominates the energy density    during that epoch.
     A phase of early   universe     inflation       not only solves
      several longstanding cosmological
  problems such as the horizon, flatness   or potential  monopole problems,
  but also provides a natural mechanism for the generation        of
  density perturbations. The resulting predictions for the anisotropies of the cosmic microwave background (CMB)
  have been    nicely        confirmed by recent
  experiments such as the   Wilkinson microwave anisotropy probe (WMAP)  \wmap.

 In view of this
observational         evidence,    it is now tempting to include
inflation  in the list of desirable properties any fundamental
theory of our world should be able to explain or at least reproduce.
String theory aspires to be such a fundamental theory, and should
therefore  be confronted with  this      evidence. Embedding inflationary
models in string theory touches upon several problems:

\noindent (i)    Time-dependence:   Describing cosmological
spacetimes in string theory in general means dealing with
time-dependent backgrounds, which are necessarily non-supersymmetric
and hence, harder to control.

\noindent (ii) Acceleration:  Reproducing longlasting accelerating spacetimes
from string compactifications is particularly non-trivial.

\noindent (iii)      Identification of the inflaton:   The inflaton
should be identified with a (possibly effective)    scalar field
in string theory. The moduli in string compactifications naturally
provide scalar fields, but a priori, it is not clear, which, if any,  of these
moduli fields should be identified with the inflaton.

\noindent (iv) Moduli stabilization: If one direction of the moduli
space is identified with the inflaton, the scalar potential along
this direction should be sufficiently flat, whereas          all the
other directions should be stabilized with sufficiently large masses
                 in order not to interfere with the slow-roll of
                 the   inflaton. This is in general  an issue of fine-tuning.

\noindent (v) End of inflation: If the inflaton is a modulus, how
can the end of inflation and reheating be understood geometrically?
Are there any unwanted relics after inflation, and if so, what is
their stringy origin?

As for problem (ii), no-go theorems against accelerating spacetimes
 have been  proven for compactifications of {\it classical} 10D or 11D supergravity
on {\it smooth}, {\it compact} and {\it time-independent}  internal
spaces \nogo. By relaxing any of the         highlighted
assumptions, this no-go theorem might possibly   be circumvented.
For example, allowing for a time-dependent hyperbolic
internal space \town, \neupane\  or a time-dependent flux compactification
\gutparuli, \wohl, \roy, \cornalba, \emparan, \neupane\   opens up the possibility of having transient periods of
inflation. These solutions are in fact related to S-branes \gutper.
Such transient periods of       accelerated expansion might be
marginally compatible with the present acceleration \gkal, but they
are far too short to describe the required 50 - 60 efolding of early
universe inflation \wohl,\gutparuli, \ceng\ (see also \teo\ for some related recent work). 

From the effective 4D point of view, this is related to the fact that
the volume modulus generically enters the scalar potentials of string
compactifications in terms of an exponential factor with a coefficient
that makes the potential too steep for slow-roll inflation. There are
essentially two ways around this problem: Either there are additional,
non-classical contributions to the scalar potential that could lead to
a sufficiently flat regime \renand, or the volume modulus should be
fixed during inflation altogether. In the latter case, the inflaton
has to be another scalar field.

This latter possibility is entertained in ``brane inflation''
models \henry, where the inflaton is identified with an open
string modulus that describes the relative position or orientation
of a D-brane with respect to another (system of) D-brane(s). A
necessary ingredient of these models is that the brane
configuration breaks supersymmetry so that the branes attract and
move towards one another. In brane/anti-brane systems \strinf,
supersymmetry is not just mildly broken, and the interaction
potential would be too steep if the branes were too close to one
another. It was pointed out in \qued, \kklmmt\ that the finite
size of a compact space   puts limits on the possible
interbrane distance, and hence, the possible flatness of the
brane/anti-brane potential that are in general incompatible with
slow roll inflation. This problem has been solved by putting the
anti-brane in a warped throat region \kklmmt, \tliz.

Another way to get reasonably flat brane interaction potentials is
to consider brane configurations that break supersymmetry more
gently than brane/anti-brane pairs. Examples for such
configurations include intersecting branes with a small relative
angle \shafi, \carlos\ or branes with small supersymmetry breaking
worldvolume fluxes.  The prime example for the latter possibility
is provided by the D3/D7 system with non-primitive fluxes on the
worldvolume of the D7 brane \dtds,\carlos. Classically the
interaction potential between the D3 and the D7 brane is
completely flat, but one-loop corrections due to the broken
supersymmetry induce a logarithmic slope. An explicit realization
of this model in a concrete string compactification was given in
\dtds, where the compact space was chosen to be $K3\times
T^2/\IZ_{2}$. This background was chosen because it is
non-trivial, but still well under theoretical control with a well
understood F-theory description. The effective low energy theory
in four dimensions is given by a hybrid D-term \dvhal\ (or, more
generally, ``P-term'' \pterm) inflation model\foot{In a different context,
low energy effective actions of D3/D7 systems with fluxes were recently studied in 
\lowaction.}.

As we will frequently refer to various aspects of this hybrid
inflation model, let us briefly recall its basic features for
later reference.  There are three different pictures we will
sometimes use for this model.  The first is the original picture
of hybrid inflation \lindehy, which can be viewed as the ``core''
of these hybrid inflation models in the sense that it only uses
the minimal field content and does not refer to any underlying
more fundamental theory. The second picture is the supergravity
P-term action of \pterm, where hybrid inflation is related to
${\cal N}=2$ supergravity.  The third picture derives from the
stringy analysis of the D3/D7 inflationary model \dtds,\carlos,
\semil.  The original picture of Linde \lindehy\ is based on the
following potential written in terms of two scalar fields, $\phi$
and $\psi$: \eqn\potlin{V(\phi, \psi) = {1\o 4\lambda} (M^2 -
\lambda \psi^2)^2 + {m^2\o 2} \phi^2 + {g^2 \o 2} \phi^2 \psi^2,}
where $M, m, \lambda$ and $g$ have been defined in \lindehy. In
this model, $\phi$ is the inflaton which is slowly rolling down
its potential, while $\psi$ is temporarily trapped in its false
vacuum with $\psi=0$. When $\phi$ reaches a critical value, $\psi$
becomes tachyonic, and the ``waterfall'' stage begins, where
$\psi$ quickly relaxes to its true vacuum state, thereby ending
inflation.

This construction can be embedded in the slightly less minimal
P-term model developed by Kallosh and Linde \pterm.  It is
basically an ${\cal N} = 2$ supergravity potential with FI terms
\foot{Strictly speaking, the ${\cal N}=2$ supersymmetry is broken
to ${\cal N}=1$ by the coupling to supergravity due to the
presence of the Fayet-Iliopoulos terms. Only in the rigid limit of
global supersymmetry, the full ${\cal N}=2$ supersymmetry is
restored.}, and has the following potential:
\eqn\potpterm{V(\Phi_{\pm}, S) = 2g_1^2 \left( \vert S \vert^2
(\vert\Phi_+\vert^2 + \vert \Phi_-\vert^2) + \left\vert \Phi_+
\Phi_- - {\zeta_+ \o 2} \right\vert^2 \right) + {g_1^2\o 2}
(\vert\Phi_+\vert^2 - \vert\Phi_-\vert^2 - \zeta_3)^2,} where $S$
is the scalar in an ${\cal N}=2$ vector multiplet and ($\Phi_+,
\Phi_-$) form the scalars of an ${\cal N} = 2$ hypermultiplet. In
terms of ${\cal N} = 1$ language $\Phi_+$ and $\Phi_-$ form the
scalars of two chiral multiplets. The coupling constant is denoted
by $g_1$ and ($\zeta_{\pm}, \zeta_3$) form a triplet of FI terms
\pterm.

The third and final picture which we will use derives from the
D3/D7 inflationary model \dtds. The potential is written in terms
of the BI and CS terms of the D3 as well as the D7 branes \semil\
and is given by \eqn\potdthdse{V(S, \chi) = c_o \int_{K3} {\cal
F}^- \wedge \ast {\cal F}^- + 2 (g_3^2 + {\tilde g}_3^2)\vert S
\vert^2 \vert \chi \vert^2 + {g_3^2 + {\tilde g}_3^2 \o 2}
(\chi^\dagger \sigma^A \chi)^2,} where $\sigma^{A}$ are the Pauli matrices,  
 $g_3$  denotes   the coupling of
the D3 brane, and ${\tilde g}_3$ is the effective four-dimensional
coupling of the D7 branes. We have also normalised the action in
such a way that the kinetic term of $S$ is simply $\vert \del
S\vert^2$. The coefficient $c_o$ is related to the volume of the
internal space and the seven-brane coupling constant. In writing
this potential, we have ignored the degrees of freedom that
determine the center-of-mass motion of the system.

A detailed comparison between \potpterm\ and \potdthdse\ can be found
in \semil, so we will be brief here. If we take the FI terms in
\potpterm\ as ($\zeta_\pm, \zeta_3$) = ($0, \zeta_3$), then $\zeta_3$
can be identified with ${\cal F}^-$. The precise identification is
\eqn\iden{ c_o \int_{K3} {\cal F}^- \wedge \ast {\cal F}^-  \to g_1^2 \zeta_3^2, ~~~~ S  \to S, ~~~~ \chi_1  \to \Phi_+, ~~~~ \chi_2 \to \Phi_-, ~~~~
g_3^2 + {\tilde g}_3^2 \to g_1^2.}

From the above identifications, it is now easy to write the gauge
fluxes ${\cal F}^-$ in terms of the variables of \pterm. We can
use any of the harmonic ($1,1$) forms $\Omega_i$ ($i =
1...h_{11}$) on the $K3$ manifold to represent the gauge flux as
\eqn\gaugeflux{ {\cal F}^- =  \sum_{i =1}^{h_{11}} c_i~\Omega_i,
~~~~~~~~~{\rm such~that} ~~~ \int_{K3} \Omega_i \wedge \ast
\Omega_j = \delta_{ij}.} Written in terms of the harmonic forms
one can easily show that the $c_i$ form a surface in an $h_{11}$
dimensional space. The equation for the surface can be written in
terms of the variables of \pterm\ and is explicitly given as
\eqn\surface{ c_1^2 + c_2^2 + ..... + c_{h_{11}}^2~ = ~
c_o^{-1}~g_1^2~ \zeta_3^2.} Taking the $c_i$ to be real, the above
equation basically defines a sphere $S^{h_{11}-1}$, with radius
$r \equiv c_o^{-{1\o 2}}g_1~\zeta_3$. Therefore, one simple
solution for the gauge flux would be to take the radius of
$S^{h_{11}-1}$ as the antiselfdual flux, i.e ${\cal F}^- = r$.
With this choice, the two models \potpterm\ and \potdthdse\ can be
easily identified.

To compare \potdthdse\ (or \potpterm) with \potlin, one should
first identify which fields in \potpterm\ become tachyonic. The
mass of the hypers is given by $m^2_{\rm hyper} = g_1^2 \vert
S\vert^2 \pm g_1 \zeta_3$, which   becomes tachyonic when
  $\vert S \vert < \vert S
\vert_c$, where $\vert S \vert_c = \sqrt{\zeta_3 \o g_1},$.
  It is now easy to make the comparison with
\potlin. The mass of the scalar field $\psi$ is given by $m^2_\psi
= -M^2 + g^2 \phi^2$. This becomes tachyonic when $\phi < \phi_c$,
with $\phi_c = {M \o g}$.  Therefore, one sees that $\vert S
\vert$ is related to $\phi$ while $\vert \chi \vert$ is related to
$\psi$. The coupling constant and the FI terms of the D3/D7 system 
  are
then related to $g$ and $M$ of \lindehy\ respectively. This
concludes our little review of hybrid inflation in the context of
the D3/D7 inflationary model\foot{We should briefly mention yet
another model that also shows hybrid inflation. This is the model
studied by Guth-Randall-Soljacic \randall, which in some regions,
show properties identical to the original hybrid inflationary
model \lindehy. The potential for inflationary behavior in this
model is given by $V(\phi,\psi) = M^4 {\rm cos}^2 \left({\psi \o
f}\right) + {1\o 2} m^2 \phi^2 + g_4^2 \phi^2 \psi^2,$ where the
coefficients are explained in \randall. A detailed comparison
between this model and \potlin\ has already appeared in \lindeBH\
so the readers can pick up details from there. We should simply
point out that in the region $\psi < f$ this model and \potlin\
are more or less identical.}. As we have seen, a certain
supergravity variant of the hybrid inflation potential arises
naturally in the D3/D7 system with non-selfdual fluxes on the
D7-brane worldvolume.  The nice feature of this model is that the
dominant contribution to the vacuum energy derives from a D-term
in the supergravity potential. This circumvents the well-known
eta-problem of supergravity models of inflation that are based on
F-terms.  In F-term inflation, the appearance of the K\"{a}hler
potential in an exponential factor of any F-term potential in
supergravity generically leads to the problematic relation, $\eta
\sim 1$, where $\eta$ is the second slow-roll parameter,
$\eta=M_P^2 V^{\prime\prime}/V$ \lidlit, \barrowlid.
Unfortunately, the currently known methods for volume
stabilization are based on F-term potentials from non-perturbative
superpotentials \kklt.  Thus, in general, even if the primary
inflationary potential is a D-term potential, the supergravity
eta-problem reenters through the back door via volume
stabilization.  This can be avoided if the K\"{a}hler potential
posesses an inflaton shift symmetry \Kawa, \hsushift\ (see also
\shandera). Interestingly, it was shown in \hsushift\ that, at
tree level, the D3/D7 model on $K3\times T^2/\IZ_{2}$ precisely
has such an inflaton shift symmetry, which is related to the
continuous isometries of the two-torus. This inflaton shift
symmetry is broken by one-loop corrections, already at the field
theory level, and these one-loop corrections are precisely needed
to slightly lift the classically flat direction of the inflaton
potential.  However, as  pointed out in \mcall,
  the stringy one-loop corrections computed in
\bergg\ also violate the shift
symmetry in a manner that the eta problem generically reappears.
On the other hand,   it can in general be removed by a modest ($\sim 10^{-2}$)
fine-tuning of the fluxes or other parameters \mcall. The necessity of
such a modest fine tuning (which might not be such a big problem
after all, as the models with sufficient efolding of inflation are
``rewarded'' by a much larger amounts of the spacetime they fill
out) seems to be a generic feature of many stringy models of
inflation, whether they are based on closed or open string moduli
\mcall. In retrospect, this does not seem very surprising, as the
very idea of all but one modulus having large masses while the
remaining one gives rise to a comparatively flat potential does
not, intuitively, look like a very generic situation.  It is
nevertheless interesting to explore how deeply hidden such
fine-tunings can be in different models, and the D3/D7 model is
certainly one of the models in which this is   
least obvious\foot{Since the time we submitted our paper, an important
update regarding
the  moduli stabilization in the $K3 \times T^{2}/\IZ_2$
model  appeared in \AspKal, where  moduli stabilization
due to Euclidean D3-brane instantons instead of gaugino condensation
is discussed. The results of \AspKal\ should have interesting
applications for the D3/D7 inflationary model in that background.}.

Leaving this issue of modest and probably not too worrysome
fine-tuning aside for a moment, it is worth emphasizing that the
brane inflation models have generally  been studied in a way that is
technically different from that of the models of transient
accelerating cosmologies in \town, \gutparuli, \wohl, \roy,
\cornalba, \emparan. More precisely, whereas in those  papers the
whole system is  studied from a genuinely 10D or 11D point of
view, the brane inflation models recover inflating
four-dimensional spacetimes from an effective four-dimensional
action that derives from an, a priori static, string
compactification in which quantum effects and non-perturbative
objects such as branes also play a role. This usage of effective
field theories is sometimes criticized \bankcrit, among other
things, because the 4D effective actions are often derived from
compactifications on static backgrounds, which are, in general,
not related by small perturbations to the time-dependent
cosmological solutions they are ultimately used for. This would
not be an issue if one were dealing with a {\it consistent
truncation} instead of a mere low-energy effective action. \foot{For a
consistent truncation, any solution of the lower-dimensional
theory can, by definition, be uplifted to a valid solution of the
higher-dimensional theory, but for a mere effective
low-energy theory, this need not necesarily be true.
  A particularly nice illustration of
the power of consistent truncations is provided by 5D, ${\cal
N}=8$ gauged supergravity \gunah, \pernici, which is believed to
be a consistent truncation of the compactification of IIB
supergravity on the five-sphere to the lowest lying Kaluza-Klein
modes. Because of this, domain wall solutions of this
five-dimensional theory are guaranteed to have a well-defined
up-lift to ten dimensions \ctheorem\ and can thus reliably encode information on
renormalization group flows of the deformed 4D, ${\cal N} = 4$ super 
Yang-Mills theory.}  It should be
emphasized, however, that so far there hasn't been any clear
evidence that the usage of 4D effective field theory is
inappropriate in this context.

Another issue of many brane inflationary models is that they tend
to produce cosmic strings or other defects at the end of
inflation.  The energy density contributed by such defects should
of course be small enough so as not to upset the successful
predictions of inflation such as the CMB power spectrum.  On the
other hand, these defects can often be identified with inherently
stringy objects such as the fundamental string itself or various
types of (possibly wrapped) D-branes. This opens up an exciting
possibility that, provided the abundance and energy density of
such objects turn out to be in the right ballpark, one could study
genuinely stringy objects by direct observations ``at the sky''.
For more details on the recent work on this subject,
   see, e.g., \coupling, \matsuda.

In this paper, we turn our attention to some of the aspects of
D3/D7 brane inflation that have to do with the higher-dimensional
and stringy nature of this scenario.  More precisely, we make a
first step towards the uplift of the full time-dependent 4D
cosmology of the D3/D7 scenario to the full higher-dimensional
spacetime picture. The purpose of this exercise is to better
understand how precisely the no-go theorems of \nogo\ are
circumvented, and which effects are the most dominant. A
successful uplift would also help to get some better insight into
the validity of 4D effective field theory actions in this context.
Doing this directly, however, would obviously require dealing with
the seven-branes, three-branes, different types of fluxes and
non-perturbative effects directly in IIB string theory or
F-theory. This is clearly a highly non-trivial problem, and we
will therefore use the slightly simpler uplift to the
eleven-dimensional M-theory\foot{Recently cosmological models
directly from M-theory were also studied in \maeda. However the
emphasis of these models were to get four-dimensional cosmology
{\it without} any brane inflation from M-theory. Thus our analysis
is completely different from \maeda.}. This has the advantage 
that the seven-branes (including the the non-perturbative ones
that might result from orientifold planes) dissolve into
background geometry. More precisely, they enter as singularities
of the torus fibration of the fourfold, which as a whole,
however, becomes a smooth eight-dimensional manifold. Furthermore,
in the M-theory picture one has to deal with only one type of
flux, namely that of the four-form field strength.  Still, a
complete uplift of the D3/D7-scenario represents a challenging
problem, and in this first part, we     limit ourselves to setting up and
identifying all the necessary ingredients as well as to   the study of
various toy cosmologies. A more refined treatment is postponed to
a later publication \work. One of the main results in the present
paper is that accelerating spacetimes in this scenario seem to
require the inclusion of the higher derivative terms of the
M-theory action and that the assumption of a single time-dependent
warp factor for the internal manifold in the M-theory picture
is probably too simplistic.

One of the drawbacks of the manifold $K3\times T^{2}/\IZ_{2}$ is
that its metric is not explicitly known, except in the orbifold
limit of the $K3$.  We will therefore also construct a non-trivial
background whose metric can be explicitly written down. This
background is a compactified version of a non-Calabi-Yau deformed
conifold and has a well-defined F-theory description in terms of a
Calabi-Yau fourfold\foot{Notice that this is different from the
Klebanov-Strassler model \kleb\ in the sense that it is compact,
non-supersymmetric, non-K\"ahler and, as we will show later, has
seven branes.}.  This background also has a non-trivial
three-cycle around which D3-branes can wrap. Such three-cycles are
absent in the $K3\times T^2/\IZ_{2}$ background and allow for
certain charged massive black hole configurations. The excitation
of such black holes during or after inflation are generally
expected to be negligible, and in general, this is confirmed by
our considerations. This, however, is not necessarily true for
cosmic strings which, in the generic D3/D7 model, can easily ruin the 
otherwise successful predictions of inflation. 
In \semil, a mechanism has been proposed by which
the production rate of such cosmic strings can be sufficiently suppressed 
without going to unnatural regions of the parameter space. In this
mechanism, one replaces the single D7-brane of the D3/D7 scenario
by a stack of two or more coincident D7-branes. This results in an
additional (approximately) global symmetry, and the resulting
cosmic strings are the so-called semilocal strings \anaV,
\hindmarsh\ instead of the conventional Abrikosov-Nielsen-Olesen
(ANO) strings \ano, which have a much lower formation rate
\borill, \anaV, \anaren.  In this paper, we take the
non-perturbative effects of F-theory seriously and consider the
possibility that the non-perturbative exceptional symmetries of
certain F-theory configurations can give rise to more general 
semilocal strings
based on non-unitary symmetry groups, such as, e.g., symplectic or 
exceptional groups. 
The possibility of such semilocal
strings was entertained in \hindvac, but no explicit field theory
realization of such strings is known to exist so far. Here we will 
provide a construction of these defects
as non-topological solitons on the D3 brane.

\subsec{Organization of the paper}

The organization of this paper is as follows: In sec. 2 we  set up 
the basic picture to study cosmological solutions  for   the D3/D7
system. We do not assume any particular embedding, but, a priori, 
   allow all
possible backgrounds for the branes to propagate.  Our strategy
here will be to study the system directly from M-theory, where the
D3 branes will become M2 branes and the fluxes will become
G-fluxes. In sec. 2.1 the background equations of motion, given by the 
Einstein  and flux equations, are spelled out completely.
We provide some of the possible quantum terms at this stage. Later
on we will specify more quantum terms as the given set of terms
would not suffice. In   that   section we also specify the criteria to
obtain accelerating cosmologies, such as de Sitter  spacetimes. 

The full analysis of the background equations  is   worked out in
sec. 2.2. We start with a very generic   ansatz   for the background
metric that would allow three warp factors: one for the 2+1
dimensional spacetime and two for the internal manifold. The internal
manifold for our case will be a complex fourfold which is a torus
fibration over a six dimensional base, with possibly different warp factors
along the fiber and along the six dimensional base.  We will discuss
how the curvature tensors etc should be written
 with three warp
factors. Because of  the large number of components   of  the 
  Einstein tensor, the
equations naturally   get very involved. To simplify the subsequent analysis we
consider only the two warp factor cases here in detail and discuss the
three warp  factor   case only briefly. Taking two warp  factors   simplifies 
things a little bit {
 but is still sufficiently interesting.} 
  In  section 2.2    many of the
interesting effects that will form the basis of our subsequent analyis
are pointed out: supersymmetry breaking by non-primitive fluxes and
the resulting   motion of the M2 branes, the
constraint coming from anomaly equations and the corresponding warp
factor dependence, an   analysis of every component    of  the
  Einstein tensor, as well as 
the effects coming from membrane and quantum terms and some
surprising simplifications that would happen if some special relations
are considered for the warp factors.

We then give many toy examples in sec 3. In most of these examples,  we
take some limit of the background fluxes, the   size of the fourfold and the
membrane velocities so that we can have some control over the quantum
corrections. In this section we also point out the relevance of higher
derivative terms that are in {\it addition} to the quantum terms that
we took into account  in sec. 2. These higher derivative terms (for G-fluxes and
curvatures) are necessary to overcome the no-go theorem for warped
compactification on a compact manifold \nogo.

 In sec. 3.1 we take very
small internal G-fluxes on an essentially non-compact fourfold with a
slowly moving membrane. This way,  one   can study   a background by ignoring
the quantum corrections (without violating the no-go theorem).

In sec. 3.2 we consider a little more elaborate analysis by taking
arbitrary fluxes. We find solutions {\it only} when higher
derivative terms in G-fluxes (and possibly the   curvatures) are
incorporated here. Our solutions are thus valid to some order in
quantum corrections. When the G-fluxes are not very large, and the
fourfold is sufficiently big, the quantum effects may be kept
under control without destabilising the solution.

In all the earlier examples above, we give cosmological solutions for
which the fourfold have time independent warp factors.  In sec. 3.3 we
give yet another cosmological solution in type IIB theory with two
warp factors, but this time take both of them as time dependent. We
again provide the complete background (up to the possible
approximations that we consider), which turns out to be a
radiation-dominated cosmology in four dimensions. In this context, we
also briefly discuss some interesting questions on the interpretation
and validity of cosmological solutions to supergravity in the presence
of higher derivative terms.

In all the previous examples, it was difficult to generate an
accelerating cosmology.  In sec. 3.4, we come back to the three warp
factor case.  With three warp factors, a de-Sitter type cosmological
solution in type IIB might, in principle, be possible.  We show that
the higher derivative terms in G-fluxes may not be a big problem for
this case.

In all the above cases, the internal fourfold was kept completely
arbitrary. In sec. 3.5 we consider a new background that is different
from the  originally  considered   $K3 \times \IP^1$ background
of \dtds.  Just as the $K3 \times \IP^1$ background, the
  new background allows D3 branes and also D7 branes
along with G-fluxes. There is also a detailed F-theory description of
this manifold in terms of a   Weierstrass equation that we specify
completely. To derive this background, we use the trick of {\it
geometric transition} that was developed for the supersymmetric case
in \bdkt,\realm. The background that we eventually get is a
non-supersymmetric and non-K\"ahler compact manifold that has some
resemblance to the deformed conifold in some regime. Globally the
manifold is quite different from a Calabi-Yau deformed conifold, and
allows three cycles as well as four-cycles. We show that the metric of the
manifold can be completely determined and present the precise {\it
unwarped} metric. This should be contrasted with the $K3 \times \IP^1$
background whose
metric can only be determined when the K3 is at its orbifold
point.

In sec. 3.6 and 3.7 we use some of the geometric properties of this
manifold to study consequences on four dimensional cosmology from
branes wrapped on cycles of the manifold. In sec. 3.6 we discuss
wrapped D3 branes and study the four dimensional non-supersymmetric
black holes they give rise to. We show that these non-supersymmetric black holes in
type IIB theory can be mapped to a string inside a deformed {\it
brane-box} configuration again in type IIB theory.  We show that
the deformation of the brane box appears from a $B_{NS}$ field
that seems to originate from the non-K\"ahler nature of the
underlying six dimensional manifold.  We also show that it might
be possible to study the full supergravity background for the
black hole using the   brane-box/geometry correspondence.

In sec. 3.7 we study cosmological effects from brane-antibranes
annihilations. We discuss many interesting effects here:
separation of brane-antibranes  due  to the   inflationary  expansion
  and the
resulting   production of black holes, brane-antibranes
annihilating to smaller branes and the resulting   production of
defects including black holes, etc.

In sec. 4,  we classify all the possible solutions that one  could, in principle,    get
  by 
using our ansatz.   We   show that with two warp factors one
could get de-Sitter cosmologies in type IIB theory at most   if  the warp
factors are all time-dependent. With three warp factors,   we show
that it  might, a priori, be   possible  to get a   de-Sitter space 
with a completely
time-independent internal warp factor. We discuss possible
consequences from moduli stabilization and interpret some of the
recent advances in cosmological studies using our scenario.

In sec. 5 we use the results of sec. 3 and 4 to study primordial
black holes in the D3/D7 system. In sec. 4 we showed that a 
de-Sitter kind of background might possibly exist
   using  three warp factors in M-theory. In 
   sec. 3.5,   we gave the explicit   metric for  a particular 
  internal space. We
combine these observations and the results coming from wrapped
branes and brane-antibrane annihilations to study possible
primordial black holes. We argue that if the background can create
copious numbers of brane-antibrane pairs, then one could possibly
relate this to  primordial   black hole production in this set-up. 
 In general, the   number of
such black holes, however, turns out to be almost negligible, and
therefore it does not pose any cosmological problem. We use our
earlier techniques to compare black  hole   formation in two
different scenario: the original $K3 \times \IP^1$ background and
the new background that we derived  in  section 3.5.  
  The former case  gives   zero
productivity rates and the latter gives very small productivity
rates (i.e., they are   exponentially suppressed).

In sec. 6 we go to a different regime of our D3/D7 model. This
regime  need not    necessarily correspond to
  an inflationary one\foot{In other words, we will consider the regime where we can have more than two scalars. 
Although having more than two scalars do not necessarily rule out an inflationary description of the system, we will
not concentrate too much on this aspect here.}.  
Here we combine the
background that we derived in sec. 3.5 with the F-theory picture
to study semilocal defects on the D3 brane(s). We show that our
model is rich enough to give us semilocal defects with almost all
allowed global symmetries, including exceptional ones. Earlier
studies of such defects were done at  a {group }   theoretical level in
\hindvac, where it was conjectured that defects with exceptional
global symmetries should also exist. Here we show that our model
might   realize them as non-topological solitons of the theory. We will be able to provide
explicit constructions of these defects using various techniques of algebraic
geometry. 
Another interesting output of this analysis is a physical way to
obtain the certain {\it homogeneous} quaternionic K\"ahler
manifolds (in fact the ones that are symmetric spaces)
directly from F-theory. This classification was done in
the mathematical literature long ago \wolf, \alex, and completed in
 \dWVP.    Here we see
their relevance to the existence of semilocal defects in the   D3/D7
system.

This concludes   the list of topics that we study in this paper. A
more detailed discussion of cosmological solutions that are
accelerating (possibly   including  de-Sitter type   spacetimes)    is 
  left for the
sequel to this paper \work.

\newsec{Cosmological solutions from the $D3/D7$ system}
\noindent
In this section, we study cosmological solutions for the D3/D7 system
from the full higher-dimensional viewpoint of ten- or
eleven-dimensional string- or M-theory. In the earlier papers
\dtds,\semil, this higher-dimensional viewpoint was only partially
addressed, and the focus was instead more on the derivation of a
four-dimensional effective action that describes the dynamics from a
four-dimensional point of view. More precisely, it was shown that the
presence of non-primitive fluxes on the world volume of D7-branes that
are wrapped on a K3 manifold in a $K3 \times \IP^1$ space would cause
a D3-brane to move towards these D7-branes. The corresponding 4D
effective theory was identified with a hybrid D-term (or, more
generally, P-term) inflationary model, where the inflaton field, $S$,
corresponds to the D3/D7 interbrane distance and the scalar field,
$\vert \chi\vert$, whose condensation ends inflation, can be
identified with a D3-D7 string mode that becomes tachyonic at a
critical value of the D3/D7 interbrane distance. In order to study the
full uplift of this scenario to ten or eleven dimensions, one now has
to consider the following important points:

\noindent $\bullet$ A possible time dependence of the internal
space.

\noindent $\bullet$ The noncompact 3+1 dimensional space as
approximately a de-Sitter space.

\noindent $\bullet$ Moduli stabilization, at least, for the static
case. This also includes radius stabilization so as to avoid the
Dine-Seiberg runaway problem \dsrun.

\noindent To account for  the first point,  we have to allow an
ansatz that could have time dependent warp factors. However, having
a time dependent warp factor would in general  mean that the
dynamics that depends on the volume of the internal spaces will now
 acquire a time dependence as well. This aspect can be used to put some
strong constraints on the warp factors of the internal manifold.

The second point is the aspect which we would like to concentrate
on the most. When the D3-brane starts moving towards the
D7-branes, the initial background would approximately be a de
Sitter   space. In the 4D field theory, this corresponds to the
slow-roll stage, where $\vert \chi \vert$ is trapped in its false
minimum, and the inflaton field, $S$, is slowly rolling down its
potential. At a critical point, $S=S_c$,   when the D3 is very
close to the D7-branes, $\vert \chi \vert$ becomes  tachyonic and
starts the waterfall stage where it  quickly rolls  down from its
false vacuum  to the final supersymmetric minimum, thereby ending
inflation.

This picture is simpler and maybe a bit more transparent when
viewed from M-theory. For the backgrounds we consider, this would
be an M-theory compactification on a fourfold with G-fluxes. The
initial condition is fixed by choosing G-fluxes that are not
primitive. The final state of primitive G-fluxes (which in the
language of inflation would be the final state after the waterfall
stage) is obtained from the contributions of G-fluxes that are
localized at the singularities of the fourfold (see \semil\ for
details).

The third  issue, the  moduli  stabilization,   is by far the most
difficult one. It is known that
switching on  fluxes can  stabilize all complex structure moduli
\gvw,\sav, \kachruone\ at tree level. The K\"ahler moduli,
including the radial modulus, on the     other hand, are   not
stabilized by such  fluxes. The first example, where the radial
modulus was shown to be stabilized, was the heterotic
compactification on a non-K\"ahler manifold \bbdp. For the type
IIB case, it was realized that non-perturbative effects can
stabilize the radial modulus \kklt. For the other K\"ahler moduli,
the proposal seems to be that a mixture of perturbative and
non-perturbative effects can fix all the K\"ahler moduli \denef\
(see also \savrobbins\ for the problems related to models with one
K\"ahler modulus)\foot{Since we submitted our paper, important  
progress has been made in the construction
of models with all moduli fixed. We refer to  \Deneftalk\
for a very  recent summary of this new work and many original
references. Some earlier work in which some aspects of models with all moduli fixed
were discussed include \landscape.}.

But this is not all. In type IIB (or in  the heterotic string)
there are the vector bundle moduli. In type IIB they appear from
the D7-branes, and they satisfy the Donaldson-Uhlenbeck-Yau
equations. In M-theory, they would appear from the moduli of the
G-fluxes that are localized at the orbifold singularities. It was
shown in \bbdg\ that near these singularities, the G-fluxes
decompose as a product of two two-form fluxes, one of them being a
normalizable (1,1) form. Combining this with the primitivity
condition of the G-fluxes, one reproduces precisely the DUY
equations \bbdg.

In any case, it seems to be highly non-trivial in general  to construct
complete
models with  all string theory moduli fixed at phenomenologically
 useful values (see e.g. \burt\ for some early attempts in the
heterotic case). 
Therefore, the first step would be to fix all the moduli at least
for the static case. We will briefly discuss the issue of moduli fixation
later in sec. 4. First let us try to understand the basic cosmology of
these  models.

\subsec{Basic cosmology of the model}

Let us  start with some standard ingredients from  cosmology (see e.g. 
\paddy\ \carroll\ \vijay ). Homogeneous and isotropic cosmological 
scenarios in $d+1$ dimensional spacetimes are based on
Friedmann-Lemaitre-Robertson-Walker (FLRW) type metrics. A spatially
flat \foot{For simplicitly, we restrict ourselves to spatially flat
cosmologies.} $(d+1)$-dimensional FLRW line element has the form
\eqn\RW{ds^2=-d\tau^2 +a^{2}(\tau)dx^i dx^i.} 
The sign of the Hubble parameter, $H\equiv {\dot{a} \o a}$,
determines whether the universe is expanding $(H>0)$ or contracting
$(H<0)$. An expanding universe is accelerating if ${\ddot{a}\o a}$
is positive and decelerating if ${\ddot{a} \o a}$ is negative.
During inflation, we have, by definition, \eqn\ind{ H>0,~~~~~~
{\ddot{a} \o {a}}>0.} Note that, for  the proper evaluation of $H$
and ${\ddot{a}\o a}$, \RW\ should be the         $(d+1)$-dimensional
{\it Einstein frame} metric. \RW\
is conformally flat and can be written in terms of conformal time,
$t\equiv x_0$, as
\eqn\conf{ ds^2=\eta^2(t)\left(-dt^2 +dx^i dx^i\right) =
\eta^2(t)~\eta_{\mu\nu}dx^{\mu}dx^{\nu},} where $t$ and $\eta(t)$
are defined by \eqn\etatau{\eta^2(t)dt^2 = d\tau^2, ~~~~
\eta^2(t(\tau)) = a(\tau)^2.} Eq. \etatau\ implies $\eta dt =\pm
d\tau$, and using this, one derives \eqn\Hallo{\eqalign{& H\equiv
{1\o a}\cdot {da \o d\tau} ~ = ~ \pm {1\o \eta^{2}}\cdot{d\eta\o dt}
\cr & {1\o a}\cdot{d^2 a \o d\tau^{2}}~ = ~ {1\o \eta^2}\cdot {d^2
{\rm log}~ \vert \eta \vert \o d t^{2}}}} In other words, $H$ can
always be chosen positive by choosing the appropriate sign {
(which just reflects the invariance of the Einstein equations under
time reversal),  } but the sign of the acceleration parameter
${\ddot a\o a}$ is independent of this choice and given by the sign
of ${d^{2} {\rm log}~ \vert \eta \vert \o d t^2}$. An important
special case, which we will also encounter in some of our models below, is
a simple power law behavior of the form \eqn\power{
\eta(t)=ct^{\gamma}} with some, possibly fractional, power $\gamma$.
It is easy to see from \Hallo\ that an expanding universe is
accelerating precisely if \eqn\acce{ \gamma<0 ~~~~ \Longrightarrow
~~~~ {\ddot a \o a} >0} with a de Sitter space corresponding to the
special case $\gamma=-1$.

In this section we are interested in uplifting accelerating
spacetimes of the above type to compactifications of string or
M-theory,  having in mind a precise higher-dimensional realization
of the D3/D7-brane inflationary scenario of type IIB string theory 
containing higher derivative terms in the effective action.

Note that from a four-dimensional point of view, one could ask
questions on the possible kinds of effective matter field
configurations that give rise to such cosmological solutions.
For instance, a perfect fluid with equation of
state $p=w\rho$ gives rise to a time-scaling $a(\tau)\sim
\tau^{[2/3(1+w)]}$ for the scale factor in \RW\ \ (see e.g. \paddy\ 
\carroll), with the energy density scaling as $\rho\sim a(\tau)^{-3(1+w)}$. 
In particular, this means $w=-1$ is a negative pressure fluid
corresponding to a cosmological constant. Note also $w=+1$
corresponding to $p=\rho$ has been argued for as the ``stiffest''
equation of state \banks. In this light, it is conceivable that some
of the cosmological solutions obtained from the higher dimensional
lift naively exhibit exotic behaviour from the point of view of
standard four-dimensional cosmology. Care must be used on the
interpretation and validity of an effective four dimensional
description of the cosmology as opposed to the full higher dimensional
description (as we will discuss in Sec.~(3.3)).

The construction of cosmological scenarios from higher dimensional
theories has been addressed in many earlier papers \chodos, \buwal.
The basic idea followed in these papers \chodos\ is to consider a
compactification of a ($1+d+D$)-dimensional theory and study the time
dependent warped metric. The metric ans\"{a}tze that have mostly been
considered in the literature are of the form \foot{ More general cases
with different time-dependent warp factors for different factors of
the internal space are of course also possible and have also been
considered in various papers.  We will return to this more general
case with several internal warp factors in Section 2.2. In the present
subsection, we limit ourselves to the simplest possible case with only
one internal warp factor, in order to illustrate the basic features of
these models with the minimal amount of complexity.}

\eqn\metansz{g_{MN} = \pmatrix{-1&0&0\cr \noalign{\vskip -0.20
cm} \cr 0& e^{2\tilde A} \delta_{ij}&0 \cr \noalign{\vskip -0.20 cm} \cr
0&0& e^{2\tilde B} g_{mn}(y)},} where $x^i, x^j = 1, ..., d$ and $y^m, y^n =
d+1, ..., D$, and  the warp factors $\tilde A$ and $\tilde B$ have usually
been assumed to be functions of  cosmological time $\tau$ only.
In this paper, we choose to work with the conformal time $t\equiv x^0$ and,
in general,  also allow for a $y$-dependence of the warp factors,
i.e., we will use ans\"{a}tze of the form
\eqn\genmetanzzs{ds^2 = e^{2A(y, t)}
\left(-dx_0^2 + \sum_{i = 1}^d ~dx_i dx^i \right) + e^{2B(y,t)}
g_{mn}(y) ~dy^m dy^n} where $g_{mn}(y)$ is still not been specified, and
$A(y,t)$ and $B(y,t)$ are the new warp factors that we shall use throughout
(unless mentioned otherwise).
In this form there is, an overall conformal factor in front of the
($d+1$)-dimensional part of spacetime. The total spacetime dimension,
($1+d+D$), will either equal 10 or 11.
For the special case of only time-dependent warp factors,
\eqn\wart{A(y, t) ~ \equiv ~ A(t), ~~~~~~~~ B(y,t) ~ \equiv ~ B(t),}
the non-vanishing components of the Ricci tensor for the metric ansatz
\genmetanzzs\ are given by
\eqn\riccioo{R_{00} = - d
\ddot A - D(\ddot B + {\dot B}^2 - \dot A \dot B)}
where the dots
are used to represent derivatives with respect to $x_0$, and
\eqn\rmunu{\eqalign{&R_{ij} =  \left[ \ddot A + D \dot A
\dot B + (d - 1) {\dot A}^2 \right] \delta_{ij} \cr & R_{mn} =
R_{mn}^{(g)} + e^{2(B - A)}\left[ \ddot B + (d-1) \dot A \dot B + D
{\dot B}^2 \right]g_{mn},}} where $R_{mn}^{(g)}$ is the Ricci tensor
for the unwarped internal metric.
 Using   $R = R_{MN} g^{MN}$,   it is straightforward to work out the Einstein
tensor $G_{MN} \equiv R_{MN} - {1\o 2} g_{MN}~R ~$    from the
above equations and to equate it to the energy momentum tensor
$T_{MN}$. Using the ansatz \genmetanzzs, the  curvature  scalar is
given by \eqn\ricscalr{R = e^{-2B} R^{(g)} + 2 e^{-2A} \left[D
\ddot B + d \ddot A + {1\o 2} D(1+D) {\dot B}^2 + D(d -1) \dot A
\dot B + {1\o 2} d(d-1) {\dot A}^2 \right]} where $R^{(g)}$ is the
curvature   scalar for the unwarped internal metric. For a Ricci
flat internal manifold, the  curvature  scalar will only have
contributions from the warp factors $A$ and $B$. For the analysis
above we took the warp factors to be  time-dependent only. In
general, they will be functions of $y^m$ and $t$, and the Ricci
tensors will get additional contributions, which we will display
later. For the purpose of this paper, we will assume that the
time- and $y^{m}$-dependences of the warp factors $A$ and $B$
factorize in the following way: \eqn\abfactor{ A(y,t) = {1\o
2}\left[{\rm log}~f(y) + {\rm log}~p(t)\right], ~~~~~~ B(y,t) =
{1\o 2}\left[{\rm log}~k(y) + {\rm log}~h(t)\right].} The
factorization of $A(y,t)$ obviously separates out a
($d+1$)-dimensional FLRW metric written in  conformal time as in
\conf, and is therefore desirable. It is also easy to see that in
the case of a supersymmetric compactification, the time
independence, $p(t) = h(t) = 1$, brings the warp factors in the
form given in \rBB.  Another argument in favor of \abfactor\ can
be given for the case where $h(t) = 1$. In this case, as we will
discuss in the next section, there is an anomaly condition that is
consistent {\it iff} the metric ansatz is of the form \abfactor.
For non-trivial   $h(t)$, the choice of the warp factors
\abfactor\ can isolate the time dependences from all the
background equations of motion, as will become  clear soon. Hence,
the assumed factorization \abfactor\ is highly desirable.

Obviously, a non-trivial   $h(t)$ would mean that the internal
manifold has some time dependence. As {mentioned earlier}, there
would then in general be a subtle issue in defining the concept of
moduli stabilization (especially the volume modulus). We will come
back to this point    later. Instead, let us first recall another
important consequence of such a non-trivial time-dependence. Given
\abfactor\ in \genmetanzzs, the metric simplifies   to
\eqn\factorized{ ds^2=f(y)p(t)~\eta_{\mu\nu}dx^{\mu} d x^{\nu} +
k(y) h(t)~g_{mn}(y)     dy^m dy^n.} If \factorized\ is in the
$(1+d+D)$-dimensional Einstein frame, then the
$(1+d+D)$-dimensional Einstein-Hilbert action can be integrated
over the $D$ compact dimensions to reduce  to an action in $(1+d)$
spacetime dimensions of the form \eqn\integralaction{ K\int
d^{d+1} x~ h^{D\o 2}~\sqrt{-{\rm det}(g_{\mu\nu})}~{R}^{(d+1)} +
.....} where $R^{(d+1)}$ is the curvature   scalar of   the metric
$g_{\mu\nu}\equiv p(t) \eta_{\mu\nu}$  in $(d+1)$ dimensions, and
$K$ is an integral over the internal $D$ dimensions. The dotted
terms  denote   the action for various fields that come from
dimensional reduction of the Einstein term. In the notations used
in \factorized, $K$ is given explictly by: \eqn\kr{K=\int d^{D}y ~
f^{d+1 \o 2}~ k^{D\o 2} ~\sqrt{ {\rm det}(g_{mn})}.}
$K$ can always be absorbed by an overall rescaling of the
coordinates $x^{\mu}$ and is irrelevant for the signs of $H$ and
${\ddot{a}\o a}$. The occurence of $h^{D\o 2}$ in \integralaction,
however, has to be properly taken care of by switching to the
$(d+1)$-dimensional Einstein frame metric, \eqn\einstframe{
g_{\mu\nu}^{E}= h^{D \o d-1}~ g_{\mu\nu} = h^{D\o d-1}~ p(t)~
\eta_{\mu\nu}~~~~~ \Longrightarrow ~~~~~ \eta^2(t)=h^{D \o
d-1}~p(t), } where $\eta^{2}(t)$ is now the proper scale factor of
the  proper  $(1+d)$-dimensional   Einstein frame metric \conf.
Furthermore, when the higher dimensional action has dilaton
factors with the Einstein term, the metric in lower dimension will
also pick up dilaton dependences. Combining \einstframe\ with the
dilaton factors will reproduce the right Einstein frame metric for
our case. However, for most of the analysis in this paper, the
dilaton will in fact be zero, and therefore we can ignore the
dilaton dependences altogether. In case the functions $p(t)$ and
$h(t)$ are simple powers, \eqn\simpowres{ p(t)=at^\alpha,~~~~~~~~
h(t)=b t^\beta,} we have $\eta^2(t)=c^2 t^{2\gamma} = c^2
t^{\alpha + {\beta D \o d-1}}$, and hence, \eqn\reasf{ {\beta D \o
d-1}+\alpha <0 ~~~~~~~ \Longrightarrow ~~~~~ {\ddot{a} \o {a}} >0}
will be our condition for an accelerating universe. We shall use
this in sec. 3 and 4 when we study some toy cosmologies in lower
dimensions.

In the following, we will naturally be interested in the $3+1$
dimensional cosmology which corresponds to a compactification of
the type IIB theory to four spacetime dimensions. As we will see,
however, the uplift to M-theory turns out to provide some
simplifications and requires some different values for $d$ and
$D$. Furthermore, we will allow for a fairly general time
dependence of the internal manifold.  This is partially also
motivated by the uplift to M-theory, where the internal manifold
generically may have a time dependence even if the IIB internal
manifold is completely static\foot{Other scenarios in which the
internal manifold is time-dependent by construction include \town\
or the models of \jarv\ or the racetrack inflation model of
\renand\ where the inflaton is actually the volume modulus. In a
slightly different vein, time dependent seven dimensional internal
manifold that is used to study four-dimensional cosmologies has
recently been addressed in \maeda.}.

So far, we have only concentrated on the metric, which we have
treated in a way general enough to encompass all string theories
as well as eleven-dimensional M-theory.  However, to study
promising cosmological models, one needs to switch on fluxes too.
In M-theory, the only available fluxes are {fluxes from the three
form potential} $C$.  In type IIB, by contrast, one can have a
variety of fluxes: the RR and NS three forms, $H_{NS}$ and
$H_{RR}$, as well as the five form, $F_5$, so unlike the metric,
the flux part should be analyzed case by case.
In this paper, our basic model corresponds to the D3/D7 system of type
IIB string theory on a six dimensional manifold with $H_{NS}$ and
$H_{RR}$ fluxes switched on.

There are, however, different approaches how this model can be
studied. For example, one can study the cosmology of this model
directly in type IIB theory. Alternatively, one can lift the whole
configuration to M-theory and study the cosmology from there. The
latter approach doesn't mean that one takes $d = 3$ and $D = 7$.
In fact, the lift to M-theory described in \dtds\ corresponds to
$d = 2$ and $D = 8$. This would mirror the type IIB dynamics with
$d = 3$ and $D = 6$. The M-theory approach has some clear
advantages over the type IIB case. In M-theory all the D7-branes
and O7-planes become pure orbifold singularities {of a torus
fibration}. The type IIB fluxes, on the other hand become G-fluxes
in M-theory. In M-theory, one would therefore have to concentrate
only on the dynamics of a {complex} fourfold with
G-fluxes\foot{One might worry whether type eleven
dimensional supergravity remains a good description when the fiber
torus becomes very small. As we discuss soon, we will consider
various higher derivative corrections that depend on the
curvatures and fluxes in M-theory. Since the fiber can be made
conformally flat over a large region of a six dimensional base,
the quantum corrections coming from torus size can be
controlled.}.  M-theory on such a fourfold with G-fluxes has been
well studied starting with \rBB, \sav.  In \rBB, it was noted
that, if we allow an ansatz \genmetanzzs\ for the metric with
$d=2$ and $D=8$, then the allowed form of the G-flux would be
\eqn\allgflux{G ~ = ~ \del_m~e^{3D}~dy^m \wedge dx^\mu \wedge
dx^\nu \wedge dx^\rho ~ + ~ {1\o 4!}
\del_{[\alpha}~C_{npq]}~dx^{\alpha} \wedge dy^n \wedge dy^p \wedge
dy^q} where $D(t,y^m)$ is another warp factor\foot{In this form,
the ansatz is slightly different from the one chosen by \rBB\ in
the static and supersymmetric case. More precisely, we made the
simplest extension to the non-supersymmetric case, namely, the
fields $C_{012}$ and $C_{mnp}$ are now allowed to be functions of
($y^m, t$) instead of just $y^m$, as in the supersymmetric case. A
more generic ansatze in which we switch on other components like
$C_{\mu mn}$ or $C_{\mu\nu m}$ probably could also be entertained
but we will not do so here.}, $x^\alpha \equiv (x^0, y^m)$, and
$C_{npq}$ is the three form completely in the internal space.  One
might wonder whether one could have a simple configuration where
this flux is switched off. It turns out, however, that if one does
not allow any M2 branes, then we cannot have vanishing three form
for a compact fourfold \rBB, \svw\ as there would otherwise be an
uncancelled anomaly from the non-trivial Euler number of the
fourfold. For the non-compact case, this is no longer true. This
also means that in type IIB we necessarily need the internal three
form(s) (as we plan to keep only a small countable number of D3
branes). Another important point is that supersymmetry would
require the internal G-fluxes to be primitive \rBB. On the other
hand, if we do not require supersymmetry, then the G-fluxes
\allgflux\ are not constrained by primitivity. This will be clear
soon when we discuss the background equations of motion.

With this in mind, we now consider the low energy limit in eleven
dimensions with leading quantum corrections and an $M2$ brane (i.e
the membrane) source term. This is the direct lift of the type IIB
background to M-theory. As is well known, the precise map of the
type IIB data to M-theory are\foot{Some aspect of this was
discussed earlier in \bbdgs. One of the author (KD) would also
like to thank C. Herdeiro and S. Hirano with whom related
investigation of these backgrounds were performed without taking
quantum corrections into account.}

\noindent $\bullet$ The threefold of type IIB becomes a fourfold in M-theory.
The fourfold has non-zero Euler number because the torus fibration is
non-trivial.

\noindent $\bullet$ The NS and RR fluxes of
type IIB will simply become G-fluxes
in M-theory. These G-fluxes are spread over the full fourfold.

\noindent $\bullet$ The seven-branes and orientifold   seven-planes {all
become singularities of the $T^2$ fibration of the fourfold, i.e., the
$T^2$-fiber degenerates over the six dimensional base. Note, however,
that the total fourfold itself is a {\it smooth} eight-dimensional
manifold.}

\noindent $\bullet$ The seven-brane gauge fluxes become {\it localised}
G-fluxes in M-theory. These localised fluxes appear at the
singularities {of the $T^{2}$-fibration} only and have zero
expectation values away from the singularities \bbdg.

\noindent $\bullet$ The D3-brane will become an M2 brane in
M-theory. The cosmology of the model will therefore consist of M2
brane(s) moving towards orbifold singularities of the {torus fibration
of the fourfold}.  Supersymmetry is broken by choosing non-primitive
G-fluxes.

To study this in detail, we need the full action in M-theory.  The
action will consist of three pieces: a bulk term, $S_{\rm bulk}$,
a quantum correction term, $S_{\rm quantum}$, and finally a
membrane source term, $S_{M2}$. The action is then given as the
sum of these three pieces: \eqn\action{S = S_{\rm bulk} + S_{\rm
quantum} + S_{M2}.} The individual pieces are:
\eqn\indivone{S_{\rm bulk} = {1\o 2\kappa^2} \int d^{11} x
\sqrt{-g} \left[R - {1\o 48} G^2 \right] - {1\o 12 \kappa^2} \int
C \wedge G \wedge G,} where we have defined $G = dC$, with $C$
being the usual three form of M-theory, and $\kappa^2 \equiv 8\pi
G_N^{(11)}$.  This is the bosonic part of the classical
eleven-dimensional supergravity action. It is clear that this is
not enough, as we require quantum corrections to the system,
otherwise we cannot put fluxes or branes in this scenario (see
\rBB, \svw\ for more details). The leading quantum correction to
the action can be written as: \eqn\quanta{S_{\rm quantum} = b_1
T_2 \int d^{11} x \sqrt{-g} \left[J_0 - {1\o 2} E_8 \right] - T_2
\int C \wedge X_8,} where the expressions for $J_0, E_8$ and $X_8$
were  already given in \rBB, \dtds\foot{Although not required for
this paper, the definition of $J_0, X_8$ and $E_8$ are: $$J_0 =
3\cdot 2^8 \left(R^{MNPQ} R_{KNPS}R_M^{~~~TUK}R^S_{~TUQ} + {1\o 2}
R^{MNPQ}R_{KSPQ}R_M^{~~~TUR}R^S_{~TUN}\right)$$
$$E_8 = {1\o 6}\epsilon^{ABCN_1...N_8}\epsilon_{ABCM_1...M_8} R^{M_1M_2}_{~~~~~~N_1N_2} R^{M_3M_4}_{~~~~~~N_3N_4} R^{M_5M_6}_{~~~~~~N_5N_6} R^{M_7M_8}_{~~~~~~N_7N_8}$$
$$ X_8 = {1\o 3 \cdot 2^9 \cdot \pi^4}\left[ {\rm tr}~R^4 - {1\o 4} ({\rm tr}~ R^2)^2 \right].$$
Thus, $E_8$ is the eleven dimensional generalization of the Euler
integrand.  Furthermore, the epsilon symbol appearing here is a
tensor and not a tensor density. Therefore, with all its indices
upper, this would be {proportional to} ${1\o \sqrt{-g}}$ while
with all its indices lower, it will be {proportional to}
$\sqrt{-g}$. This will be crucial later when we have to extract
the warp factor dependences of various components.}.  The
coefficient $T_2$ is the membrane tension. For our case, $T_2 =
\left({2\pi^2\o \kappa^2}\right)^{1\o 3}$, and $b_1$ is a constant
number given explicitly as $b_1 = (2\pi)^{-4} 3^{-2} 2^{-13}$. The
M2 brane action is now given by: \eqn\mtwo{S_{M2} = -{T_2\o 2}
\int d^3\sigma \sqrt{-\gamma}\left[\gamma^{\mu\nu} ~\del_\mu X^M
\del_\nu X^N g_{MN} - 1 + {1\o 3} \epsilon^{\mu\nu\rho} \del_\mu
X^M \del_\nu X^N \del_\rho X^P C_{MNP} \right],} where $X^M$ are
the embedding coordinates of the membrane (recall that the
membrane is along the 2+1 dimensional spacetime, and therefore is
a point on the fourfold).  The world-volume metric
$\gamma_{\mu\nu}, \mu,\nu = 0, 1, 2$ is simply the pull-back of
$g_{MN}$, the spacetime metric. {Due to the last term, the} motion
of this M2 brane is {obviously} influenced by the background
G-fluxes. The geodesic motion of the M2 brane \eqn\geomot{\quabla
X^P + \gamma^{\mu\nu}\del_\mu X^M \del_\nu X^N \Gamma^P_{MN} =
r_o,} where $\quabla$ is the Laplacian operator and \eqn\rzero{r_o
\equiv {1 \o 3!} ~\epsilon^{\mu\nu\rho} \del_\mu X^M \del_\nu X^N
\del_\rho X^Q G^P_{~MNQ},} is the standard one when $r_o = 0$. In
that case $\Gamma^P_{MN}$ would be the standard Christoffel
symbols. However because of the background G-fluxes, this is not
the case and $r_o$ becomes non-zero and will then govern the motion of
the M2 brane in this scenario.

\noindent The background field equations consist of two sets of
equations: \eqn\eien{\eqalign{ & R^{MN} - {1\o 2} g^{MN} R =
T^{MN} ~~~~~~~~~~~~~~~~~~~~~~ {\rm ~:Einstein~Equation}\cr & d\ast
G = {1\o 2} G \wedge G + 2\kappa^2 (T_2 X_8 + \ast J) ~~~~~~~~~~
{\rm :G-flux~Equation}}} The energy-momentum tensor appearing in
the Einstein equation has three pieces. The first is the standard
contribution from G-fluxes, given by: \eqn\tmnG{T^{MN}_G = {1\o
12}\left[G^{MPQR}G^N_{~~PQR} - {1\o 8}g^{MN}
G^{PQRS}G_{PQRS}\right].} The second contribution comes from the
quantum corrections. This is also easy to work out, and is
given by \eqn\tmnQ{T^{MN}_Q = {b_1 \kappa^2 T_2\o \sqrt{-g}}
{\del\o \del g_{MN}}\left[\sqrt{-g}(2J_0 - E_8)\right],} where all
the quantities have been defined earlier. The third, and  final,
contribution to the energy-momentum tensor comes from the membrane
term. Since the membrane is localised at a point on the fourfold,
we would require delta functions to describe   its position. The
membrane term is given explicitly by \eqn\tmB{T^{MN}_B =
-{\kappa^2 T_2 \o \sqrt{-g}} \int d^{3}\sigma~\sqrt{-\gamma}
\gamma^{\mu\nu}\del_\mu X^M \del_\nu X^N ~\delta^{11}(x - X),}
where we used $\delta^{11}(x - X)$ to denote the position of the
M2 brane on the fourfold. Since we expect the membrane to be
moving, these delta functions will themselves be time dependent
via $\delta^{11}(x -X(t))$. We will discuss the implications of
this when we study the solutions explictly.

\noindent The second equation in \eien\ is the equation for
G-fluxes. This can be written in terms of components as
\eqn\eientwo{\eqalign{D_M (G^{MPQR}) =~ &
{\epsilon^{PQRM_1....M_8}} \left[{1\o 2\cdot (4!)^2}
G_{M_1...M_4}G_{M_5...M_8} + {2\kappa^2 T_2\o 8!}
(X_8)_{M_1....M_8}\right] \cr & + {2\kappa^2 T_2 \o \sqrt{-g}}
\int d^{3}\sigma~\sqrt{-\gamma}~\epsilon^{\mu\nu\rho}\del_\mu X^P
\del_\nu X^Q \del_\rho X^R ~\delta^{11}(x - X)}} where the last
term is the definition of $J$  \foot{$J$ should  not be confused
with the quantity $J_{0}$.}
   that we used in \eien, and $D_M$ is the covariant derivative. Furthermore,
in addition to \eientwo, there would be the Bianchi identity for the G-fluxes\foot{The equations that we presented here will get further corrected by
higher derivative terms, which are in {\it addition} to the quantum terms that we mentioned here. The necessity of these terms will become clear when we will
try to construct explicit solutions to these equations.}.

It is worth mentioning that spacetime metrics of the form \conf,
\genmetanzzs\ are necessarily only applicable in an effective
supergravity description, which can potentially break down in some
regions. For instance, singularities in the warp factors $A(t),
B(t)$ can potentially develop at specific locations in the time
$t$-coordinate, where one must necessarily resort to a more
microscopic description. Obvious locations for such singularities
are the endpoints of the time $t$-interval where such a spacetime
metric is treated as valid: e.g. we expect a breakdown towards the
end of the slow-roll phase when the D3-brane makes contact with the
7-branes, signalling the dominance of possible stretched open
string tachyonic modes (in the Type II description). But also
noteworthy is the potential breakdown of the above solutions when
(in the M-theory lift) the M2-brane is far from any singularities
of the fourfold, since the higher derivative terms in the M-theory
lift are likely to be less important here and the M2-brane only
senses the local approximately trivial local geometry of the
fourfold\foot{For example, as we will encounter later in sec. 3, 
for early times (small t) we may have some components of the metric vanishing 
and $G_{0mnp}$ diverging with $C_{012}$ vanishing. While for late times (large t), 
we may have divergent metric warping with $G_{omnp}$ vanishing and $C_{012}$ diverging
(for instance, this example will appear in sec. (3.1)). 
Both ends at the level of the supergravity solution seem unreliable or 
imcomplete but the M-theory physics is not singular (just the two ends of 
the M2-brane motion). This of course doesn't mean that supergravity description is breaking down,
rather that we have to be somewhat careful in interpreting the physics here.}.    
Thus the gravity solutions we will find in what follows
are necessarily to be understood within these regimes of validity
and this will in fact be borne out in the specifics below.

\subsec{Analysis of the background equations of motion}

\noindent
Having set up the equations, it is now time to consider a
little more realistic model. We have already mentioned several general
features of the lift of the  $D3/D7$ system
to M-theory. For the particular choice of  $K3\times T^2/\IZ_{2}$ (or, more generally, $ K3\times\IP^1$ if one is away from the orientifold limit)
   as the    six-dimensional   compact
space in type IIB theory with  D3- and D7-branes, this lift to
M-theory was made more explicit in \dtds.  There it was argued
that the M-theory compactification will be on an eight-dimensional
manifold which is $K3 \times K3$. The fluxes in the $D3/D7$ system
that are responsible to break supersymmetry in the Coulomb phase
become G-fluxes in M-theory. The D7-brane that is wrapped on the
base K3 and is a point on the $\IP^1$ simply disappears in
M-theory and is absorbed in the other K3. Recall that a K3
manifold is a non-trivial $T^2$ fibration over a $\IP^1$ base.
This $\IP^1$ is the same $\IP^1$ on which our D7-brane in the IIB
picture   is a point. The  axion-dilaton, which is seeded by the
D7-brane(s),
  becomes the complex structure of the  $T^2$-fiber and degenerates precisely at the loci of the seven-branes on the $\IP^1$.  This way,
we can account for the complicated brane system of the  type IIB
framework by using   a simple (smooth)   compact manifold. The D3-brane, as we already mentioned,
becomes an M2 brane.

This is of course only one out of many possible  backgrounds. In sec. 3.5
we will also  discuss
another consistent background with D3/D7-branes that allows for certain black hole configurations from wrapped branes.
 The M-theory lift of this is another fourfold and will be   discussed in section 3.5.
The manifold  $K3 \times K3$ has $b_3 = 0$ whereas the other background
has a non-vanishing third Betti number   $b_3 \ge 1$. Both of these backgrounds have explicit F-theory
descriptions and therefore they are consistent static (and hence
supersymmetric) supergravity solutions, that may even be
quantum-mechanically exact.

The time dependent case, on the other hand, is a  slightly more   tricky issue.
The reason is that  the warp factors
 will not only depend on time $t$, but also on the internal
coordinates $y^m$. The metric ans\"atze that we want to consider for
this case will be slightly more generic than \genmetanzzs\ in the
sense that we will allow for three warp factors $A = A(t, y^m), B
= B(t, y^m)$ and $C(t, y^m)$, i.e \eqn\netvix{ds^2 = e^{2A}
~\eta_{\mu\nu}~dx^\mu dx^\nu + e^{2B}~g_{mn}~dy^m dy^n + e^{2C}
~\vert dz \vert^2,} $\mu, \nu = 0...2$, $m,n = 4...9$, and $dz =
dx^3 + \tau dx^{11}$. Here,  $dz$ is the coordinate of the fiber
torus with complex structure $\tau$, and the coordinates of the
fourfold are: $y^{4,5...9}, x^3, x^{11}$, where $x^3$ comes from
T-dualizing the third coordinate in the IIB picture and
$x^{11}$ is the  M-theory circle. The Ricci tensors will now
depend on both the time and the space indices. The unwarped metric
$g_{mn}$ is a function of $y^m$ only and will be taken to be
independent of time $t$. {Similarly, $\tau$ will depend on $y^m$
only.}  In other words,
 the only  time-dependence  is  assumed to    come from the warp factors $A, B$ and $C$.
The case where some components of the {unwarped}  metric
themselves are  time-dependent,   is equivalent to having a different ansatz for the metric (than the one that we  made   above in \netvix) where also more
time-dependent   warp factors for the internal space could exist
(for example three or more warp factors for the internal space).
Of course such an even  more general  ansatz cannot be ruled out and a
detailed
discussion will be presented in the sequel {to  this paper}.

The above ansatz with  three warp factors can be understood as a special case of the  two warp factor case   \genmetanzzs\ if the internal, unwarped
metric in \genmetanzzs\ is allowed to depend on time as well.  More precisely,  defining   $y^{a,b} = x^3, x^{11}$, we can rewrite the warp factors as
\eqn\warred{e^{2C}~\vert dz\vert^2 = e^{2B}~g_{ab}~dy^a dy^b,}
which is nothing more than saying that $g_{ab}$ is time independent. Using the above relation, physics with two warp factors $A, B$ in \genmetanzzs\ and three
warp factors $A, B, C$ in \netvix\ will be exactly the same. An alternative way to say this would be to observe that as long as
\eqn\cdefb{C = B + {\rm log}~\sqrt{g_{33}}, ~~~~~~{\rm and}~~~~~~ \del_0 C = \del_0 B}
there is no difference between \genmetanzzs\ and \netvix. The difference comes when the time evolution of $C$ and $B$ are not the same. In that case the theory will have
three different warp factors. This will happen, for example, when $g_{33}$
itself is time-dependent (which is equivalent to saying that we have a
different warped torus fiber). More generic discussion will be presented below.

When the internal manifold is a flat torus i.e when $g_{mn} = \delta_{mn}$ with a square fiber  two-torus,
and the warp factors only  {depend on the internal  space coordinates}, the Ricci tensors (with two warp factors) have been worked out in
\duff. For the generic case {of an additional time dependence (and with three warp factors)}  the Ricci tensor along the non-compact
  spacetime directions  is given by:
\eqn\rmununow{{\cal R}_{\mu\nu} = -\eta_{\mu\nu}e^{2A-2B}\left[
\quabla A + 3\del_m A \del^m A + 4 \del_m A \del^m B + 2 \del_m A
\del^m C \right] + R_{\mu\nu},} where $R_{\mu\nu} = R_{00},
R_{ij}$, ($i, j = 1,2$) are more complicated than the ones that we
derived earlier in \riccioo\ and \rmunu, respectively, due to the
presence of the third warp factor. They are now given by
\eqn\rijoon{\eqalign{&R_{ij} = \left[\ddot A + 6 \dot A \dot B +
\dot A^2 + 2 \dot A \dot C\right]~\delta_{ij} \cr & R_{00} = - 2
\ddot A - 6(\ddot B + \dot B^2 - \dot A \dot B) - 2(\ddot C + \dot
C^2 - \dot A \dot C).}} The Ricci tensor along the $m,n$
directions will likewise pick up both time and space derivatives
and is given by \eqn\rmnnowk{\eqalign{& {\cal R}_{mn} =3 \left[ 2
\del_{(m}A \del_{n)}B - \del_m A \del_n A - g_{mn} \del_k A \del^k
B \right] + 4 \left[ \del_m B \del_n B - g_{mn} \del_k B \del^k B
\right]~+  \cr & ~~~~~~~~  - 3 D_{(m} \del_{n)} A - 2 D_{(m}
\del_{n)} C + 2 \left[ 2 \del_{(m}C \del_{n)}B - \del_m C \del_n C
- g_{mn} \del_k C \del^k B \right]~+ \cr
 & ~~~~~~~~ -4 D_{(m} \del_{n)} B - g_{mn} \quabla B + R_{mn}^{(g)}
 + e^{2(B-A)} \left[ \ddot B + \dot A \dot B + 6 \dot B^2 + 2 \dot C \dot B \right] g_{mn}}}
where $R^{(g)}_{mn}$ is the ``unwarped''
Ricci tensor of the six dimensional base
   of the fourfold   and $D_m$ is the covariant derivative  with respect to   the unwarped internal metric $g_{mn}$.
An easy check of the above relations is to plug in
\eqn\plugin{g_{mn} = \delta_{mn}, ~~~ \tau = i, ~~~ \dot A = \dot
B = \dot C = 0, ~~~ B = C} assuming that the warp factors depend
only on the internal coordinates, which reproduces the Ricci
tensors of \duff\ with two warp factors. In addition to the above
set of components, there are now also  components of the Ricci
tensor that do not appear when the warp factors are independent of
time or when we have two warp factors  as in \genmetanzzs. These
are the Ricci tensors ${\cal R}_{0m}$ and ${\cal R}_{ab}$. They
are given by \eqn\rome{\eqalign{& {\cal R}_{0m} = 7 \dot B \del_m
A  + 2 \dot B \del_m C + 2 \dot C \del_m A - 2 \dot C \del_m C - 2
\del_m \dot A - 5 \del_m \dot B - 2 \del_m \dot C\cr & {\cal
R}_{ab} = -\delta_{ab}e^{2(C-B)}\left[ \quabla C + 3\del_m C
\del^m A + 4 \del_m C \del^m B + 2 \del_m C \del^m C \right] + \cr
& ~~~~~~~~~~~~~~~ + e^{2(C - A)} \left[ \ddot C + \dot A \dot C +
6 \dot C \dot B + 2 \dot C^2 \right]}} where we have taken $\tau
= i$ to evaluate ${\cal R}_{ab}$. Its easy to generalize to
arbitrary complex structures.
 One notes               that
the ${\cal R}_{0m}$ components will not be visible for the case
when we take the warp factors to be independent of $y^m$ but
dependent on time $t$. Similarly, if there are only   two warp
factors,  the terms $\dot B \del_m C - \dot C \del_m C$ in ${\cal
R}_{0m}$ would cancel when $B = C$. Therefore such terms are only
visible for more than two warp factors. With four warp factors
there would likewise  be even  more terms. The Ricci scalar is now
easy to evaluate from the tensors given above. It is given
explicitly by \eqn\reese{\eqalign{{\cal R} = ~& -e^{-2B} \left[ 10
\quabla B + 6 \quabla A + 2 \quabla C + 20 \del_m B \del^m B
\right] - 3 e^{-2B} \left[ 4 \del_m A \del^m A +6 \del_m A \del^m
B \right] + \cr & ~~~~~~ -2 e^{-2B} \left[ 3\del_m C \del^m C + 8
\del_m B \del^m C + 6 \del_m A \del^m C\right] + e^{-2B}~R^{(g)} +
\cr & ~~~~~~~ + 2 e^{-2A}\left[6 \ddot B + 2 \ddot A + 2 \ddot C +
21 \dot B^2 + 6 \dot A \dot B + 12 \dot C \dot B + 2 \dot A \dot C
+ \dot A^2 + 3 \dot C^2\right]}} where $R^{(g)}$ is the Ricci
scalar of the six dimensional base  {with metric $g_{mn}$}. Since
the fiber torus is a square one, it does not  degenerate in local
neighborhoods and therefore has vanishing curvature. This also
means that the seven-branes are kept far away. Again, this is not
the most generic picture, but   it is simple enough to illustrate
the basic structure.

Let us now come to the G-fluxes. We have
 decomposed   the spacetime coordinates as $x^M = [x^\mu, y^m, y^a]$. The G-fluxes would consequently have to be divided along those directions because of the
different warp factors. The G-fluxes are non-trivial functions of
the fourfold coordinates as well as time, just as  the warp
factors. As a simple special case, we will, in sec. 3.1, consider
in greater detail   warp factors that depend only  on time $t$.
But before we come to this special case,  let us first  take a
general   look at    the most generic case. For that it is
convenient to write everything in terms of unwarped metric and
unwarped fields. From the ansatz \netvix, we can easily see that
the warped and the unwarped fields are related as follows:
\eqn\warunwar{\eqalign{ & G^{012m}\to {G}^{012m} ~e^{-6A-2B}, ~~~~
G^{012a}\to {G}^{012a} ~e^{-6A-2C}, ~~~~{\rm det}~g \to e^{6A
+12B+4C}~{\rm det}~ g  \cr & G^{0mnp} \to {G}^{0mnp}~e^{-2A-6B},
G^{0mna} \to {G}^{0mna}~e^{-2A-4B - 2C}, G^{0mab} \to
{G}^{0mab}~e^{-2A-2B - 4C} \cr & G^{mnpq} \to {G}^{mnpq}~e^{-8B},
~~~~~ G^{mnpa} \to {G}^{mnpa}~e^{-6B - 2C}, ~~~~~ G^{mnab} \to
{G}^{mnab}~e^{-4B- 4C}}} where the unwarped fields are on the
right. We will also concentrate on the case where the world volume
coordinates $\sigma^{0,1,2}$ are identified with $x^{0,1,2}$. The
embedding coordinates $y^m$ and $y^a$, however,
 are no longer constants, as we expect the M2 brane to move
on the fourfold. Thus,
\eqn\gammamu{\gamma_{\mu\nu} = e^{2A} \eta_{\mu\nu} + e^{2B}~ \del_\mu y^m~ \del_\nu y^n ~ g_{mn} + e^{2C}~\del_\mu z ~\del_\nu \bar z,}
where we have isolated the unwarped metric components.

We remind the reader  that the M-theory solution is only to mimic the real type IIB solution that we are really  interested in.
From the choice of the metric \netvix\ in M-theory, the
type IIB metric can be easily worked out. For our case, let us define a vector ${\rm x} = \pmatrix{x^3\cr x^{11}}$, which is
the fiber coordinate, and denote   the corresponding metric of
the fiber torus as $g^t$. With these definitions, one has
\eqn\defmetnb{ e^{2C}~\vert dz\vert^2 \equiv  e^{2\tilde C}~ d{\rm x}^\top g^t ~d{\rm x} ~~ \Longrightarrow~~  C = \tilde C +
{\rm log}~\sqrt{{\rm tr}~(g^t\sigma_1)}, ~
\tau = {{\rm tr}~(g^t\sigma_4) + i~ \sqrt{{\rm det}~g^t} \o {\rm tr}~(g^t\sigma_1)},}
where $\sigma_i$ are {the  Chan-Paton matrices}    (see for example eq. (3.24) of \beckerD)\foot{In this language
${\rm det}~g^t =  {\rm tr}~(g^t\sigma_1)\cdot {\rm tr}~(g^t\sigma_3) - {\rm tr}~(g^t\sigma_2)\cdot {\rm tr}~(g^t\sigma_4)$.}. Observe also that
\defmetnb\ is more precise than \warred\ and \cdefb, as it encodes the information of 
non-trivial complex structure (i.e the fiber
torus is not a square one) in a concise way. Furthermore, writing the metric in
terms of warp factor $\tilde C$ instead of $C$ has the advantage
that $\tilde C = B$ will be the case of two warp factors. We will
eventually work out the two warp factor case in detail later. For
the present purpose, it will be instructive to write the type IIB
metric  we get from M-theory with three warp factors \netvix. This
type IIB metric turns out to be: \eqn\metrora{ds^2 = e^{2A + C }
\vert \tau \vert~\left(-dx_0^2 + dx_1^2 + dx_2^2 \right) +
{e^{-3C} ~\vert \tau \vert\o ({\rm Im}~\tau)^2} ~dx_3^2 + e^{2B +
C} \vert\tau \vert~g_{mn} ~dy^m dy^n ,} which tells us that the
metric of the internal space {in M-theory}  could in principle be
time dependent {even when the internal metric in IIB string theory
is not.} Both the axion $\tilde\phi$ and the dilaton field $\phi$ for this background
will in general not vanish as we have not removed  all
singularities far away. In fact the axion-dilaton ($\tilde\phi, \phi$) is generated by
the non-trivial complex structure of the fiber torus as
\eqn\axdil{\varphi = \tilde \phi + i e^{-\phi} = {\tau \o \vert
\tau\vert^2},} which tells that that for a square torus, $\tau =
i$, there would be no axionic field\foot{It is easy to see that when the torus is 
non-trivially fibered over the base (with coordinates $y^m$), we expect to get a 
cross-term $g_{3m}$ in the metric \metrora. Existence of 
such cross terms have been proposed  recently in a paper \gidm\ (although with only two 
warp factors) that appeared after our work. Here, in this paper, we will stick with 
the simplest possible ansatze for finding cosmological solutions. In the sequel to this
paper, more generic ansatze will be taken.}. 
In addition to this, the
G-fluxes can allow NS fluxes in Type IIB theory, and this would
break supersymmetry. Thus \metrora\ will be the expected metric
that we would get here, and along with the M-theory G-fluxes
\allgflux\ $-$ which will give us the non-primitive threeforms $-$
will specify the full background.

From the type IIB metric \metrora\ we observe something interesting. The metric along the
$x^{0, 1, 2}$ directions, in general,  behaves
differently from the metric along the
 $x^3$ direction. The difference comes from the three different warp factors that we {have allowed for   the M-theory   metric}.
We can therefore have the following four scenarios:
\eqn\abrela{\eqalign{& (1)~~A, B ~ = ~{\rm arbitrary}, ~~~~ C = B +
{\rm log}~\sqrt{{\rm tr}~(g^t\sigma_1)}\cr & (2)~~ A + 2 C + 2~ {\rm
log}~ {\rm Im}~\tau ~ = ~ 0, ~~~~ C = B+ {\rm log}~\sqrt{{\rm
tr}~(g^t\sigma_1)} \cr & (3)~~ A, B, C ~ = ~{\rm arbitrary} \cr &
(4)~~ A + 2C + 2~{\rm log}~ {\rm Im}~\tau ~ = ~ 0, ~~~~ B ~ = ~ {\rm
arbitrary}.}} The first two are basically two warp factor cases, and
the next two are {genuine}   three warp factor cases. The first and the third
case arise when we study the cosmology of the model without taking
all quantum corrections into account. The second case, will tell us
that we have a 3+1 dimensional space that has an overall warp factor
of ${e^{-3C} ~\vert \tau \vert \o ({\rm Im}~\tau)^2} $. As will be
clear soon, this case is an exact solution for the supersymmetric
background. All the background equations show miraculous
cancellations when we apply this condition. Whether this condition
persists for the time dependent case is still not clear. {This  case might}  arise when we take the full quantum corrections
into account. We will provide some evidence for this {in this paper}. A more
detailed analysis will be left for the sequel to this paper. The
fourth case, with a special relation between $A$ and $C$, is much less
constrained than case 2. It has an advantage over case 2 from the
fact that this might keep the internal six manifold in type IIB to
be     time-independent.    On the other hand, an increase in number of
warp factors makes the analysis of background equations of motion
much more   difficult.   Furthermore, it is also not clear
whether   the simple ansatz   for      the               G-flux with two warp factors \allgflux\
can carry over to the three warp factors case. In this paper,
therefore, we will be mostly concerned with two warp factors,
although we will perform some analysis with three warp factors
also.
For the two warp
factor cases, it would simplify quite a bit if we also move the
seven-branes far away and only study the motion of the D3-brane(s)
in a non-primitive background. This means {that the axion vanishes, i.e., }  we consider the fiber
torus to have a metric $g^t = {\rm diag}~[g_{33}~~g_{11,11}]$. For this
approximation, the  type IIB metric will become:
\eqn\metiib{ds^2 = e^{2A+ B} \sqrt{g_{11,11}}~\left(-dx_0^2 + dx_1^2 + dx_2^2 \right)
+{e^{-3B}\o g_{33}\sqrt{g_{11,11}}} dx_3^2 + e^{3B}
\sqrt{g_{11,11}}~g_{mn}~dy^m dy^n} with an almost vanishing
axion-dilaton\foot{The dilaton, or $e^{-\phi}$, is given by ${\rm Im}~\tau$
\axdil. If we choose the fiber components such that $g_{11,11} \approx g_{33}$,
then the dilaton can also be made very small.}.
Note   that for constant metric
components of the $T^2$ fiber of the fourfold, the second condition
of \abrela\  (with $g^t = \II$) will become $A \approx -2B$.
   This simplification will
have some important consequences as we will soon see.

Let us  first  study the background G-fluxes. The equations of
motion for the   G-fluxes are given by \eientwo. To study them, we
will allow the following non-trivial components of G-fluxes
\eqn\notrc{G_{012m}, ~~~~~~ G_{0mnp}, ~~~~~~ G_{mnpq}} where, {by
a slight abuse of notation,} $m,n$ denote {\it all} the
coordinates of the internal space\foot{These are the coordinates
of the fourfold and hence $m,n= 4...9,3,11$ unless mentioned
otherwise. When we go to type IIB, $m,n = 4...9$. It should always
be clear from the context whether we mean a fourfold or a six
dimensional base. Similarly the metric $g_{mn}$, Ricci-scalar
$R^{(g)}$ etc. are either fourfold or six-manifold data depending
on whether we are in M-theory or type IIB.}. Using the scaling
relations given in \warunwar\ and the first condition of \abrela,
the equation governing the component $G_{mnpq}$ is given by
\eqn\mnpqone{\del_0\left(e^{A+2B}G^{0mnq}\right) =
D_p\left[e^{3A}(G^{mnpq} - (\ast G)^{mnpq})\right] -{2\kappa^2
T_2\o 8!} \epsilon^{mnp a_1...a_8} (X_8)_{a_1...a_8}} where
$a_1...a_8 = 0,1,2...$ and therefore involve coordinates outside
the fourfold\foot{We have kept $X_8$ as a function of the {\it
warped} metric, as we are not very concerned about its scaling
behavior  with respect to   the warp factors $A,B$ in this paper.
Everything else is written in terms of the   unwarped metric.
Furthermore, the identity $\int X_8 = -{1\o 4!(2\pi)^4}\chi$ with
$\chi$ being the Euler number of the fourfold, continues to hold
because the warp factors are all globally defined variables and
therefore do not affect the integral.} that are essentially
conformally flat. Therefore they have vanishing curvature tensors.
Now since the $X_8$ term is measured w.r.t. to the curvatures, this
can be made arbitrarily small and we can ignore it completely. On
the other hand, when  $X_8$ has all the components inside the
fourfold, this will no longer be the case because of non-trivial
curvatures, and then $X_8$ will contribute.

The covariant derivatives and the Hodge star in \mnpqone\
are all measured  with respect to   the unwarped metric.
We now observe a few interesting
details from the above equation.  The first   is that the coefficient of $G^{0mnq}$ is $e^{A + 2B}$. Using our simplifying
second condition of \abrela\ for a square torus, this would simply go away.
And therefore \mnpqone\ will be an equation for $\del_0 G^{0mnq}$. The second observation is that, for the case when the warp factors and the G-fluxes are
independent of time $t$,  the fluxes satisfy the self-duality condition
\eqn\sedu{ G_{mnpq} ~ = ~ (\ast G)_{mnpq}}
which is exactly the primitivity condition. As is known from \rBB, having primitivity is equivalent to having supersymmetry. Therefore \mnpqone\ tells us
how we could break supersymmetry: a small deviation from primitivity \sedu\ will break supersymmetry for our case. We therefore define
\eqn\devfro{G_{mnpq} - (\ast G)_{mnpq} \equiv e^{-3A}\gamma_{mnpq} \ne 0}
as our condition for breaking supersymmetry. The quantity $\gamma$ would therefore determine the scale of susy breaking here. The above consideration actually
imposes the following two conditions on the G-flux components $G_{0mnq}$:
\eqn\confo{D_m G^{0mnq} ~ = ~ 0, ~~~~~~ \del_0 G^{0mnq} ~ = ~ D_p\gamma^{mnpq}.}
The first condition implies covariant constancy of $G^{0mnq}$, and the second implies susy breaking by G-fluxes. Thus, choosing a covariantly constant function
on the fourfold will give the spatial part of $G^{0mnq}$. The temporal part of $G^{0mnq}$, i.e the susy breaking parameter $\gamma$, on the other hand can be
related to the
other components of the  G-fluxes, the warp factors as well as the membrane velocities. The precise relation can be worked out with some effort and is
given by:
\eqn\temporeal{e^{-6B} G^{mnpq}G_{0npq} + {12 \kappa^2 T_2 e^{-6B} \o \sqrt{-g}}
\int d^3\sigma ~\dot y^m ~\delta^{11}(x - X) + 6e^{-6B} \del_0 \del^m e^{6B} - G^m = 0}
where we have defined $G^m \equiv  \gamma^{mnpq} G_{0npq}$. We have also taken $D = -2B$ in \allgflux.
The equation \temporeal\ can be used to relate the membrane velocities to the warp factors, when $\gamma$ and $G_{mnpq}$ are known. Therefore, to determine the membrane
velocities,
we need to know the warp factors completely as they
are again
  intertwined with G-fluxes. The equation for the warp factor can be determined from \eientwo\ by taking the equation of motion for the component
$G_{012m}$, as:
\eqn\warpff{ - \quabla e^{6B} = {1\o 2\cdot 4!} G_{mnpq} (\ast G)^{mnpq} + {2\kappa^2 T_2\o \sqrt{-g}} \left[ {\delta^8 (y - Y)} +
{X_8 \o 8!} \right]}
\noindent where $\quabla$ in the LHS comes from applying condition (2) of \abrela\ to \eientwo\ and \warunwar. 
Integrating the above equation reproduces the anomaly relation for the fourfold, implying that we can only put fluxes on a compact fourfold if we also
consider quantum corrections. For the non-compact fourfold, this poses no problem of course.

So far, we have found that  the component $G_{0mnq}$ can be determined
  by knowing the susy breaking function $\gamma$. The other component $G_{012m} = \del_m e^{3D}$
will be known from the solution of \warpff\ and the metric
$g_{mn}$ once we specify the relation between $D$ and $B$. From
the relation \devfro\ it is clear that once we determine the Hodge
operator $\ast$ we can use \devfro\ to fix the value of
$G_{mnpq}$. The equations that relate the metric components
$g_{mn}$ with the fluxes and warp factors are the Einstein
equations \eien. These are very complicated second order
equations, and therefore let us tackle them component by
component.   We start with   the 00 component: $G_{00} = T_{00}$.
The generic form of $G_{00}$ is given by:
\eqn\gooc{\eqalign{G_{00} = ~e^{2(A - B)}& \Bigg[d(2-D)\del_m A
\del^m B - {d(1+d)\o 2} \del_m A \del^m A - {(D-2)(D-1)\o 2}
\del_m B \del^m B + \cr & - d ~\quabla A + {{R^{(g)}} \o 2} - (D
-1) \quabla B \Bigg] + {d(d-1)\o 2} \dot A^2 + {D(D-1)\o 2} \dot
B^2 + D d \dot A \dot B}} where $d =2$ and $D = 8$. The scalar
curvature of the fourfold is given by $R^{(g)}$. The Einstein
equation will become \eqn\eienoo{\eqalign{ & e^{2(A - B)} \left[
{R^{(g)} \o 2} -12 \del_m A \del^m B - {21\o 4} \del_m A \del^m A
-21 \del_m B \del^m B - 2 ~\quabla A - 7~\quabla B \right] + \dot
A^2 + \cr & ~~~~~~~~~ + 28 \dot B^2 + 16 \dot A \dot B = {e^{-6 B}
\o 4!} G_{0mnp}G^{0mnp} + e^{2A - 8B} \Bigg( {1\o 4 \cdot 4!}
G_{mnpq} G^{mnpq} ~ + \cr & ~~ - {\kappa^2 T_2 e^{-A}\o \sqrt{-g}}
\int d^3 \sigma \sqrt{-\gamma} ~\gamma^{00} ~\delta^{11} (x - X)
\Bigg) + {\kappa^2 b_1 T_2 e^{-3A-8B}\o 2\sqrt{-g}} {\del \o \del
A} \Big[\sqrt{-g}(2J_o - E_8)e^{\rm f}\Big]}} where $b_1$ is a
constant defined earlier, ${\rm f} \equiv a_1 A + a_2 B$ where
$a_{1,2}$ are constants that can be determined from the metric
components and $\gamma^{00}$ is a non-trivial function of the
membrane velocities. The term ${21\o 4}~\del_m A \del^m A$ in
\eienoo\ comes from the contribution of the energy momentum tensor
$G_{012m}G^{012m}$, and the last term in \eienoo\ comes from
quantum corrections. If we define $E_q \equiv \sqrt{-g}(2J_0 -
E_8)e^{\rm f}$, then the quantum corrections to the energy
momentum tensor fall into the following three categories:
\eqn\enercat{ {\cal T}_1 = {\del E_q \o \del A}, ~~~~~  {\cal T}_2
= {\del E_q \o \del B}, ~~~~~  [{\cal T}_3]_{mn} = {\del E_q \o
~\del g_{mn}}.} Therefore even for $g_{mn} = \delta_{mn}$, there
would be contributions from the quantum corrections coming from
the warp factors $A$ and $B$. For the specific choice of our
metric, all the ${\cal T}_i$ are non-zero.

Let us now analyze the equation \eienoo. We can plug in the values
of $G_{0mnp}$ and $G_{mnpq}$ which we determined earlier. The
equation \eienoo\ therefore intertwines $A, B, \dot y^m$ and
$g_{mn}$, where $\dot y^m$ enters through the definition of
$\gamma$ and $\gamma^{00}$ given earlier in \gammamu.
 The scalar curvature is of course zero for a  Ricci flat fourfold. Observe also that we haven't yet imposed the relation
$A = -2B$ on the above equation. Before we do that, we can
re-arrange this equation a little bit by using the warp factor
equation \warpff. The warp factor equation is written in terms of
$\quabla e^{6B}$ on the LHS. In terms of $A$ and $B$, the LHS of
\warpff\ can be replaced by \eqn\lhswa{ \quabla e^{6B} ~
\leftrightarrow ~ - 3 ~D_m \left( e^{6B} \del^m A \right)} with
the RHS of \warpff\ unchanged. Plugging \lhswa\ into \warpff, we
can add the equations \eienoo\ and \warpff\ in the following way:
\eqn\adding{ 2\Big[{\rm eqn}~\eienoo\Big] - e^{2A-8B} \Big[{\rm
eqn}~\warpff\Big],} so that the final equation for our background
will become: \eqn\eienfi{\eqalign{& e^{2(A - B)} \left[R^{(g)} -
42\left(\del_m A \del^m B + {1\o 4} \del_m A \del^m A + \del_m B
\del^m B\right) - 7(\quabla A + 2~\quabla B)\right] + 2\dot A^2 +
\cr & + 56 \dot B^2 + 32 \dot A \dot B = {e^{-6 B} \o 2\cdot 3!}
G_{0mnp}G^{0mnp} + {e^{2A - 8B} \o 2 \cdot 4!}
\left[G_{mnpq}G^{mnpq} - G_{mnpq} (\ast G)^{mnpq}\right] ~ + \cr &
+ {2e^{2A - 8B} \kappa^2 T_2\o \sqrt{-g}} \Bigg[\left({b_1
e^{-5A}\o 2} {\del E_q \o \del A} - {X_8\o 8!}\right) - \int
d^3\sigma \left(e^{-A} \sqrt{-\gamma} \gamma^{00} +
1\right)\delta^{11}(x-X)\Bigg].}} The above equation may look even
more complicated than the one that we started with, i.e \eienoo.
However if we look closely, we see that once we impose the
condition $A = -2B$ the equation \eienfi\ shows some amazing
simplification. Putting $A = -2B$ in \eienfi\ results in the
following equation: \eqn\eiensec{{1\o 12} G_{0mnq}G^{0mnq} + {1\o
48}~ G_{mnpq}\gamma^{mnpq} +  e^{-6B} T_2 {\cal Q} ~ = ~ 0} where
${\cal Q}$ is the last term of \eienfi,  which is a combination of
the quantum and the membrane term. In deriving this we have
strictly used $A = -2B$. This is true only when $g_{33}g_{11,11}
=$ constant. This however is not true generically. Therefore we
will get additional warp factor dependent terms in \eiensec\ when
we use $A = -2B - {\rm log}~\sqrt{g_{33}g_{11,11}}$ as the
relation between the warp factors.

In any case, the above equation \eiensec\ can only tell us the $y^m$ dependent terms of the warp factor. For the complete description we need the time dependences of the
warp factor. This can in principle come from the other Einstein equations. The $G_{ij}$, $i,j = 1,2$ components can be written as:
\eqn\gijcol{\eqalign{G_{ij} ~ = ~ & \delta_{ij}~e^{2(A - B)} \Bigg[d(D -2) \del_m A\del^m B + {d(d+1)\o 2} \del_m A\del^m A + d ~\quabla A + (D -1)~\quabla B + \cr
& ~~~~~~ - {R^{(g)}\o 2} +  {(D-2)(D-1)\o 2}\del_m B\del^m B \Bigg] + \delta_{ij} \Bigg[(1-d)\ddot A  - D\ddot B + \cr
& ~~~~~~~~ -{D(D+1)\o 2} \dot B^2 - {(D-1)(D-2) \o 2} \dot A^2  - D(d-2) \dot A \dot B \Bigg].}}
Putting in the values of $d = 2$ and $D = 8$, we can get the Einstein equation $G_{ij} = T_{ij}$, for $i, j = 1,2$. Since $x^1, x^2$       are on   the same footing, the
equation for any of these components is:
\eqn\eienij{\eqalign{&e^{2(A - B)}\left[12 \del_m A \del^m B + {21 \o 4} \del_m A \del^m A + 2~\quabla A + 7~\quabla B + 21 \del_m B \del^m B - {R^{(g)}\o 2}\right]
 - \ddot A \cr & ~~~~~ - 8 \ddot B - 36 \dot B^2 = {e^{-6B}\o 4!} ~G_{0mnq}G^{0mnq} - {b_1 \kappa^2 T_2 e^{-3A - 8B} \o 2\sqrt{-g}}~{\del E_q\o \del A} ~+ \cr
& ~~~~~~~ - {e^{2A-8B}\o 4\cdot 4!}G_{mnpq}G^{mnpq} - {\kappa^2 T_2 e^{A - 8B}\o \sqrt{-g}}\int d^3 \sigma \sqrt{-\gamma} ~\gamma^{ii} ~\delta^{11} (x - X).}}
Comparing \eienij\ with \eienoo\ we see some crucial differences. The space derivatives of the warp factors are exactly negative of each other, and so is the
quantum correction. On the other   hand,   the   $G_{0mnq}^2$ term comes with the same sign, and the membrane term has $\gamma^{ii}$ (no sum over $i$) instead of
$\gamma^{00}$ as in \eienoo.
 This tells us that we can add these two equations to get the following simple relation:
\eqn\eienadd{\eqalign{ \left[{\rm eqn}~\eienoo\right]~ + ~
\left[{\rm eqn}~\eienij\right] ~ = ~ & \dot A^2 - 8 \dot B^2 + 16
\dot A \dot B - \ddot A - 8 \ddot B ~ = ~  {e^{-6B} \o 12}
~G_{0mnq}G^{0mnq} ~ + \cr &~~ -  {\kappa^2 T_2 e^{A - 8B} \o
\sqrt{-g}} \int d^3 \sigma \sqrt{-\gamma} ~(\gamma^{00} +
\gamma^{ii}) ~\delta^{11}(x - X)}} which  relates the time
derivatives of the warp factors to the G-fluxes $G_{0mnq}$ and the
membrane velocities. We now make three observations related to
\eienadd.
   The first   is that in the static gauge (where we identify the membrane
coordinates with $x^{0,1,2}$),
$\gamma^{00} + \gamma^{ii} = 0$ (again no sum over $i$ indices), and therefore the membrane contribution would vanish. For our case this term is non vanishing
precisely because of the membrane velocities. Secondly, \eienadd\ will be one equation where the quantum effects are cancelled out. We will soon exploit this
property to get some cosmological solutions. And finally, since the LHS of \eienadd\ involves only time derivatives the $\sqrt{g_{33}g_{11,11}}$ term in
$A = -2B - {\rm log}~\sqrt{g_{33}g_{11,11}}$ will drop out, and therefore the LHS will become $-6(\ddot B + 6\dot B^2)$. This will be useful soon.

Most of the above analysis has focussed on the time and space
derivatives of the warp factors $A$ and $B$ with the explicit
metric of the internal fourfold appearing only through higher
order corrections or/and through possible membrane terms. A more
direct way to get the metric would be to concentrate on the
$G_{mn}$ term. This is given explicitly as
\eqn\gmnterms{\eqalign{& ~~~~~~~~~~ G_{mn} = G^{(g)}_{mn} + (d+1)
\left[2 \del_{(m}A \del_{n)}B - \del_m A \del_n A -
D_{(m}\del_{n)}A \right] + \cr & + (D-2) \Bigg[\del_m B \del_n B +
D_{(m}\del_{n)}B \Bigg] + g_{mn}\Bigg[(d+1)\Big(\quabla A + (D -
3) \del_k A \del^k B + \cr & ~~~~~~~~~~~~  + {d+2\o 2} \del_k A
\del^k A\Big) + (D-2)\left(\quabla B +{D -3\o 2} \del_k B \del^k
B\right)\Bigg] \cr &  - e^{2(A - B)}g_{mn}\left[(D-1)\ddot B + d
\ddot A + {D(D-1)\o 2} ~\dot B^2 + (D-1)(d-1) \dot A \dot B
+{d(d-1)\o 2}~\dot A^2 \right]}} where $G^{(g)}_{mn}$ is the
Einstein tensor for the unwarped metric $g_{mn}$. For the fourfold
in consideration, the equation of motion resulting from the above
will be: \eqn\eienmn{\eqalign{& G^{(g)}_{mn} -
3\left[D_{(m}\del_{n)}A + 2D_{(m}\del_{n)}B \right] +
6\left[\del_{(m}A \del_{n)}B + {1\o 4} \del_m A \del_n A + \del_m
B \del_n B \right] ~ +  \cr &~~~~~~ + g_{mn} \Bigg[3~\quabla A +
6~\quabla B + 15\left( \del_k A \del^k B + {1\o 4} \del_k A \del^k
A + \del_k B \del^k B \right)\Bigg] ~+ \cr & - e^{2(B - A)}g_{mn}
\left(7\ddot B+ 2 \ddot A + 28 \dot B^2 + 7 \dot A \dot B + \dot
A^2\right)
 = {e^{-6B}\o 12} \left[G_{mpqr}G^{~~pqr}_n - {g_{mn} \o 8}G_{pqrs}G^{pqrs}\right] \cr
& ~~~~~~ - {e^{-2A-4B}\o 4}\left[G_{0mpq}G^{~~0pq}_n -{g_{mn}\o 6}
G_{0pqr}G^{0pqr}\right]  + {b_1 ~\kappa^2 ~T_2 e^{-3A - 10B} \o
\sqrt{-g}} \Bigg[ {1\o 2} (g^{-1})_{mn} {\cal T}_2 + \cr & ~~~~~~
+ [{\cal T}_3]_{mn} \Bigg] - {T_2 \kappa^2 \o \sqrt{-g}} e^{-3A -
4B} \int d^3\sigma ~\sqrt{-\gamma}~ \gamma^{00}~\dot y^p~ \dot
y^q~g_{mp}~ g_{nq}~\delta^{11}(x - X)}} where the ${\cal T}_i$
are defined in \enercat. We see that all the ${\cal T}_i$ are
responsible for the total energy momentum tensor in this scenario
(because of the non-trivial warp factors and metric components).
Thus along with the membrane contribution, \eienmn\ will determine
the unwarped metric components once all others are known.

Looking at \eienmn\ we see that the equation can get drastically
simplified once we impose the relation between the two warp factors $A
\approx -2B$. For the Ricci flat fourfold, all the spatial derivatives
of $A$ and $B$ go away completely in the limit $A = -2B$. Of course
this simplifying relation is not true generically, and therefore we
will eventually be left with some spatial derivatives. For the time
being, as above, let us concentrate only on $A = -2B$. The LHS of
\eienmn\ will now become $-3 g_{mn}~ e^{6B}~ (\ddot B + 6 \dot
B^2)$. Observe that the time derivatives of $B$ is exactly the same as
we got earlier for \eienadd. On the other hand \eienadd\ doesn't have
any quantum terms whereas \eienmn\ does. Therefore by contracting
\eienmn\ with metric $g_{mn}$ we can have non-trivial relations
between the quantum (and possibly the membrane) terms.  It will be
interesting to relate the resulting equation to \eiensec.  We will
however not exploit these details in this paper and leave the rest of
the discussion for the sequel. The cosmological solutions that we will
study in the next two sections will ignore some of the quantum
effects.

The above set of equations would have sufficed if the warp factors
were only space or time dependent. This is obviously not the case
here and therefore we will have yet another equation that would
relate the space-time derivatives of the warp factors. Recall that
so far we had either space derivatives or the time derivatives (or
the sum of both) of the warp factors i.e $\del_m\del_n A$ or
$\ddot A$ etc. Now we expect equations that relate $\del_m \dot A$
terms. In fact there is a non-trivial Einstein tensor that does
the job for us. This is $G_{0m}$ and is given by \eqn\gom{G_{0m} =
(d+D-1) \dot B \del_m A - d \del_m \dot A - (D-1) \del_m \dot B,}
where $m$, as usual, are {all} the internal coordinates. The
Einstein equation resulting from this can be easily derived from
the requisite energy-momentum tensor.  It is given explicitly as:
\eqn\eienom{\eqalign{9 \dot B \del_m A - 2 \del_m \dot A - 7
\del_m \dot B ~ =~& {e^{-6B}\o 2\cdot 3!}~G_{0pqr}G_m^{~~pqr} ~ +
\cr &  - {e^{-A-6B} \kappa^2 T_2 \o \sqrt{-g}} \int d^3\sigma
\sqrt{-\gamma}~\gamma^{00} ~\dot y^n~g_{mn}~\delta^{11}(x - X)}}
which has vanishing quantum terms because of the absence of cross
terms in the metric of the form $g_{0m}$. In the presence of such
cross terms the above equation \eienom\ will have an additional
quantum term. When we consider $A = -2B$ the LHS of \eienom\
becomes $-3(\del_m\del_0 B + 6 \del_0 B \del_m B)$ which is then
related to the G-fluxes and the membrane velocities via \eienom.

The equation \eienom\ can now be compared to \temporeal. Both
these equations relate the membrane velocities to the warp factors
and G-fluxes. To see whether there is any connection between the
two equations, we have to rewrite \temporeal\ in a more suggestive
way, as: \eqn\tempoagain{3~\del_0(e^{6B}~\del^m A) ~ = ~ {1\o
3!}~(\ast G)^{mnpq} G_{0npq} + {2\kappa^2 T_2 \o \sqrt{-g}} \int
d^3\sigma ~\dot y^m ~\delta^{11}(x-X).} In this form this equation
is close to \eienom. In the limit $A = -2B$, the LHS of
\tempoagain\ becomes $-6(\del_m\del_o B + 6 \del_o B \del_m B)$.
Thus, \eqn\tempom{\eqalign{\left[{\rm eqn}~\tempoagain\right] -
2e^{6B} \left[{\rm eqn}~\eienom\right] ~ & =  ~{1\o 3!} ~G_{0npq}
\left[ (\ast G)^{mnpq} - G^{mnpq} \right] \cr & = ~ {\kappa^2 T_2
\o \sqrt{-g}}~e^{2(B-A)} ~\dot y^m ~\vert \dot y
\vert^2~\delta^8(y-y(t)).}} Comparing this equation with \devfro,
we see that if the LHS is zero, then the membrane velocities have
to vanish. This is perfectly consistent with what we know, namely:
when \devfro\ vanishes, we have {\it primitive} G-fluxes.
Therefore supersymmetry will be preserved, and so the membrane
cannot move. Breaking primitivity via switching on a non-zero
$\gamma_{mnpq}$ will immediately trigger the motion of the M2
branes from \tempom. Therefore we need a non-zero $G_{mnpq}$
satisfying \devfro\ with non-vanishing $\gamma_{mnpq}$ to start
the motion of the M2 brane in our setup. The non-zero $G_{mnpq}$
on the other hand should be (a) quantized, and (b) time
independent. The former condition is the requirement first
proposed in \wittenflux, and later elaborated on in numerous
papers including \sav; and the latter condition is from the warp
factor equation \warpff. It is encouraging to see how the expected
conditions appear from the background equations of motion.

With this set of equations, the background can, in principle, be
completely determined. As we saw above, the background equations of
motion are pretty involved not only because they are second order
differential equations, but also because they involve quantum
corrections that are fourth order in curvature. Solving these
equations therefore becomes complicated. In the next few sections we
will try to solve these equations by ignoring some of the quantum and
the membrane terms. Again because of the anomaly constraint \warpff\
it is not always possible to ignore quantum corrections. We have to
consider things case by case and see how far we can ignore these
corrections. The cases we will concentrate on here are the ones given
in \abrela\ which considers only two warp factors $A$ and $B$. As
mentioned earlier, this is by no means a generic choice.  In fact we
will see that with two warp factors we get cosmological examples, that
are interesting, but not quite a de-Sitter one.  Therefore, it could
be that the dynamical evolution of the fields in this theory is given
by three or more warp factors.  We will provide some evidence here
that with three warp factors we can get close to a de-Sitter
cosmology, but the precise equations will not be worked out here.
These and other detailed scenarios will be investigated in the sequel
to this paper.

Another interesting question is to ask if there are general
restrictions on the possible cosmologies (i.e. on the warp factors)
that are allowed when the higher derivative terms are included, in
particular stemming from the energy conditions that they impose (see
Sec.~(3.3) for more on this). Furthermore it is tempting to draw
potential parallels between this effective theory involving higher
derivative terms in this cosmological setting and the microscopic
entropy counting matches \senBH, \Atish\ and string-corrections
\maloney, \senhetbh\ of 2-charge BPS black hole solutions in
supergravity theories with higher derivative terms \dewit, including
questions of possible field redefinitions therein \senhetbh.

\newsec{Some toy cosmologies in detail}

\noindent
In the previous section, we have laid out the full set of equations
that govern cosmological solutions of M-theory on a fourfold with
time-dependent warp factors, fluxes and M2-branes. In this section we
make a first attempt towards solutions of these equations by
considering some simplified situations that reduce the complexities of
the equations of motion. These toy cosmologies are not always meant to
correspond to realistic cosmologies, but rather serve as means to
discuss various important features of the more realistic scenarios in
a simplified context and to gain some better intuition for these more
realistic solutions.

Before we study these toy models, however, it is worth emphasizing an
important difference to the cosmological solutions that have mostly been 
studied before.  In order to see this, recall that our framework of
M-theory on an fourfold (corresponding to $d=2$ and $D=8$) is chosen
only as a tool to study the cosmology of the D3/D7 system in type IIB
theory on a threefold (corresponding to $d=3$ and $D=6$), which is
what we are really interested in.  This distinguishes our approach
from direct studies of $(3+1)$-dimensional cosmologies (i.e., the case
$d=3$) in compactifications of six- or seven-dimensional spaces (i.e.,
$D=6$ or $D=7$) in string or M-theory, respectively, which appeared
for example in \chodos.  Typical among those examples are Kasner type
solutions or Jordan-Brans-Dicke type solutions \kasner\ (see also
\buwal\ for further work on cosmological solutions in M- or string
theory).  For the type of solutions studied in \chodos, the typical
behavior of the warp factor is
\eqn\wabehave{e^A ~=~ \sum_n ~a_n ~t^n, ~~~~~~~~~ e^B ~ = ~ \sum_m ~b_m~t^m, ~~~~~~ {\rm with}~~ n_{\rm max} > m_{\rm   max},}
i.e., three spatial dimensions expand much faster than the remaining
six/seven dimensions.  In our case with $d=2$ and $D=8$, the situation
is a bit more complicated, as one of the internal dimensions of the
fourfold should correspond to the third expanding dimension in the
type IIB picture. The interpretation of the different warp factors
therefore requires some care, as we will soon illustrate.

In the following examples we will study cosmologies when we switch on
small internal fluxes, and also arbitrary internal fluxes. We will
give a new six dimensional space that satisfies all the equations of
motion and whose metric is completely determinable; along with
cosmological effects from wrapped branes and brane-antibranes
annihilations on this geometry. All these cases will mostly be with
two warp factors. We shall give evidence that with three warp factors
it might be possible to get a de-Sitter type IIB cosmology.

\subsec{{Example 1}: Small Background Fluxes}

\noindent
In order to reduce the complexity due to the $y^m$- and $t$-dependence
of the warp factors, we will start with a toy example where the warp
factors only depend on time $t$. We will further assume that we have
only two warp factors, i.e., we will consider case 1 of \abrela\ with
warp factors $A(t)$ and $B(t)$ as we are not putting any relation
between the warp factors {\it a-priori}.

For this case, the M2 brane as usual will move on the fourfold. The
equation of motion for the M2 brane is \geomot\ with
$\gamma_{\mu\nu}$ given by \gammamu.  In the case where the
embedding coordinates of the membrane {only} depend on time $t$,
the above relation \gammamu\ simplifies to \eqn\simea{\gamma_{00}
= - e^{2A} + e^{2B} ~\dot y^m ~\dot y^n ~g_{mn}, ~~~~ \gamma_{11}
= \gamma_{22} = e^{2A}, ~~~~ \gamma_{12} = 0,} and therefore the
world volume metric is no longer trivial {as it would be in} the
static supersymmetric case. The motion of the M2 brane is
triggered by non-primitive G-fluxes via \tempom. Therefore, we
need a non-zero $G_{mnpq}$ which is quantized and time independent
that satisfies \devfro, and most importantly \warpff. For a G-flux
to satisfy \warpff, we have to {\it necessarily} impose quantum
corrections, {because} without quantum corrections, $G_{mnpq} =
0$. We also have to solve \mnpqone\ for the component $G_{0mnp}$,
and in general we would require $G_{0mnp}$ to be time dependent.
On the other hand, $G_{mnpq}$ has to be time independent so as to
satisfy \warpff, and also \devfro, to break supersymmetry.
Furthermore, the integral of $G_{mnpq}$ over {\it any} four-cycle
has to be quantized as an integer or as a fraction depending on
the type of background one chooses. This leads us to the following
ansatz for the background three-form field in the internal space:
\eqn\cmnp{C = \left[\psi (t) + \lambda(y)\right]_{mnp}~dy^m \wedge
dy^n \wedge dy^p.} The G-fluxes consequently will be: $G_{0mnp} =
\del_0 \psi_{mnp}$ and $G_{mnpq} = 4\del_{[m}\lambda_{npq]}$. We
also require \eqn\greq{{1\o 2\pi}\int_{4-{\rm cycle}}
G_{mnpq}~dy^m\wedge dy^n\wedge dy^p\wedge dy^q ~ = ~ {\IZ\o 2},}
where $\IZ$ is an integer. If we assume that all the four-cycles
are large in the cosmological setting, then we can break
supersymmetry by very small expectation values of the $G$ fluxes.
This means in particular that $\gamma$ in \devfro\ is $\gamma \sim
c_3 e^{3A}$ with $c_3 \to 0$. The $C$ field will now take the
following form: \eqn\cnowf{ C = \left[\psi(t) + \epsilon\cdot
y\right]_{mnp} ~dy^m \wedge dy^n \wedge dy^p, ~~~~~~~ \epsilon \to
0} and $\epsilon$ and $c_3$ can be related by \devfro. The case
with $\gamma$ not small will be dealt with  later.

With this ansatz for the $C$ field, the value of $G_{0mnp}$ can be
easily determined from \mnpqone. The RHS of \mnpqone\ has $D_p
\gamma^{mnpq}$, and this vanishes because it is only a function of
time (via the warp factor $A(t)$).  In general, therefore, the
internal components of the G-flux are determined solely by the warp
factors appearing in the metric and a possible quantum correction as:
\eqn\one{G^{0mnp} ~ = ~e^{-A-2B}~\int dt~\left(D_p \gamma^{mnpq}\right) +  e^{-A- 2B} ~(c_o +  c_1)^{npq}}
where $c_o$ is independent of time and $c_1$ is the quantum correction coming
from the $X_8$ term in \mnpqone.
Since some of the directions of $X_8$ are  along $x^{1,2}$, $c_1 =
0$ for our case\foot{This quantum correction may not always vanish
for the case when all components are inside the fourfold. But we
can make this small by taking the fourfold as an orbifold, with
orbifold singularities. The quantum corrections there will be
determined from the number of singularities and will in fact be
close to zero away from the singular points. If we take the base
of the fourfold to be large and assume that the orbifold points
are also shifted far away, then this can be made very small. This
way we can keep the $x^3$ and $x^{11}$ directions of the fiber to
be small and still get a good type IIB cosmology.}.

Consider now the motion of the M2 brane on the fourfold. We expect
the membrane to move slowly on the fourfold towards one of the
orbifold singularities. Once we isolate a bunch of singularities
at a point and keep the other singularities far away, the membrane
will eventually fall into the singularities that are nearby. With
the assumption that we  made  above, this is possible to realize.
It is easy to see   that if we ignore the quantum correction to
the energy-momentum tensor $T^{0m}$ \eqn\tom{ {\del \o \del
g_{0m}}~(2J_0-E_8) \approx 0} and use \eienom\ with $G_{mnpq} \to
0$ and \one, then the velocity of the membrane on the fourfold
will be constrained as \eqn\velofm{{\kappa^2 T_2 \o {\rm
V}_8}~{\dot y^m \o \sqrt{1- v^2}} = {1\o 2\cdot 3!}~
G_{0npq}~G^{mnpq},} where ${\rm V}_8$ is the volume of the
fourfold, $v$ is the warped velocity of the membrane  given by $v
= e^{B-A}~\vert \dot y\vert$, and $G^{mnpq}$ is proportional to
$\epsilon$ as discussed above. Thus, the internal flux $G_{mnpq}$
is a vanishingly small quantity. However, the RHS of the above
equation would not vanish when the warp factors are functions of
the internal coordinates, or/and if we switch on large expectation
values for $G_{mnpq}$. We will discuss these cases later, but for
the time being let us assume \velofm\ to hold. One possible way
for \velofm\ to hold without violating any of the conditions would
be to assume a very slow velocity of the membrane on a fourfold of
very large volume ${\rm V}_8$, that is\foot{The issue of large
size fourfold is a little subtle, because it deals with moduli
stabilization. We will discuss this later.}. \eqn\condme{ \dot y^m
~ \approx ~ 0, ~~~~~~~ {\rm V}_8 ~ \to~ \infty.} This way we can
take care of \velofm. The above equation would also mean that the
contribution to the energy momentum tensor coming from the
velocities of the membrane is almost negligible and therefore we
can ignore it for the time being. Thus, \condme\ will be our slow
roll condition. For the case when $A = -2B$, the slow roll
condition can be equivalently put as \eqn\slonow{{\kappa^2 T_2 \o
{\rm V}_8}~\dot y^m ~v^2 = -{1\o 3!}~G_{0npq}~\gamma^{mnpq}.} So
far, we have the following situation: we have a large radius
fourfold with a time dependent $C_{mnp}$ flux switched on (the
space dependent part of $C_{mnp}$ is vanishingly small although
the integral $\int G \wedge G$ over the full fourfold is a
non-vanishing integer satisfying \warpff). The membrane on the
fourfold is moving very slowly towards the orbifold fixed point
that we have isolated from the other bunches of singularities. The
fluxes break supersymmetry and satisfy equations \mnpqone,
\temporeal, \tempom\ and \warpff.   We   haven't discussed the
issue of moduli stabilization as yet, and neither did we    solve
for the warp factors $A$ and $B$, which we took to be explicit
functions of time. From the analysis of the previous section, we
know all the equations for the warp factors, and we will now study
them one by one. The first non-trivial equation for the warp
factors will be \eienoo. From the ansatze of $A,B$ and the
solution \one, we see that \eienoo\ simplifies. However, \eienoo\
also has membrane and quantum terms that we cannot ignore.
Furthermore, the membrane terms are written in terms of
$\delta^{8}(y-y(t))$ where $y^m(t)$ denotes the position of the
membrane on the fourfold at a given time $t$
 moving with a velocity $\vert \dot y\vert^2 = \dot y^m \dot
y^n g_{mn}$. But the issue of membrane and the quantum terms can
actually be ignored if we consider \eienfi, where we used \adding\
to get from \eienoo\ to \eienfi. In \eienfi\ we see that the quantum
terms come as a  {\it  difference} between two terms.  We  can, therefore,
assume that the difference can be made small even though the
individual terms may not be very small\foot{If the warp factors are
not too big, then ${\del E_q \o \del A}$ and ${\del E_q \o \del B}$
could be made small in principle. Also the $X_8$ terms are sensitive
to orbifold singularities. Therefore if we shift the singularities
far away (this means the base of the fourfold is of very large radius)
then we
can also make $X_8$ small locally. In this case the difference $T_q$
can indeed be very small.}. Therefore the membrane and the quantum
terms $T_m$ and $T_q$ respectively goes to zero as:
\eqn\memqua{\eqalign{&T_q ~\equiv~ {b_1 e^{-5A} \o 2} ~{\del E_q\o
\del A} - {X_8 \o 8!} ~ \to ~ 0\cr & T^{(1)}_m ~\equiv~ {1\o 2}
~e^{2(B-A)}~\vert \dot y \vert^2 ~\delta^8(y - y(t)) ~ \to ~ 0,}}
where we used \condme\ to derive the second equation. Under the
above assumptions, we get the first non-trivial equation for the
warp factors: \eqn\warpone{\dot A^2 + 28 \dot B^2 + 16 \dot A \dot B
= {e^{-6B} \o 24} \vert G \vert^2 - {1\o 2} e^{2(A - B)} ~{R^{(g)}}
+ {e^{2A-8B}\o 4\cdot 4!} ~{\cal O}(\epsilon^2),} where $R^{(g)}$ is
the Ricci scalar of the fourfold, and we have defined $\vert G
\vert^2 = G_{0mnp} G^{0mnp}$. Notice also that we have put in the
contributions from the $G_{mnpq}$ terms that are of the form
${e^{2A-8B}\o 4\cdot 4!} ~{\cal O}(\epsilon^2)$. A more precise
computation, which we will do next, will have to incorporate this
term, as well as the membrane and the quantum terms $T^{(1)}_m$ and
$T_q$, respectively.

The next equation for the warp factor is \eienij. We see that this
also has non-trivial membrane and quantum terms. However \eienadd\
gives us an equation where the quantum terms have cancelled out
completely and the membrane term would go to zero as:
\eqn\memze{ T^{(2)}_m ~\equiv ~ e^{2B}~\vert \dot y \vert^2 ~\delta^8(y-y(t)) ~ \to ~ 0,}
giving us the second non-trivial equation for the warp factors:
\eqn\warptwo{\dot A^2 - 8 \dot B^2 + 16 \dot A \dot B  - \ddot A - 8 \ddot B = {e^{-6B} \o 12}~\vert G \vert^2.}
The above set of equations can be simplified a little bit by taking
into account the fact that the unwarped metric of the fourfold is
Ricci flat. As we saw earlier, this of course doesn't mean that the
six-dimensional base also has to be Ricci flat. In fact, generically,
the six dimensional base is never a Calabi-Yau manifold or even a
K\"ahler manifold. Ricci flatness of the fourfold implies that
\eqn\ricciflat{{G^{(g)}}_{mn} ~ = ~ 0, ~~~~~~ {R^{(g)}} ~ = ~ 0.}
This simplifies \warpone\ and \warptwo.
Furthermore we see from \warpone\ and \warptwo\ that one possible
ansatz for $A$ and $B$ could be that they are logarithmic functions of
time $t$. In other words, we will consider the following ans\"{a}tze
for $A$ and $B$:
\eqn\ansab{A(t) ~ = ~\alpha~{\rm log}~(t - t_o), ~~~~~~~~~ B(t) ~ =~ \beta~{\rm log}~(t - t_o)}
along with the background G-flux $G_{0mnp}$ as in \one\ but with $c_1= 0$.
This means that the warp factors are\
$e^A=(t-t_o)^{\alpha}, e^B=(t-t_o)^{\beta}$. These are approximations
to the general series solutions \wabehave\ valid for large $t\gg t_o$,
with $\alpha, \beta$ here playing the role of $n_{max}, m_{max}$ of
\wabehave\ \ \ (note that in general, one could imagine the existence of
cosmological solutions to the system of higher derivative term equations,
represented by the complete $t$-series \wabehave\ with nontrivial
coefficients).

In the analysis above, observe that we have not ignored the quantum
corrections completely. The only constraint we are putting on the
unwarped metric will be \memqua. The scalings of the G-fluxes are
now important. The component $G_{mnpq}$ is time independent, but could
depend on the fourfold coordinates $y^m$. In this example, it is in
fact a vanishingly small constant number. On the other hand
$G_{0mnpq}$ is a function of time $t$ via the warp factors $A$ and
$B$. From the form of $G_{0mnp}$ in \one, we see that it involves an
integral over $D_p\gamma$ and therefore would change the
$t$-dependences. A generic ansatz for the $G_{0mnp}$ component should
then be
\eqn\goom{G_{0mnp} = \left[f(y)~e^{aA + bB}\right]_{0mnp},}
where $a, b$ are constants that we will soon determine; and $f(y)$ is
a function on the fourfold.

Plugging \ansab\ and \goom\ in the first equation \warpone, and taking
the limit where $G_{mnpq} \to 0$, we immediately get a relation
between $\alpha$ and $\beta$. This relation comes from comparing the
powers of ($t - t_o$) on both sides of the equation. The relation is
\eqn\firstrel{ (3-b)\beta  -  a\alpha ~ = ~1.}
In fact, it is easy to see that the same relation would come from both
the equations \warpone\ and \warptwo\ as the powers of the exponential
are the same in all the equations. When there is a non-trivial
$G_{mnpq}$ component the above scaling will not work.

Having gotten a relation between $\alpha$ and $\beta$, we now use
the two equations, \warpone\ and \warptwo, to determine these
values. Observe that due to the presence of $\vert G \vert^2$ in
all the equations, there will be another unknown $f(y)$ from the
choice of $G_{0mnp}$ in \one\ (recall that the quantum correction
$c_1 = 0$). The relations now are: \eqn\two{\eqalign{& \alpha^2  +
28 \beta^2  +  16 \alpha \beta ~  =~  {f^2 \o 24} + {\cal
O}(\epsilon^2) \cr & \alpha^2  -   8 \beta^2 +  16 \alpha \beta +
\alpha  +  8 \beta ~ = ~ {f^2 \o 12}.}} These two are simple
algebraic equations and therefore the solutions can be easily
determined. Combining \two\ with \firstrel,  the solution set is
given by \eqn\albeso{ \left({k\beta -1\o a}, \beta \right),
~~~~~~~~k = 3-b,} with $\beta$ being the root of a quadratic
equation. The explicit value of $\beta$ would be
\eqn\betavalue{\beta = -{1\o 2(1+a)}\left[2(k+8a)+(8+ka)
\mp\sqrt{(8+ka)^2 + 32(k+8a)(1-a^2)}\right], ~ a \ne -1.} On the
other hand, when $a = -1$,  the solution set  is \eqn\solset{
(1,0), ~~~~~~~ \left({1\o a}\cdot{8\o k-8}, {1\o k-8}\right).} One
can ignore the fractional solution as it gives imaginary values
for the background fluxes. We can now plug in \solset\ in \two\ to
get $f = \pm \sqrt{24}$. This way the background G-fluxes
$G_{0mnp}$ and the warp factors seem to be completely determined.

However, this is not enough. We still have another equation to
go. This is \eienmn. For a Ricci flat manifold, this equation
simplifies to
\eqn\eiennowt{\eqalign{& ~~~~~~~~ (7 \ddot B + 2 \ddot A + 28 \dot B^2 + 7 \dot A \dot B + \dot A^2)g_{mn} = {e^{-6B} \o 4} \left(G_{0mpq} G_n^{~~0pq} -
{g_{mn} \o 6} \vert G \vert^2 \right) + \cr
& - {\kappa^2 T_2 e^{-6B} \o \sqrt{-g}}\left[
{b_1 e^{-A - 4 B}} {\del E_q \o \del(g_{mn}e^{2B})} + ~g_{mp}~ g_{nq}~\dot y^p ~\dot y^q ~
\left(1 + {1\o 2} e^{2(B-A)} \vert \dot y \vert^2 \right) ~\delta^8(y-y(t))\right].}}

From this, we see that we cannot quite ignore the quantum terms if the
manifold is compact\foot{If we ignore the quantum terms, then the only
solution to all the set of equations \warpone,
\warptwo, \eiennowt\ and \warpff\ is $A = B = G_{0mnp} = 0$ when the internal
 manifold is compact.  This is of course the no-go theorem for warped
compactifications \nogo. As is well known including higher order
corrections or taking a non-compact fourfold invalidates the premises
that went into the no-go theorem.}.  We might be able to ignore the
contributions from the membrane terms that depend on powers of $\dot
y^m$.  This equation therefore relates the unwarped metric with the
membrane velocities and the quantum terms.

We now face the following puzzle. Once we cannot ignore the quantum
terms in \eienmn, then taking the trace of this equation relates the
warp factors to the membrane and the quantum terms.  Furthermore, the
membrane and the quantum terms come with different powers of the warp
factors.  This would mean that the time parameter $t$ does not scale
out of the equations completely. The equation \eiennowt\ will also
have contributions from $G_{mnpq}$ terms that are typically of order
${e^{2A-8B}\o 12}~{\cal O}(\epsilon^2)$.  Thus the only possible way
to get a solution here would be to consider an essentially non-compact
manifold (so that ignoring quantum terms do not cause problems with
the no-go theorems) and a very slow velocity of the membrane (so that
the velocity dependent contributions of the membrane is small).  This
is, in fact, the precise content of the slow-roll condition \condme\
that we proposed earlier!  Therefore the solution set is given by:
\eqn\solid{(\alpha,~\beta,~a, ~b, ~k, ~f)~~ = ~~(1, ~0, ~-1,~-2,~5, ~\pm\sqrt{24}).}
This solution set will determine our background metric and the G-fluxes
completely, with the assumptions that the membrane motion is very slow
\condme\ and the quantum corrections are small. With these conditions,
we solve all the equations of motion to get our final result as:
\eqn\metG{\eqalign{& ds^2 = (t - t_o)^2 ~\left(-dx_o^2 + dx_1^2 + dx_2^2 \right)~ + ~ g_{mn}~ dy^m dy^n \cr
& G^{0mnp} = \pm {\sqrt{24} \o t - t_o}, ~~~ C_{012} = (t-t_o)^3, ~~~ G_{mnpq} = [\pm\epsilon]_{mnpq}.}}
with $\epsilon \to 0$ for an essentially non-compact fourfold. Once we
know the velocities of the membrane from \eiennowt, we can use
\tempom\ to determine the values of $G_{mnpq}$.  Knowing the
background G-fluxes, \warpff\ will tell us how many membranes we have
to put in to saturate the anomaly equation.  In this way we can
determine the complete solution for the system.

The above solution is very interesting in the sense that the
internal fourfold is completely time independent\foot{This however
{\it doesn't} mean that the volume of the fourfold is stabilized!
Volume stabilization here will be related to the equivalent
problem that we face for the supersymmetric case, namely: the
Dine-Seiberg runaway \dsrun. Therefore having gotten a time
independent internal fourfold, our next exercise would be to fix
this volume against any runaway by non-perturbative effects.
Details on this will be presented in the sequel of this paper.}!
This would mean that any dynamics that we get in 2+1 dimensional
space from the fourfold will in fact be independent of time. While
this time-independence of the compact space can of course be
potentially lifted once we move away from the approximations here,
it is encouraging to see that the toy model that we study here
comes close to something we are interested in. We also repeat here
that the regime of validity of this solution is for $t\gg t_o$, as
in the general discussion at the end of Sec.(2.1) on potential
singularities in these effective gravity solutions.

The question now is whether in type IIB it is also possible to keep
the six dimensional internal space independent of time.  The type IIB
metric \metiib\ immediately tells us that the internal space will
indeed also be time independent. Both the dilaton and the axion for
this background vanish locally, as we have shifted any kind of
singularities far away. The G-fluxes can allow NS fluxes in Type IIB
theory\foot{To see this explicitly, take the background C-fields in
\cnowf\ to be oriented along $C_{mn,11}$, where $x^{m,n}$ lie on the
four-dimensional subspace of the six-dimensional base of the fourfold.
If the six dimensional base is $K3 \times \IP^1$ then $x^{m,n}$
are the coordinates of $K3$.  The $C_{mn,11}$ field will then
become a $B^{NS}_{mn}$ field in type IIB. Eq.  \cnowf, when
$\epsilon \to 0$, then gives a time dependent non-commutativity on
the seven-brane world volume that wraps the four-dimensional
internal space.  Such time-dependent non-commutativity has been
addressed earlier in \robbins, \hash\ in a different context. For
our case, the situation is more involved than a simple time
dependent non-commutativity because of the $y^m$ dependent in part
on $B^{NS}$ which, although small, has the effect of breaking
supersymmetry.}, and this would break supersymmetry. The type IIB
metric has the following explicit form: \eqn\extiib{ds^2 =
(t-t_o)^2 ~\sqrt{g_{aa}}~\left(-dx_0^2 + dx_1^2 + dx_2^2 \right) +
{dx_3^2 \o g_{33}\sqrt{g_{aa}}} + ~ g_{mn} ~dy^m dy^n} where we
used \metiib\ to write the type IIB metric. The range of the
indices $m,n,\ldots $ is $m, n = 4 ... 9$, and we have absorbed
all numerical constants in the definition of the coordinates. The
above metric is not really a de-Sitter like solution {that we are
ultimately interested in}, but as we have made many simplifying
assumptions, we really do not expect the metric to look like a
de-Sitter one.  Most importantly, we see that the $x^3$ direction
does not have any time dependent warp factor because $B = 0$.
Therefore, the $x^{0, 1, 2}$ directions expand differently than
$x^3$ direction (which doesn't expand at all). In fact, the $x^3$
direction is a small circle with vanishing radius, and therefore
the type IIB theory is effectively a 2+1 dimensional space as we
are in the large radius limit of M-theory \condme\ (plus the
internal six dimensional space).

Now that we have the M-theory background with fluxes (and also the
type IIB background), it is time to study the motion of the M2
brane. We have already assumed a slow roll condition for the membrane,
\condme. The precise equation of motion for the membrane will be
\eqn\eommemb{ {d^2 y^p \o dt^2} + \gamma^{00} ~\dot y^m \dot y^n~\Gamma^p_{mn} - {1 \o 2} ~{e^{-2B}} \dot y^m~ G^p_{~m12} ~ = ~ 0,}
where $\Gamma^p_{mn}$ are the Christoffel symbols measured with respect to
the {\it warped} metric.
From the choice of the background, \metG, $G^p_{~m12} =0$  and therefore
the acceleration of the membrane can be determined from the velocities alone.

Therefore to summarize: in this example we took very small
internal G-fluxes $G_{mnpq}$ on an essentially non-compact
fourfold with a slowly moving membrane, so that the quantum
corrections could be ignored without violating the no-go theorem.
This essentially means that the slow-roll condition has in a sense
naturally imposed hierarchies on the G-flux components (which
translates to effective non-compactness of the underlying
fourfold), ensuring that nontrivial cosmological solutions valid
classically at leading order in some of the G-fluxes do not
violate equations such as the G-flux anomaly equation. Similar
strategies will be seen to apply in the Type IIB discussion of
Example 3 in Sec.~(3.3). It is conceivable that other nontrivial
cosmological solutions exist to the full system of higher
derivative equations when the G-flux components are of comparable
strength: finding such solutions is potentially harder however,
since there is no obvious way to approximate them by a leading
classical solution.

\subsec{{Example 2}: Arbitrary   background fluxes}

\noindent In deriving the above background, we made  many assumptions.
Although  these are   not inconsistent,
it   would   be nice to see if  one  can have an ansatz for the background
that   could
(a) allow non-trivial function for $G_{mnpq}$ instead of the small value that
we took earlier,
(b) allow warp factors to depend on the fourfold coordinates $y^m$ also in
addition to time $t$, and (c) allow membrane and quantum corrections.
These are many requirements, and it turns out that this is only possible if
we consider the following conditions:

\noindent $\bullet$ Since the $G_{mnpq}$ has to be time-independent, the
$C_{mnp}$ field will have to follow the ansatze given in \cmnp. This means
that there is a separate $t$ dependent part $\psi(t)$, and a separate $y^m$
dependent part $\lambda(y)$.

\noindent $\bullet$ Since $G_{0mnp}$ has to satisfy \mnpqone, this component
of the flux   will in general have to be time dependent. From \cmnp, this
will be determined by $\dot \psi(t)$.

\noindent $\bullet$ The warp factor $B$ has to be time independent. This is
obvious from the warp factor equation \warpff\ and \lhswa\ as the RHS of
\warpff\ is time independent. On the other hand,
$A$ could depend on time $t$. The dependence of time $t$ on  the warp factor
can again be inferred from \lhswa: $A(y,t)$ should be a linear combination
of some function of time $t$ and some other function of fourfold coordinates.

\noindent $\bullet$ For the equations to make sense, {\it all} the time
dependences should scale out. This also tells us that one possible
time dependence of the warp factor $A$ could be a logarithmic function of
time. This way there is a possibility of scaling out $t$ dependences from
all the equations.

\noindent It turns out that there is a simple   ansatz   that takes care of
all the points raised above. The ansatze can be written in terms of functions
of the internal fourfold that have to be determined, and is given by:
\eqn\ansatzeforbg{\eqalign{&{\rm Warp~ factor}:~~~ A = f_1(y) - (1+\epsilon)~{\rm log}~(t - t_o), ~~~~~ B = f_2(y), ~~~\epsilon \to 0 \cr
& {\rm G-flux}: ~~~G_{0mnp} = -{c_o \o t-t_o}, ~~~~~ G_{mnpq} = h_{mnpq}(y), ~~~~~G_{012m} = {\del_m e^{3f_1} \o (t-t_o)^3}\cr
& {\rm C-field}: ~~~C_{mnp} = - c_o~{\rm log}~(t - t_o) + c_{mnp}(y), ~~~~~C_{012} = {e^{3f_1}\o (t-t_o)^3}, ~~~~~ h = dc \cr
& {\rm Membrane~ velocity}: ~~~~ \vert \dot y \vert = \sqrt{\dot y^m \dot y^n g_{mn}} = {\alpha \o t-t_o}, ~~~~ v = \alpha ~e^{f_2-f_1}}}
where the tensor indices in the constant quantity $c_o$ is implied,
and $f_1(y), f_2(y)$ are scalar functions whereas $h_{mnpq}$ and
$c_{mnp}$ are tensors on the fourfold. The quantity $\alpha$ appearing
in the velocity is a constant number, and $v$ is the warped velocity
defined earlier. Observe also that we have given a very small shift to
the time dependence for the warp factor $A$.  The shift is very small
but {\it not} zero.  The reason for this will be clear soon.

To compare this solution to what we presented earlier, observe
that the $C_{mnp}$ field takes the expected ansatz that we
presented earlier in \cmnp. The tensor $\lambda_{mnp}(y)$ should
be identified with $c_{mnp}(y)$ now. This should now be a generic
function of $y^m$ and is not constrained to be small.  Furthermore
$G_{mnpq}$ is quantized as in \greq\ as before.

Its also clear that the choice of function $f(y)$ in \goom\ cannot be
arbitrary. The function $f(y)$ has to be chosen in such a way that
\eqn\quaf{ {1\o f_2} ~\left({\rm log}~{c_o \o f} - f_1\right)~ = ~ \IZ, ~~~~ {\rm and} ~~ a = 1,}
where the integer $\IZ$ specify the value of $b$ in \goom. This
ambiguity in defining $b$ is reflected in the simple choice of warp
factors that we made earlier, namely: the warp factors to depend only
on time $t$. The correct ansatze for the background therefore should
be \ansatzeforbg.

For the moving membrane, the velocity involves a constant $\alpha$.
Our slow roll condition corresponds to $\alpha$ being small. As
expected, the world volume metric $\gamma_{\mu\nu}$ is no longer time
independent. The metric changes with time in the following way:
\eqn\wvmet{\gamma_{\mu\nu} = {1\o (t-t_o)^2}\pmatrix{\alpha^2e^{2f_2}-e^{2f_1}&0&0\cr 0&e^{2f_1}&0\cr 0&0&e^{2f_1}}}
and therefore picks up different values at different points on the fourfold.
The total velocity is given by \ansatzeforbg, and the individual components
will be determined later.

Notice also the fact that the ansatz \ansatzeforbg\ does not require
us to keep a Ricci flat manifold. In fact it is easy to see that $-$
if we denote the temporal part of $A(y,t)$ by $A_o$ $-$ all equations
have a scale factor of $e^{2A_o}$ for every term. This way we can
scale away the $(t-t_o)$ dependent parts completely. Therefore the
equations would relate only the $y^m$ dependent parts, which we could
use to solve for them. However, we will soon see that solutions to
these equations exist {\it if and only if} we incorporate higher
derivative terms. In the absence of higher derivative terms, the
background goes back to the standard supersymmetric solution.

With this in mind, let us study the equations carefully now. We will
typically take $\epsilon \to 0$ in \ansatzeforbg. The necessity of
non-zero $\epsilon$ will arise for some of the cases that we shall
illustrate soon. First let us us consider the tensor $G^{0mnq}$. From
\ansatzeforbg, this is determined by $c_o$.  The non-primitivity can
now be expressed via $c_o$ and the warp factors in the following way:
\eqn\comnpq{\eqalign{& (1)~D_p \left[h^{mnpq} - (\ast h)^{mnpq}\right] = \left[2c_o e^{2(f_2-f_1)}\right]^{mnq} \cr
& (2)~D_m\left[e^{3f_1} D_p\left(h^{mnpq} - \ast h^{mnpq}\right) \right] = 0}}
which would be easily determined by knowing the warp factors $f_1, f_2$ and
the tensor $h^{mnpq}$. Indeed, there is a relation that connects them. This
is the anomaly equation that combines the warp factors and the $h$-fluxes as:
\eqn\anowah{ 3~D_m(e^{6f_2} \del^m f_1) = {1\o 2\cdot 4!}~h_{mnpq}(\ast h)^{mnpq} + {2\kappa^2T_2 \o \sqrt{-g}} ~\left[\delta^8(y-Y) + {X_8\o 8!}\right].}
For a compact fourfold, this will give the anomaly relation connection the
$h$-fluxes with the number of membrane and Euler characteristics.

The next relation would be the Einstein equation \eienfi. The $t$
terms of course filter out of this equation, and therefore this will
connect the warp factors with the membrane and the quantum terms,
$T^{(1)}_m$ and $T_q$ respectively. The main problem, however, comes
when we analyse the next equation i.e \eienadd.  Plugging in the
ansatz \ansatzeforbg\ with non-zero $\epsilon$, we get the following
relation:
\eqn\addeien{{\alpha^2 \kappa^2 T_2 e^{-6f_2} \o \sqrt{1-\alpha^2 e^{2(f_2-f_1)}}} + {1\o 12} \int d^8 y~\sqrt{-g}~\vert c_o\vert^2 e^{-6f_2} +
{\cal O}(G^n R^m) ~ = ~ \epsilon {\rm V}_8,}
where we have added higher derivative terms. Observe now that (a) in
the absence of these higher derivative terms, or (b) when the quantity
$\epsilon = 0$, the LHS of \addeien\ is a sum of positive definite
quantities and would therefore vanish! This means the membrane
velocity vanishes ($\alpha = 0$), and $G_{0mnp} =0$, resulting in a
standard supersymmetric compactification with no de-Sitter like
solution. However incorporating higher derivative terms this will not
be the case. Thus we have our first non-trivial conclusion: {\it only
in the presence of higher derivative terms} the set of equations would
have solutions with time dependences.

The existence of these higher derivative terms has been argued in
the literature starting with \highd. These are typically of the
form of $\int \sqrt{-g}~[c_{mn} G^m R^n + d_{mn} (\del G)^m R^n]$,
with $c_{mn}, d_{mn}$ being various combinatorial factors that
come with different signs. The significance of these terms were
realized recently in \robbins, where an M-theory lagrangian was
used to derive the full action for a 6+1 dimensional
non-commutative theory. It was observed that the existence of
these higher derivative terms are absolutely {\it necessary} to
make sense of the Seiberg-Witten (SW) map \swmap\ for the
non-commutative theory.  In \robbins, a few terms of the SW map
were derived from the low energy effective action. It was
conjectured there that the full M-theory lagrangian with all the
higher derivative terms would give the complete SW map from
M-theory. Once the higher derivative terms are known, one could
easily solve \addeien\ to get a relation between membrane
velocity, G-fluxes and the warp factors $f_1$ and $f_2$.

It turns out that the velocity components of the membrane
$\alpha^m$ can be separately determined from \tempoagain. This is
given in terms of $h$ and $c_o$, the tensors appearing in
$G_{mnpq}$ and $G_{0mnq}$, respectively, in the ansatz
\ansatzeforbg. The velocity components are \eqn\velco{\alpha^m =
{1\o 12 \kappa^2 T_2}~\int~d^8y~\sqrt{-g}~(\ast h)^{mnpq}
(c_o)_{npq},} and then \eienom\ can relate the total velocity
$\alpha$ with the warp factors and G-fluxes.

Since we have not used the restriction of Ricci flatness, we can use
other equations to determine the Ricci scalar and the Einstein tensor
$R^{(g)}$ and $G^{(g)}_{mn}$, respectively.  This will provide the
complete set of equations for our background. The final M-theory
metric will take the following form:
\eqn\mmetricf{ds^2 = {e^{2f_1}\o t^2}~\left(-dt^2 + dx_1^2 + dx_2^2 \right) + e^{2f_2}~g_{mn}~dy^m dy^n,}
where we have put $t_o = 0$ for simplicity. The G-fluxes and the membrane
will follow the form given in \ansatzeforbg\ when we plug in the values of
$f_1, f_2, \alpha, h$ and $c_o$ in those equations.

Finally, observe that the ansatz \ansatzeforbg\ can be put in the
general form $A(y,t) = -2B(y) + g(t)$ if we put the condition $f_1(y)
= -2f_2(y)$ on the warp factors. This special relation between the
$f_i$ is true for the time independent case. For the time dependent
case, one can argue that $f_1+2f_2 = k$, where $k$ is a covariantly
constant function on the fourfold. Putting $k=0$ will give the
required relation between the $f_i$, and this will simplify all the
background equations quite a bit as we saw earlier.

Therefore to summarize: with arbitrary background fluxes, we can
scale out the time dependence of the warp factor $A$ from all the
equations to get a fourfold that has time independent warp factor.
It is of course an interesting question to ask whether higher
derivative terms could spoil this property.  In the absence of any
concrete calculation, we can only say that as long as the
curvatures and the G-fluxes are reasonably small, all these
corrections could possible be made small. This way \ansatzeforbg\
can still survive as a valid solution to the system.

\subsec {{Example 3}: Towards type IIB cosmology}

\noindent In the previous examples we have considered warp factors such
that $B$ is time independent, whereas $A$ could be both time and space
dependent. This helped us to get some interesting M-theory
cosmological solutions.  The corresponding type IIB solution, on the
other hand, had time dependent warp factors only along the $x^{0,1,2}$
directions, whereas the $x^3$ direction was time independent.  It
could mean that we might need more than two warp factors to get a
reasonable type IIB cosmological solution. We will discuss this case
later.  In this example, we will try to see how far we can go to get a
type IIB cosmological solution with only two warp factors. In the 
process, we will find some interesting questions raised on the 
interpretation and validity of these solutions, that we then discuss 
briefly.

To start with, first consider a model whose unwarped metric
$g_{mn}$ is {\it not} Ricci flat. This would immediately put some
constraints on the warp factors $A$ and $B$. Our equations of
motion are of course \eienoo\ and \eienij\ (or their equivalent
simplified form \eienfi\ and \eienadd) with the warp factors given
by the ansatz \ansab. Since the Ricci scalar and Ricci tensors of
the unwarped metric are time independent, \eienoo\ and \eienij\
will hold if and only if the time dependences in both the
equations can be scaled away. The Ricci scalar $R^{(g)}$ in
\eienoo\ and \eienfi\ comes with the coefficient $e^{2(A-B)}$. On
the other hand, the time derivatives of the warp factors all go as
$t^{-2}$ if we use \ansab\ in \eienoo\ or \eienij. This gives us
the first non-trivial relation between the warp factors:
\eqn\takam{\alpha ~-~\beta ~=~ -1.} To get the other equation we
need to see how $G_{0mnp}$ or $G_{mnpq}$ scale with respect to
the warp factors. Assume now  that $G_{mnpq}$ now satisfies the
equation \devfro\ with some tensor $\gamma_{mnpq}$ that is
vanishingly small. From \mnpqone\ we see that $G_{0mnp}$ will
continue to satisfy \one\ and therefore the G-fluxes for our case
will be: \eqn\gfla{G_{012m} = \del_m e^{3A}, ~~~~~~~ G^{0mnpq} =
e^{-A-2B}~c_o, ~~~~~~~ G_{mnpq} = e^{-3A}\gamma_{mnpq}~c_2,} with
$c_o, c_2$ as some undetermined constants. Knowing the background
G-fluxes as some functions of the warp factors, we can use
\eienfi\ or \eienadd\ to get the next non-trivial relation between
the warp factors: \eqn\nextofkin{  \alpha ~ = ~ -2 \beta} where we
compared the powers of the Ricci scalar and the $\vert G\vert^2$
terms in \eienoo\ or \eienij. We see that the relation \nextofkin\
is strongly reminiscent of the second relation of \abrela!
Although \abrela\ is more complicated because of its dependence on
space coordinates $y^m$, the temporal part of this equation does
satisfy $A = - 2B$. Furthermore, having $A = -2B$ simplifies all
the Einstein equations considerably, as we saw in detail in the
previous sections. Now we see another advantage of having $A =
-2B$: the scalings of the Ricci tensors (or  the curvature scalar)
can be extracted  from the equations and therefore we may not
impose the condition of Ricci flatness for our case.

The above analysis unfortunately leads to a problem when we encounter the
anomaly equation \warpff. Since the $G_{mnpq}$ tensor has become time dependent
the anomaly equation will no longer hold because the membrane and the $X_8$
terms are in general time independent. One way out of this is to
completely ignore these terms. In other words, we may not consider any branes
(D3 or D7 branes) in the type IIB scenario\foot{Recall that the singularities
of $X_8$ are the points where seven-branes are located.}, i.e. we {could}
only study the supergravity background without any brane inflation.
Alternatively, \warpff\ can receive corrections such that the time
dependences of $\quabla e^{6B}$ can be cancelled by the flux term.

In any case, this is of course not a serious problem as generic study
of cosmological solutions have been done before without invoking brane
inflation {\it per-se}.  Without {requiring} inflation, our system
becomes {perhaps less interesting, but} nevertheless much simpler,
although one might wonder if in the absence of the $X_8$ term, there
could be any solution at all. For a {\it compact} internal fourfold
there is {\it no} solution. But when we make the manifold non-compact
and keep the $J_o, E_8$ and higher derivative terms, then solutions
would exist.

Solving for $\alpha$ and $\beta$ we get $\alpha = -{2\o 3}$ and
$\beta = {1\o 3}$. The type IIB metric can now have an overall
factor in front of $x^{0,1,2,3}$ because of its resemblance with
the second equation of \abrela. In fact the explicit type IIB
metric can be easily shown using \metiib\ to be: \eqn\iibeasy{ds^2
= {1\o t} ~\Bigg[\sqrt{g_{11,11}}(-dt^2 + dx_1^2 + dx_2^2) +
{dx_3^2 \o g_{33}\sqrt{g_{11,11}}} \Bigg] + t~g_{mn}~dy^m~dy^n,}
with $m,n = 4 .... 9$ and the other variables are already defined
in \metiib. 
To glean four dimensional physics here, note that this metric is 
of the form \factorized\ with a time-dependent warping to the six 
dimensional space: a naive dimensional reduction would yield a 
time-dependent 4D Planck constant. Thus one needs to first absorb 
this time-dependence, going to the 4D Einstein frame as discussed 
in general in \factorized-\einstframe. From there, we read off 
$p(t)\sim {1\o t},\ h(t)\sim t$ giving $\eta^2(t)=h^{6/2} p\sim t^2$ 
as the proper scale factor of this cosmology. Using \RW-\etatau, 
this gives the FRW scale factor to be $a(\tau)\sim \tau^{1/2}$, 
which is precisely a four dimensional radiation-dominated cosmology
\foot{Recall, as we have mentioned in Sec.~(2.1), the expressions 
$a(T)\sim T^{[2/3(1+w)]}$ for the time-scaling of the scale factor 
corresponding to a perfect-fluid with equation of state $p=w\rho$.}. 
Clearly this effective solution breaks down at $\tau\sim 0$ where we 
necessarily must resort to the full string theoretic description, as
we have discussed in Example 1\ (Sec.~(3.1)), and more generally at 
the end of Sec.~(2.1).

We will elaborate in more detail below, but observe here that there
are non-trivial functions of $y^m$ that are distributed in an uneven
way among $x^{1,2}$ and $x^3$ which spoils required properties of the
metric because we took $A = -2B$ instead of the second equation of
\abrela.  Therefore, first we have to search for a metric that can get
rid of these problems.

To find an ansatz for the warp factors, we have to consider a relation
between the warp factors that is of the form $A = -2B + g(y)$. With
this relation, it will no longer be possible to keep $B$ independent
of time $t$. As discussed above, this is only possible (at least for
this scenario) when we ignore the membrane and the seven-brane
singularities. This means that the internal manifold is essentially
non-compact. The ansatz now will be the following:
\eqn\ansanowp{\eqalign{& {\rm Warp~factor}: ~~~ A = -2b~{\rm log}~(t-t_o) + 2g_1(y) + g(y), ~~~~~ B = b~{\rm log}~(t-t_o) + g_1(y) \cr
& {\rm G - flux}:~ G_{mnpq} = g_3(y) ~(t-t_o)^{3b}, ~~ G_{0mnq} = {g_4(y) \o (t-t_o)^{1-3b}}, ~~
G_{012m} = {\del_m \left[e^{3g(y) + 6g_1(y)}\right] \o (t-t_o)^{6b}}}}
where the tensor indices of $g_3, g_4$ are implied. Non-trivial
solutions for $g_i$ exist only when we incorporate higher derivative
terms, as can be seen by plugging this ansatz in the set of equations
of sec. 3.2. We will not present the full solution set for the $y^m$
dependent function in this paper. It will suffice to give the value of
$b$ and the form of the $C_{mnp}$ field. This is given by
\eqn\solbc{ b = {1\o 3}, ~~~~~~~ C_{mnp} = [(t-t_o)~g_5(y)]_{mnp}}
with $g_5(y)$ being yet another function that one could easily relate to the
other $g_i(y)$. The value of $b$ will tell us the overall conformal factor.
We see that this is the same as we got earlier with a rather restrictive
choice of G-fluxes. The final metric will now only have an overall conformal
factor, thanks to \abrela, and is given by:
\eqn\metabvx{ds^2 = {c_4t^{-1}}\left(-dt^2 + dx_1^2 + dx_2^2 + dx_3^2\right) + t~\tilde g_{mn} ~dy^m dy^n}
where $c_4 = {e^{-3g_1}\o g_{33}\sqrt{g_{11,11}}}$ and $\tilde g_{mn}
= e^{3g_1} \sqrt{g_{11,11}}~g_{mn}$. As we have seen, this is a 
radiation-dominated cosmology.  It is also easy to see from \solbc\
and \ansanowp\ that $G_{mnpq}$ will become time dependent whereas
$G_{0mnp}$ will be time independent. All of them will of course now be
non-trivial functions of the internal space coordinates.

While the above solution seems quite a benign 4D cosmology, it is
important to note that this solution is effectively valid only in the
noncompact limit of the fourfold as we have discussed above. Thus the
validity of a four-dimensional description is not particularly clear:
essentially this is the description on $some$ hyperslice in the
noncompact higher dimensional spacetime.  We expect that if we can
find an embedding of such a solution in a compact space, these
solutions will get modified to possibly different four dimensional
cosmologies.

In the context of a cosmological solution in a compact setting, it is
interesting to ask if there are general restrictions imposed on the
space of cosmologies by such an M-theory lift involving higher
derivative terms (which are necessarily nonzero in the compact case),
and when the higher derivative terms can be treated as giving rise to
a sensible effective fluid-like matter field configuration. In this
case, it would be interesting to study how the energy contributions
from the higher derivative terms intertwine with known energy
conditions and cosmological bounds in four dimensions.

\subsec{{Example 4}: Three   warp    factors}

\noindent Our next example is a model with a different choice of
the warp factors for the $T^2$ fiber directions, i.e., we consider
now a metric of the fourfold and the spacetime  of the form
\eqn\metfosp{ds^2 = e^{2A} (-dx_0^2 + dx_i dx_i) + e^{2B} g_{mn}~
dy^m dy^n + e^{-A} (dx_3^2 + dx_{11}^2)} where $x^3, x^{11}$
denote the directions of the $T^2$ fiber and $i = 1,2$. In terms
of our earlier metric \netvix, this is the case when $C = -{A\o
2}$, along with a complex structure of $\tau = i$ for the fiber
torus for simplicity. Now let us consider the following generic
choices of the warp factors $A$ and $B$: \eqn\chiba{A (y, t) =
{2\o 3} ~{\rm log} ~f(y)g(t), ~~~~~~~~ B(y, t) = {1\o 2} ~{\rm
log} ~h(y)f^{1\o 3}(y) g^{1\o 3}(t).} At this point, such a choice
of warp factors may look completely arbitrary. But once we go to
type IIB we see that the metric takes the following form:
\eqn\metlu{ds^2 = f(y) g(t)~\left(-dx_0^2 + \sum_{i = 1}^3 ~dx_i
dx_i\right) + h(y) ~g_{mn}~dy^m dy^n} which is in fact the kind of
metric that we are looking for. But   this is   not all.   If we
further impose the following restriction on $h$ and $f$:
\eqn\restonhf{ h(y) ~ = ~ f^{-1}(y) ~\equiv~ \sqrt{j(y)}} then the
type IIB metric will take the standard form of a D3-brane metric
with a harmonic function $j(y)$, at least at some initial time
$t_o$ when $g(t_o) \approx 1$. For any other time $t$, $g(t)$
could be any arbitrary value.  However if we want to mimic a
de-Sitter background then we can fix some particular value of
$g(t)$ in terms of a cosmological constant $\Lambda$ as
\eqn\gtv{g(t) ~ = ~ {1\o \Lambda t^2}.} If we switch on a
non-primitive G-flux $G_{mnpq}$ on the six dimensional base of the
fourfold, then the cosmological constant can be explicitly
determined in terms of $G_{mnpq}$ as \eqn\cosmo{ \Lambda = -{1\o
12j^2} \left[ \quabla j + {2\o 3} {\vert G \vert^2}\right] +
....} where $\vert G \vert^2 = G_{mnpq} G^{mnpq}$ and the dotted
terms involve contributions from the membrane and the quantum
terms whch would eventually make $\Lambda > 0$. Taking the
G-fluxes to have legs along the $T^2$ fiber direction will in fact
make the analysis much more involved, as one can show that the
background Einstein equations become time dependent for    such a
choice of fluxes.

Of course the choice of warp factors that we took here has no {\it
a-priori} justification. For a more generic discussion we have to
start with a metric that has three distinct warp factors in
M-theory as in \netvix\ and also with a non-trivial complex
structure for the fiber torus. In fact it is easy to generalize
\metlu\ to incorporate the information about the complex
structure.  {Namely, from } \metrora, {we have to take}
\eqn\pamrod{f(y) g(t) ~ \to ~ {f(y) g(t)~ \vert \tau\vert
~\tau_2^{-1}}, ~~~~ h(y) ~ \to ~ h(y)~ \vert \tau\vert
~\tau_2^{-1}} when we choose the complex structure as $\tau =
\tau_1 + i~\tau_2$.

In a background with three warp factors $A, B$ and $C$ one has the
freedom to keep the internal manifold time independent, but the
spacetime with some requisite time dependence. However, from \abrela\
we see that the analysis of the equations of motion that we did here
will have to change, as the scalings etc have changed. Also it is not
clear whether the ansatz for G-flux \allgflux\ will remain the same
with three warp factors. But there are some issues we can study
without going to the G-flux ansatz
\allgflux:

\noindent $\bullet$ First, we observed for the two warp factor case that the
time and space dependences of the warp factors isolate. In fact keeping
$B$ only as a function of $y^m$ gave us a type IIB six manifold with
no time dependence. On the other hand keeping $B$ as $B(y)$ we were
unable to reproduce the kind of metric that we wanted in 3+1
dimensions. It is interesting to observe that the three warp factor
case can give us a 3+1 dimensional space with an overall
time-dependent warp factor and a six dimensional internal manifold
with a warp factor that is only a function of $y^m$, if we consider
the following ansatz for $A, B$ and $C$:
\eqn\abcans{A(y,t) = f_1(y) + g(t), ~~~~~ B(y,t) = f_2(y) + {1\o 4} g(t), ~~~~~ C(y,t) = f_3(y) - {1\o 2} g(t)}
where $f_i(y)$ are some functions on the internal fourfold. Whether the background equations of motion allow this ansatz still remains to be seen.

\noindent $\bullet$  The second point   is the analysis of
\addeien, where we showed that when the RHS of the equation
vanishes, this equation will have solution iff we put in higher
derivative terms in G-fluxes. When we have three warp factors, the
RHS of \addeien\ can get an additional contribution from the warp
factor $C$ which could be a function of time (although the warp
factor $B$ may not be).  Therefore \eienadd\ is now given by (no
sum over i): \eqn\googii{G_{00} + G_{ii} = \dot A^2 - \ddot A  - 2
(\ddot C + \dot C^2 - 2 \dot A \dot C) - 6(\ddot B + \dot B^2 - 2
\dot A \dot B).} Assuming now   the above ansatz \abcans\ is an
allowed one,  it is easy to see that \googii\ can be made positive
definite if the time dependence of the warp factors $g(t)$   obey
the inequality: \eqn\inequa{ \dot g^2 ~\ge ~ {4\o 3}~\ddot g} with
an equality leading to the same problem that we encountered
earlier. It will be interesting to see whether this could now have
solutions without assuming   higher derivative corrections.

\noindent $\bullet$ The third  point    concerns    the membrane velocity.
This is related to the Einstein tensor $G_{0m}$, which in turn is related to
non-primitivity. Switching on a time dependent warp factor $C$, will give us
the following contribution
\eqn\gomnew{G_{0m} = -{3\o 2}~\dot g~\del_m{\rm log}~\tau_2}
and therefore change the membrane velocity (see \eienom\ for the derivation).
A similar analysis can be done for the time dependent $B_{NS}$ in type
IIB theory, {which comes from the} $C_{mn,11}$ field as discussed
earlier. The energy momentum tensor of this field is now correlated
with $G_{11,11}$. We can also switch on $C_{mn3}$ giving rise to a
$B_{RR}(y,t)$ field in type IIB theory. These two fields together will
break supersymmetry for our case. We would now need a five form field
in type IIB to satisfy equation of motion. This in turn means that we
need a G-flux ansatz for $G_{012m}$.

Therefore to summarize: we see that with three (or more warp factors)
we can overcome some of the problems related to the higher derivative
terms that plague the two warp factors case. It would seem that, with
three warp factors, a de-Sitter type cosmology could be derived as
there are much more freedom within the range of variables.  A detailed
analysis of this is beyond the scope of this paper and left for the
sequel to this paper.

\subsec{{Example 5}: A new background for the D3/D7 system}

\noindent In all the previous examples we have kept the metric
$g_{mn}$ of the internal space (in say type IIB) undetermined.  We
will now be a bit more explicit and study some possible choices
for $g_{mn}$. In fact we already know one example, namely the case
when $g_{mn}$ is the metric for a $K3 \times T^2/\IZ_2$ background
geometry.  This background was chosen in \dtds\ because the metric
of the system $-$ at least when $K3$ is at the orbifold limit $-$
is known. Furthermore, there is also a concrete F-theory
realization of the system as a $K3 \times K3$ fourfold with G
fluxes, that we can exploit to study a system of a $D3$ brane in
the background of seven-branes. In the $\IZ_2$ {\it orientifold}
limit, we have a system of $D7$ branes and $O7$ planes that are
arranged in such a way as to allow a $D_4^4$ singularity. Away
from the orientifold point, we have generic seven-branes, and the
cosmological $D3/D7$ system was studied by keeping all the others
seven-branes far away. The choice of non-primitive $G$ fluxes were
made in such a way that it allowed a single $D3$ brane to move
towards the single seven-brane. Furthermore, in this limit the
background changed to $K3 \times \IP^1$ because of
non-perturbative corrections. The $\IP^1$ is not smooth, but has
singular points where the seven-branes are located.

Although the above compact space is simple enough to give us a
nice cosmological scenario, the background suffers from one
important drawback: away from the orbifold limit, the metric of K3
is not known, and unless we go to this orbifold limit of K3 we
cannot, in practice, evaluate any precise cosmological effects
from this scenario. Furthermore, due to the inherent orientifold
nature of this background, the polarizations of $B_{NS}$ fields
are rather restricted.  Overcoming these issues would therefore
require a new embedding for the $D3/D7$ system, which would have
the following properties:

\noindent $\bullet$  Allow $D3$ and $D7$ branes\foot{We need a
background that allows { at least } {\it two} $D7$ branes instead
of one. Having two $D7$ branes means that on the $D3$ brane we
will see a global symmetry of $SU(2)$ which would, in turn, allow
{\it semi-local strings} instead of cosmic strings.  The existence
of such semi-local strings in the $D3/D7$ system solves the
problem of the otherwise generic overproduction of cosmic strings
\anaren, \semil. Some other aspects of semi-local strings will be
discussed later in this paper.}.  In other words, should have an
explicit F-theory realization as a fourfold with $G$ fluxes. These
G-fluxes will give $H_{NS}$ and $H_{RR}$ fields in type IIB.

\noindent $\bullet$  The six dimensional base of the fourfold should have
a supergravity description. This is required to understand the motion of
the $D3$ brane towards the $D7$ brane.


\noindent One of the known six dimensional    manifolds   that would satisy all the criteria  is a deformed conifold.
The metric of the deformed conifold is exactly
known\foot{In a somewhat different
context, cosmological   models
involving the conifold transition  were recently studied in \motu.}.
What   remains,   is
to   include   the seven-branes in this   background.

To   include   seven-branes, we require  {an F-theory   }   fourfold that will have the deformed conifold as its base.
The $T^2$ fiber of the fourfold will have to degenerate at some number of points on the
base. The degeneration points are where the seven-branes   occur.   Let us
therefore first construct   this fourfold.

The fourfold that we require for our case should be compact, although we will not consider
the compactness too seriously for the subsequent analysis.
Some analysis of such a fourfold is   given   in \dotd. Taking the usual deformed conifold
equation  defined in ${\bf C}^3$,
we can compactify to a projective variety ${\cal B}_\mu$
in ${\bf P}^4$ by adding
${\bf P}^1 \times {\bf P}^1$ to the boundary of the deformed conifold, which {itself}   is
symplectically isomorphic to the cotangent bundle $T^\ast {\bf S}^3$ over the three sphere.
The quadratic equation now becomes:
\eqn\defconi{
{\cal B}_\mu:~~ z_0^2 + z_1^2 + z_2^2 + z_3^2  - \mu z_4^2 =0.}
The quadric threefold ${\cal B}_\mu$ does not develop any new singularities at
infinity and is   thus
smooth for $\mu \ne 0$ and has a conifold singularity at $(0,0,0,0,1) \in {\bf P}^4$ when
$\mu=0$. Moreover, the anti-canonical bundle $-{\cal K}_{{\cal B}_\mu} :=
\wedge^3 {\bf T}{\cal B}_\mu$ will be the restriction of ${\cal O}(3)$ of ${\bf P}^4$
to ${\cal B}_\mu$ by the adjunction formula,  and hence it is very ample. From the
Kodaira vanishing theorem, one can also show that $H^i({\cal B}_\mu, {\cal O}) =0$ for $i>0$.
We now define a fourfold
$Y_\mu$ as a subvariety in  the projective bundle
${\cal P} ({\cal O} \oplus {\cal L}^2  \oplus {\cal L}^3)$, with ${\cal L} := {\cal K}_{{\cal B}_\mu}^{-1}$. The
Weierstrass equation is given by
\eqn\weoh{y^2z_o = x^3 + f z_o^2x + g z_o^3}
where
 $z_o, x,y, f, g$ are the sections of ${\cal O}, {\cal L}^2, {\cal L}^3,
{\cal L}^4, {\cal L}^6$ respectively.
Since the anti-canonical bundle
${\cal L} = {{\cal O}(3)}\vert_{{\cal B}_\mu}$ is very ample,
we may choose $f$ and $g$ so that the fourfold $Y_\mu$ is smooth.  By the
projection formula, one can see that $Y_\mu$ is Calabi-Yau since $H^i( \CB_\mu,\CO) =0, i>0$.
By construction, the natural projection ${\cal P} ({\cal O} \oplus {\cal L}^2  \oplus {\cal L}^3) \to {\cal B}_\mu$ induces a
 fibration     $Y_\mu \to {\cal B}_\mu$ whose fibers are elliptic curves.
F-theory on the Calabi-Yau fourfold $Y_\mu$ is by definition type IIB
theory
compactified on the base ${\cal B}_\mu$ with background axion-dilaton field
$\lambda$ whose $j$-invariant is given by the usual formula with
various $(p,q)$ seven-branes appearing at the loci where the elliptic
  fibration    degenerates.
Using the  Riemann-Roch theorem   for
${\cal B}_\mu$ and integrating over the elliptic fibers, one can evaluate
the Euler-Characteristic $\chi$ of $Y_\mu$ to be 19728.

The above construction of the   fourfold gives us a way to see the
structure of the base. As we can easily see, the base is not quite
a deformed conifold as it not possible to have $c_1 = 0$ for the
base, where $c_1$ is the first Chern class. Although the metric
can be shown to be approximately a warped deformed conifold. The
warp factors will come from the backreactions of the seven-branes,
fluxes and the D3-brane. For the specific manifold that we
constructed, we require the $H_{NS}$ and the $H_{RR}$ fluxes to
satisfy \eqn\hnshrr{\int_{{\cal B}_{\mu}} H_{NS} \wedge H_{RR} =
822 - n,} where $n$ is the number of $D3$ branes. Normalizing the
NS and RR fluxes to satisfy this will be essential to have $D3$
branes in this scenario. For the simplest case we would require $n
= 1$, i.e a single $D3$ brane.

But this is not enough. The specific configuration that we require should (a) allow the $D3$ brane to move towards the $D7$ branes and (b) not allow
the problematic ANO cosmic strings. This two aspects can be easily solved by switching on $H_{NS}$ and $H_{RR}$ fluxes that are not {\it primitive}.
Once primitivity is gone, the fluxes in the background with the compact base ${\cal B}_\mu$ will break supersymmetry and the $D3$ brane will move towards
the $D7$ branes.

To get rid of the ANO strings, we have to isolate two $D7$ branes at a point on the compact space and keep all the other seven-branes far away, as discussed
in the footnote earlier.
This
configuration will therefore be similar to the one developed in \semil\ for the $D3/D7$ system in $K3 \times T^2/\IZ_2$ background. The non-primitive fluxes
will drive the $D3$ brane towards the two $D7$ branes, and this will be the Coulomb phase for our inflationary model. The story is similar to our earlier papers
\dtds, \semil\ so we will not repeat it here anymore. Interested readers may want to look up \dtds\ and \semil\ for details.

What we want to concentrate  on         in this section is the
precise metric for the system. We already know from \defconi\ that
the F-theory base of our fourfold, ${\cal B}_\mu$, has $b_3 \ge
1$, where $b_3$ is the third Betti number and therefore supports
three cycles. On the other hand we have shown above that, by
Kodiara's vanishing theorem, \eqn\kodvan{H^i({\cal B}_\mu, {\cal
O}) =0~~ ~~ {\rm for} ~~~  i>0} which implies that the fourfold
$Y_\mu$ is a Calabi-Yau. For our case however we have to change
the geometry a little bit by shifting most of the seven-branes far
away. What we require is the fourfold base to be approximately a
Calabi-Yau, or more appropriately, a deformed conifold. Since the
first Chern class of the base ${\cal B}_\mu$ is non zero, this
might seem difficult to realize in principle. In practice however
the situation can be controlled. We can shift the seven-branes in
such a way that the metric of the system is approximately a warped
deformed conifold. In fact we have to take care of the following
things:

\noindent $\bullet$ The effect of warping coming from the back reactions of branes and fluxes on the geometry.

\noindent $\bullet$  The precise metric of the {\it deformed} conifold\foot{As mentioned earlier and will also be clear soon, our six dimensional manifold is
not a Calabi-Yau and only resembles the usual deformed conifold in some local region. Thus by {\it deformed} conifold we mean a non K\"ahler deformation of
a {\it compact} Calabi-Yau metric.}, and

\noindent $\bullet$ The background three-form fluxes that are non-primitive, in other words $H_{NS} \ne \ast H_{RR}$.

\noindent In addition to these, we have to make sure that the background satisfies all the string   equations   of motion. This is now a rather complicated issue. Let us
see how far can we proceed towards getting the full answer. First, observe that the background with $D3, D7$ and fluxes on the fourfold base ${\cal B}_\mu$
is closely related to the recently studied type IIB manifolds in the geometric transitions setting of \realm.

We could then follow the procedure laid down in \realm\ to get our background metric. However one immediate problem is that the background studied in
\realm\ is supersymmetric. But what we require is actually a non-supersymmetric (and hence non-static) background. From earlier sections, we know that this
can be easily achieved by taking fluxes that are non-primitive. With a time-independent choice of the warp factors for the internal space we can in fact use the
full machinery of geometric transition to determine our final answer. Our goal therefore is to determine $g_{mn}$ in
\eqn\metmik{ds^2 = e^{A(y,t)}(-dt^2 + dx_1^2 + dx_2^2 + dx_3^2) + e^{B(y)}~g_{mn}~dy^m dy^n}
with the assumption that $A(y,t)$ and $B(y)$ to be determined by the analysis that we presented earlier.

Therefore, to start off, first let us use some simple
approximations that we used in \bdkt\ and \realm. We want to
remind the reader that these approximations are just done to
simplify the ensuing analysis. The six-dimensional manifold ${\cal
B}_\mu$ will have coordinates ($r, \phi_1, \theta_1, \phi_2,
\theta_2, \psi$), where ($\theta_i, \phi_i$)  ($i = 1,2$) are
related to two $S^2$ metrics (that doesn't always imply that we
will have two topologically non-trivial $S^2$ in our setup). The
radial coordinate will be denoted as $r$, and $\psi$ is the usual
$U(1)$ fibration over the $S^2$ bases. With this, we can now use
the same trick that was used in \bdkt\ and \realm\ to get the
final type IIB metric, namely: use the geometric transition from
type IIB to another type IIB solution. Thus our initial solution
will be related to the one in \pandoz, \bdkt, \realm\ but with
seven-branes inserted in. In other words we are choosing the
metric: \eqn\realmmet{ds^2 = h_6~dr^2 + (dz + E~dx + F~dy)^2 +
\vert dz_1\vert^2  + \vert dz_2\vert^2} with $dz_i, dz_2$ being
two-tori with two different complex structures and $E,F$ as some
functions given in \bdkt,\realm\ (see also \pandoz). The
background also had non-trivial dilaton $\phi$ and axion
$\tilde\phi$ switched on. The coordinate   redefinition   that we
are using here will be: \eqn\cord{(dx, dy, dz) \equiv \left({1\o
2} \sqrt{\gamma_1 \sqrt{h}}~{\rm sin}~\theta_1~d\phi_1, {1\o 2}
\sqrt{\gamma_2 \sqrt{h}}~{\rm sin}~\theta_2~d\phi_2, {r\o 2}
\sqrt{\gamma'_1 \sqrt{h}}~d\psi \right)} where $h$ is the overall
warp factor that comes from the   back reaction   of fluxes and
branes on the geometry. This coordinate     redefinition   is made
to the geometry {\it before} the   geometric transition (in the
language of \realm). Thus the threefold base of F-theory is
approximately a resolved conifold with the resolution parameter
being given by: \eqn\respara{ {1\o 2}\sqrt{\gamma_2 - \gamma_1}.}
The physical meaning of $x, y$ and $z$ is already  explained in
\bdkt. They are related to the tori that we use in place of the
two spheres. In other words, we will   have   two tori with
coordinates ($x, \theta_1$) and ($y, \theta_2$). The coordinate
$z$ remains as the usual $U(1)$ fibration over tori bases.
Furthermore, these tori will have non-trivial complex structures
that can be either integrable or non-integrable. What we require
for geometric transitions is non-integrable complex structures,
although, as discussed in \bdkt, integrable complex structures
lead us very close to the expected right metric in the mirror. We
will not elaborate on  these here     anymore,   and the readers
may want to see \bdkt\ and \realm\ for more details.

However, there is one subtlety that we have to mention at this stage. This is related to the configuration of seven-branes in our setup. Recall that F-theory
generically predicts seven-branes that could be either $D7$ branes and/or generic non-local seven-branes charged under both axion and
dilaton of type IIB theory.
The backreactions of the branes and fluxes tell us that the type IIB metric can be determined by two independent variables $E$ and $F$ from \realmmet\ as
\eqn\met{j_{xx} = 1 + E^2, ~~~~~~ j_{yy} = 1 + F^2}
with all other components of the metric \realmmet, $j_{\mu\nu}$,
 some functions of $E, F$ and the complex structures only.
With these choices of $E$ and $F$, F-theory predicts that the
metric along $x,y$ and $\theta_1, \theta_2$         corresponds,
in fact, to   two different tori with complex coordinates $d\chi_1
\equiv dx + \tau_1~ dy$ and $d\chi_2 \equiv d\theta_1 +
\tau_2~d\theta_2$ and complex structures $\tau_1$ and $\tau_2$
respectively. Observe that these complex structures are different
from the original complex structures of the ($x,\theta_1$) and
($y,\theta_2$) tori. The new complex structures that mixes $x,y$
and $\theta_1, \theta_2$ are given by \realm \eqn\mixedco{\tau_1 =
{1\o 1+E^2} \left[c E F + i~ \sqrt{\alpha^{-1} + (1-c^2) E^2
F^2}\right], ~~~~~ \tau_2 = i~\sqrt{\gamma_2 \o \gamma_1}} where
$\alpha$ is defined as $\alpha = (1 + E^2 + F^2)^{-1}$.

The value of $c$ in \mixedco\ is now important. As discussed in \realm, the constant $c$ can take only two values, $c = 1$ or $c = 0$. The two values of
$c$         distinguish   two possible configurations  we can have with the seven-branes, namely: {\it at} the orientifold point or {\it away} from the orientifold point.
At the orientifold point $c$ is zero and therefore the two tori ($x,y$) and ($\theta_1, \theta_2$) are square tori. On the other hand, away from the orientifold point,
the value of $c$ is typically 1. In fact this is the situation that we really require for our case, because we have to move the seven-branes far away, keeping only
two $D7$ branes at a point. According to the  F-theory analysis, this amounts to having the orientifold planes split into generic seven-branes.  As  a result the ($x,y$)
torus pick up a complex structure that can be determined from \mixedco\ as
\eqn\costra{\tau_1 = {EF + i~\sqrt{1 + E^2 + F^2}\o 1 + E^2}}
with the other ($\theta_1, \theta_2$) torus remaining a square torus.

At this point, one would like to compare our background to the one that appeared in \realm\ (at least before the  geometric transition). The configuration  in
\realm\    corresponds to   $c = 0$ and the seven-branes (with orientifold seven planes) located at points on the ($x,y$) torus. For our case, we require a situation that is
the exact
opposite of what we had in \realm. We require $c =1$ with the seven-branes (no more orientifold planes) located at points on the ($x,\theta_1$) torus.

The above story was before  the  geometric transition in the type IIB theory. Having set the value of $c$ and the required torus, 
we can do a geometric transition to get to our
final, albeit, static background. The final metric is
\eqn\finmet{\eqalign{ ds^2 = &~~~ h_1~(dz + a_1 dx + a_2 dy)^2 + h_2~(dy^2 + d\theta_2^2) + h_4~(dx^2 + h_3~d\theta_1^2)~ + \cr
& ~~ + h_5~{\rm sin}~\psi~(dx~d\theta_2 + dy~d\theta_1) + h_5~{\rm cos}~\psi~(d\theta_1d\theta_2 - dx~dy) + h_6~dr^2},}
where we can now use the variable of the metric \realmmet\ to determine the values of $h_i$ as follows (see also \realm):
\eqn\hideterm{\eqalign{& h_1 = {e^{-2\phi}\o \alpha_0 CD}, ~ h_2 = \alpha_0(C + e^{2\phi} E^2), ~ h_4 = \alpha_0(D + e^{2\phi} F^2), ~
h_5 = 2 \alpha_0 e^{2\phi} EF \cr
& h_3 = {C - \beta_1^2 E^2 \o \alpha_0(D + e^{2\phi} F^2)}, ~ a_1 = -\alpha_0 e^{2\phi} ED, ~ a_2 = -\alpha_0 e^{2\phi} FC, ~
\alpha = {1\o 1+E^2 + F^2}\cr
&  \alpha_0 = {1\o CD + (CF^2 + D E^2) e^{2\phi}},~~~~ \beta_1 = {\sqrt{\alpha_0} e^{2\phi}\o \sqrt{ e^{2\phi} CD - {(C + e^{2\phi} E^2) (1-D^2) F^{-2}}}}\cr
& C = {\alpha \o 2} \left( 1 + F^2 - {EF\sqrt{1+F^2} \o \sqrt{(1+E^2)}}\right), ~~~~ D = {\alpha \o 2} \left( 1 + F^2 + {EF \sqrt{1+F^2}\o \sqrt{(1+E^2)}}\right).}}
This way we can determine the metric \metmik\ completely, with the only unknowns being the warp factors $A$ and $B$. We believe that using three
warp factors in  the  M-theory setting, these will be uniquely determined.
This would therefore be the static background that would allow
seven-branes, $D3$ brane and non-trivial three form fluxes. The axion-dilaton distribution can be determined from the corresponding F-theory
curve. A slight deviation from the primitivity will trigger the motion of the  $D3$ brane towards the $D7$ branes. Both the $D3$ brane and the $D7$ branes are now located at
points on the ($x,\theta_1$) torus. The $D7$ branes therefore wrap directions ($y, \theta_2, z, r$) and are stretched along $x^{0,1,2,3}$. The $D3$ brane is along the
$x^{0,1,2,3}$     directions.   We can also choose the three forms $H_{NS} \equiv H$ and $H_{RR} \equiv H'$ in the following way:
\eqn\threeform{\eqalign{ & H =  \del_r B_{x\theta_1}~dr \wedge dx \wedge d\theta_1 + \del_{\theta_2} B_{x\theta_1}~d\theta_2 \wedge dx \wedge d\theta_1 ~ + \cr
& ~~~~~~~~~ + \del_r B_{y\theta_2}~dr \wedge dy \wedge d\theta_2 + \del_{\theta_1} B_{y\theta_2}~d\theta_1 \wedge dy \wedge d\theta_2  \cr
& H' = \del_{\theta_1} B'_{zy}~dz \wedge d\theta_2 \wedge dy + \del_{\theta_2} B'_{zy}~dz \wedge d\theta_1 \wedge dy ~ + \cr
& ~~~~~~~~~ + \del_{\theta_1} B'_{zx}~dz \wedge d\theta_1 \wedge dx + \del_{\theta_2} B'_{zx}~dz \wedge d\theta_2 \wedge dx,}}
where $B_{x\theta_1}(r,\theta_2)$ and $B_{y\theta_2}(r,\theta_1)$ are the $B_{NS}$ fields and
$B'_{zx}(\theta_1,\theta_2)$ and  $B'_{zy}(\theta_1,\theta_2)$ are the
$B_{RR}$ fields. Observe that these $B$ fields are functions of the radial coordinate $r$ as well as the angular coordinates $\theta_1$ and $\theta_2$.
Furthermore, since we do not require primitivity at the Coulomb stage of our model, we can take a simplified choice of the
threeforms by putting $B'_{zx} = B'_{zy} = 0$.
This would mean that we only have $H_{NS}$ in our background. This choice will be useful to study black holes in this
scenario. We will also discuss the changes that can appear when we switch on $H_{RR}$. In the figure below:
\vskip.1in

\centerline{\epsfbox{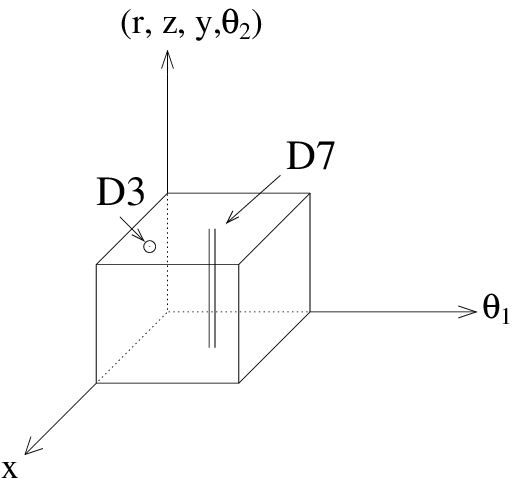}}\nobreak

\vskip.1in

\noindent the $D3$ brane is at a point on the ($x,\theta_1$) direction 
similar to the two $D7$ branes which are stretched along the
($r, z, y, \theta_2$) directions. As discussed above, the fluxes would
drive the $D3$ brane towards the two $D7$ branes. The whole system is
of course embedded in a deformed conifold background. In the figure
above, the two $D7$ branes wrap the three cycle ($z, y, \theta_2$) of
the deformed conifold. Furthermore when we switch off $H_{RR}$ to keep
only the $H_{NS}$ field, we have to consider multiple $D3$ branes
satisfying \hnshrr. In the above figure we can still have the same
configuration, but now we have to keep other $D3$ branes far away
(along with the set of seven branes). This way the cosmology of the
model will remain intact.

One might ask if there are other interesting geometries that one could
use to construct cosmological models. In this context, we recall that
\drmknmrp\ studies the physics of closed string tachyons localized to
~ $\IC^3/\IZ_N$ nonsupersymmetric noncompact orbifold singularities and the
phenomena therein using the connections between the worldsheet
orbifold conformal field theory, RG flows and the toric geometry of
these orbifolds. The geometry in codimension three is more intricate
than the lower dimensional cases \aps\ \hkmm\ \vafatach. In particular
if a more relevant tachyon condenses in the process of condensation of
a less relevant tachyon, there are flip transitions \drmkn\ between
topologically distinct tachyonic resolutions of the original
singularity (contrast with the more familiar flops in Calabi-Yau
spaces, executed by moduli). It is interesting to ask if there are
compact embeddings of these ~$\IC^3/\IZ_N$ singularities in the presence of
nontrivial background fluxes, which then give rise to a four
dimensional effective spacetime. Since these are nonsupersymmetric,
such compact embeddings are necessarily non-Calabi-Yau. We expect that
turning on fluxes would modify the bare masses of closed string
tachyons (see e.g. \evatach), and more generally the tachyonic physics
of the geometry. In the context of codimension three singularities,
this could potentially give rise to interesting nonsupersymmetric
vacua in four dimensions, with the fluxes potentially stabilizing both
tachyons and moduli in the geometry. It would be worth investigating
these ideas further.

\subsec{{Example 6}: Wrapped branes and non-supersymmetric black
holes}

\noindent The new background that we constructed as \metmik\ and
\finmet\ should more or less capture the full cosmology of our
inflationary model, including the slow roll and the exit from
inflation. A viable cosmological model, like ours, should also
open up avenues to study, for example, black holes and other
possible solitonic defects in the model.  Black holes can, in
principle, form in the early universe (known as {\it primordial}
black holes) during the slow roll or even in the later waterfall
stage of inflation \lindeBH.  An overproduction of these of course
will have disastrous cosmological problems. And therefore their
quantity should be negligible. For uncharged black holes this is
not much of a problem as most of these black holes eventually all
decay via Hawking radiation and therefore they are not there in
the present time.  However for black holes that are charged, this
is not possible as they cannot completely decay. They will radiate
till the charge is enough to sustain the mass of these black
holes. This way we might be able to see them in the present time.

In this example, we would therefore like to ask whether in our
supergravity background, it is possible to construct charged
blackhole configurations. It turns out that we can exploit the
fact that $b_3 \ge 1$ to construct such black holes using
D3-branes wrapping three cycles of the non Calabi-Yau deformed
conifold\foot{Not all wrapped branes are black holes in lower
dimensions. A generic state of mass $M$ in quantum gravity is a
black hole if its Schwarzschild radius is much greater than its
Compton wavelength, i.e. $GM >> {1 \o M}$, i.e. for $M >> M_{pl}$,
the state will look like a black hole. Now consider wrapping $N$
Dp-branes on a collection of noncontractible $p$-cycles with
homology basis $\Sigma_i$ of a compact space that admits
$p$-cycles. This has a charge given in terms of a central charge
written schematically as  $Z=\sum_i ~q_i \Sigma_i$, where the sum
includes dual cycles etc. Consider for simplicity a single cycle.
Then the mass of this state in the noncompact 4D spacetime is $M
\sim  N \cdot T_{Dp} \cdot V_p \sim  {N V_p\o g_s l_s^{p+1}}$,
where $V_p$ is the volume of the $p$-cycle and $T_{Dp}$ is the
brane tension. If this cycle is supersymmetric, the state is BPS
and we have  $M \sim \vert Z\vert$. For this state to be a black
hole in 4D, we require ${N \o g_s}\cdot { V_p \o l_s^{p+1}} >> {1
\o g_s l_s}$ i.e. $N \cdot {V_p \o l_s^p} >> 1$. Now if  ${V_p \o
l_s^p} >> 1$,  a $p$-cycle can be thought of in terms of classical
geometry $-$ worldsheet $\alpha'$ corrections are negligible. If
we have small $N \sim 1$,  this looks like a point particle BPS
state since we can potentially violate  $M >> M_{pl}$  if the
cycle is not large enough. However if $N >> 1$, then we
necessarily have  $M >> M_{pl}$ even if the cycle volume is not
too large, so that the state is indeed a black hole. If the theory
has sufficiently low supersymmetry, moduli spaces are lifted
typically and one does not need to worry about possible lines of
marginal stability where this charge-$N$ state becomes marginally
unstable to bifurcation into states of lower charge.}. However we
have to be careful here. The black holes discussed in \andyS\ are
actually BPS and therefore they have the charge-mass equality (i.e
the supersymmetry condition). Furthermore, since the masses of
these black holes are given by the size of the three cycle,
shrinking the size of the three cycle will result in  charged
massless black holes\foot{For an analysis of entropy of Calabi-Yau
black holes the reader may refer to \marinaCY.}.

On the other hand, what we {have} are non-supersymmetric black
holes. Therefore the constructions of \andyS\ are not exactly the one
{relevant here, } although our discussion will follow closely the
discussion of \andyS\ as the underlying manifold is a warped deformed
conifold, but with non-primitive fluxes switched on.  In the figure
below:
\vskip.1in

\centerline{\epsfbox{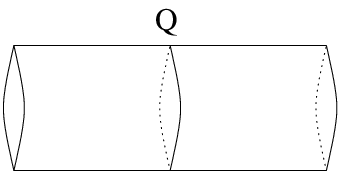}}\nobreak

\vskip.1in
\noindent the black hole is the wrapped D3 along the compact direction. The compact direction of the cylinder in the figure above
is the three-cycle and the non-compact direction of the cylinder is the four dimensional spacetime. We represent the charge of the
wrapped $D3$ as $Q$.
In the $D3/D7$ system, the full cylinder will lie on the world volume of the $D7$ branes, and the $D3$ brane will be
along the non-compact direction. Therefore from the world volume of $D3$ brane, the wrapped $D3$ on the compact direction will appear as a
black hole of charge $Q$.

The original configuration of \andyS\ is a supersymmetric one and therefore we have to see whether we can say something about the non-supersymmetric case.
The situation is not that bad, because in the absence of fluxes, the background   does   preserve the required amount of supersymmetry.   Therefore,
let us first switch off the fluxes. Then we have a static configuration with $D3/D7$ branes on a deformed conifold background, with another $Q$ number of
$D3$ branes on the compact three cycle of the deformed conifold. In the earlier sub-section we studied the supergravity solution (albeit with fluxes)
for the system, when the wrapped $D3$ branes were {\it absent}. We now put them in and, in the first approximation, remove the fluxes. Can we write the
supergravity solution for this case?

It turns out that when the size of the three cycle is vanishing we
can indeed say something about this configuration. This will be
the case where we expect to get massless black holes in four
dimensions. To study the supergravity solution, let us first
T-dualize along directions $x$ and $y$, where we have already
defined ($x, \theta_1$) as one $S^2$ and ($y, \theta_2, z$) as the
three cycle. This T-duality was first studied in  \bsv\ and was
later detailed in \karch. Our configuration will be more involved
than the one presented by \bsv\ or \karch, because we have
seven-branes in the background. In the limit where $\mu \to 0$ in
\defconi, the deformed conifold will turn into a pair of
intersecting NS5 branes and the $D3$ brane of the $D3/D7$ system
will become a $D5$ brane stretched along $x,y$ directions.
Together they form a brane box configuration given below as:
\vskip.1in

\centerline{\epsfbox{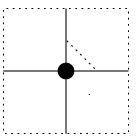}}\nobreak

\vskip.1in
\noindent where the box is along $x,y$ directions, and the NS5 branes are denoted as orthogonal lines with $D5$ on the slots of the box (see \hanany\ for a description
of brane boxes). The wrapped $D3$ on the three cycle is now a D-string. This is given as the diagonal  dotted line on the $x-y$ plane. The black dot in the
middle corresponds to the T-dual of the deformed three cycle.
This part of this configuration is in fact the
original picture of \bsv. In our case, however we will have additional branes. This will be the T-dual of the $D7$ branes, which, here will remain as $D7$ branes but
oriented along different directions (in the figure above this is rather
difficult to depict).
So combining everything, we have a brane box like configuration with additional $D7$ branes.
Question now is what happens to this configuration if we
switch on a $B_{NS}$ field in the original deformed conifold background oriented along:
\eqn\bns{B_{NS} = B_{x\theta_1}~dx \wedge d\theta_1 + B_{y\theta_2}~dy \wedge d\theta_2}
where $B_{mn}$ are defined in \threeform.
It is easy to see that this $B$ field will dissolve into metric after two T-dualities along $x,y$ directions, and therefore   deform   the brane box configuration. This
deformation is where we break supersymmetry in this configuration.

In the original case where we had primitive $B$ fields, i.e both the NS and RR three forms selfdual to each other, the T-dualities of the $B_{NS}$ still gave a
deformed brane box configuration. But the presence of $B_{RR}$     in fact   made the resulting background supersymmetric. Now that we do not have the RR field (or switch on a
non-primitive RR field) the configuration becomes non-supersymmetric.

For the supersymmetric case, the spectrum of the string between the NS5 branes was easy to evaluate using standard technique. This gave rise to a {\it single} hypermultiplet
at the intersection region (denoted as the dark circle in the figure above). When the brane box gets deformed, the mode expansion will now no longer be easy to
calculate.

On the other hand, the deformation of the brane box {itself}   is not too difficult to calculate. The metric of the intersecting NS5 branes can be written down using standard
techniques. Putting D5 branes in the slots of the NS5 branes actually convert this system to an intersecting NS5-D5 brane system. The metric of the box is therefore written
using the coordinates $dx, dy$ and $d\theta_i, dr, dz$ (we will not write this here, but the readers can easily work this out). The deformation to the box is therefore
the following: in the metric where ever we have $dx, dy$,   we   replace this by $dx - B_{x\theta_1}d\theta_1$ and $dy - B_{y\theta_2}d\theta_2$ respectively. An example of
this has recently appeared in the last sections of \realm, which the reader can look up for     a derivation.    The presence of $D7$ branes   does    not modify any of these
conclusions  (though the metric may get more involved by the back-reactions of the $D7$ branes).

Although the $D7$ branes do not modify the deformation (that we
expect in the absence of seven-branes) they do create an
obstruction in the system which hinders a simple lift to   the
type I scenario. This is where we can compare how our construction
differs from earlier studies of related scenarios. The obstruction
comes from the fact that we are away from the orientifold point
with extra seven-branes that are charged   with respect to   both
the axion as well as the dilaton. A somewhat more clear
manifestation of the obstruction is the appearance of extra $B$
fields in the brane box configuration. These $B$ fields have
components that are related to the original type IIB $B$ field
\bns\ but they would arise even if we switch off \bns. In the
original framework of \realm,   where the $z$ coordinate of the
deformed conifold was delocalized, these $B$ fields were absent
(or can be shown to be pure gauge). However when we have localized
metric one cannot gauge these $B$ fields away as they become
non-trivial functions of the inherent $U(1)$ fibration of the type
IIB metric. For the configuration of interest, the $B$ fields will
appear in the brane box configuration with the following
components: \eqn\newbf{B = \alpha_1~(dx - B_{x\theta_1}~d\theta_1)
\wedge d\theta_2 + \alpha_2~(dy - B_{y\theta_2}~d\theta_2) \wedge
d\theta_1,} where $\alpha_i$ are functions of all coordinates
except $x$ and $y$. From \newbf\ we see that even when we switch
off $B_{x\theta_1}$ and $B_{y\theta_2}$, there exist non-trivial
$B$ fields whose components are determined by $\alpha_i$.
Furthermore from the components we see that the $B$ fields do not
lie on any one of the NS5 branes. In fact they have one leg on
each of the NS5 and D5 branes.

These aspects make the mode expansion of strings (that    are   stretched between the NS5 branes) a bit difficult. The physical meaning of the $B$ field \newbf\ should
now be clear. This is related to the conifold deformation that converts a conifold geometry to a deformed conifold geometry. So starting with a
{\it deformed} brane box configuration
(which is the T-dual of a confold geometry with a probe $D3$ and $B_{NS}$ flux \bns)
we switch on a $B$ field \newbf; and this is the brane box configuration for our case. By construction
primitivity is already broken in the original type IIB framework and therefore this configuration will not preserve supersymmetry. This way we can get the supergravity
solution for our case.
Of course it still remains a complicated problem to calculate the spectrum of the strings between the two NS5 branes using first principles.

Having gotten the T-dual box configuration, we can use indirect
means to estimate the spectrum of the strings inside the box. These
strings are of course T-dual of the wrapped D3-branes in the
deformed conifold setup. In the absence of fluxes, the wrapped $D3$
brane give rise to massless (or massive, depending on whether $\mu$
in
\defconi\ vanishes or not) hypermultiplets. In the presence of fluxes and also when $\mu \ne 0$, the deformed conifold is T-dual to NS5 brane box with a $B$ field
and/or $\vert \mu \vert$ separation between the NS5 branes. As we
can see, even if the configuration gets deformed by \bns, the
string will always produce massive states. Generic study of $B$
field done earlier in somewhat related scenario showed that these
massive states are always hypermultiplets (see \ori,\marina\ for
details). These massive hypermultiplets will therefore be the
massive black holes that we expect to see in the $D3/D7$
inflationary scenario.

This completes our discussion of the basic picture of black holes from supergravity point of view. Below, we will discuss 
whether such configurations could be generated in the inflationary D3/D7 setup.

\subsec{{Example 7}: Cosmological effects from brane-antibrane annihilation}

\noindent Now that we have argued a way to get massive charged
black holes in our setup by wrapping D3-branes on three cycles,
our next question would be whether higher dimensional wrapped
branes could also be used to get strings and domain walls in four
dimensions. Naively one might think that wrapping D5 branes on
either three or four cycles should produce the required defects in
our spacetime. On the flip side, formations of domain walls and
strings would be  in conflict  with  many cosmological
observations. Thus we have to find a way to argue that such
defects will not form.

 In order to do this, we should go back to the black hole
configuration that we studied in the previous example. These black
holes are charged objects: therefore to study charged black holes
in the inflationary scenario, charge conservation dictates that the
only known mechanism is {\it pair production}.
This scenario has been addressed earlier in various papers (see for
example \ggs, \dowker, \hawking). The black holes in these papers are
pair produced as charged point particles.

This simple observation therefore gives us a way to address all the
subsequent brane related defects, namely: study them in the framework
of pair productions. Therefore we should consider pair productions of
$D5/ {\bar D5}, D3/ {\bar D3}$ and even $D1/ {\bar D1}$.  The
following three conditions would now prevail:

\noindent $\bullet$ The pair produced branes can fall back into
each other and annihilate. This will not have any observable effects, and therefore can be ignored.

\noindent $\bullet$ The pair produced branes are separated by some
external forces. The external forces could be via some external
fluxes or from inflation itself. The external fluxes that we have
are in general not sufficient enough to do this job unless they
are very large or have large fluctuation, and therefore the only
other mechanism could be inflation itself. As discussed for the
point particle cases in \hawking, the process of inflation can
pull these particle-antiparticle pairs so that they do not
annihilate any more. This process will keep, say, the particle
(i.e the charged black hole) in our cosmological event horizon and
hurl away the anti-particle far away so that there will be no
causal contact between them any more.

\noindent $\bullet$ The pair produced branes do not annihilate completely. This is a new phenomenon that is possible in our set-up because of the special
configuration of NS fluxes. There is no counterpart of this effect in the point particle cases and, as we will discuss below, can happen only for some specific
branes. Thus it is not universal.

We now begin with the discussion of the second case. In all the
cosmological examples {we have in mind},   the part which is
inflating is only along the D3-brane world volume, i.e a de-Sitter
metric along $x^{0,1,2,3}$ directions. To an observer along the
D3-brane world volume, the brane-antibrane pairs will appear as
particle-antiparticle pairs. Therefore the inflationary expansion
will be equally responsible here to separate the branes from the
anti-branes.

Therefore first consider the creation of $D3/{\bar D3}$ pairs. The pairs
wrap the three cycle that is oriented along ($y, \theta_2, z$) and that is  orthogonal to the radial direction $r$ and $S^2$ with coordinates ($x , \theta_1$).
In the figure below:
\vskip.1in

\centerline{\epsfbox{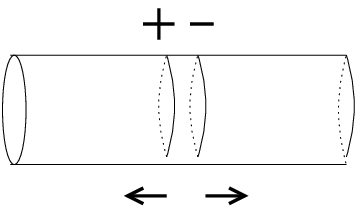}}\nobreak

\vskip.1in \noindent we denote the $D3/{\bar D3}$ pairs by the two
cycles with charges $\pm Q$. As before the cylinder is oriented
along the three cycle (the compact circle in the figure above) and
the space time direction (the non-compact direction of the cylinder
in the figure above). The inflationary drag separates   the wrapped
D3-branes from the ${\bar D3}$ branes, so that we only   see, say,   the
wrapped D3-branes.   They   appear as charged black holes in four
dimensional spacetime.

The above analysis did not consider other orientations for the
$D3/{\bar D3}$ pairs. There is no reason why the pairs should only
be along the $S^3$ direction. Since $b_1 = 0$ i.e the first Betti
number vanishes, the possibility of getting domain walls from the
pairs is not there. It is easy to see that
other configurations will not lie on the D7-brane world volume and
therefore will not form any wrapped states. Thus the only things that
could come out from these pairs are some charged black holes (albeit
in very small numbers, which will clear when we apply this mechanism to study primordial black-holes in sec. 5).
The above considerations also rule out any
cosmological effects from $D1/{\bar D1}$ pairs, as in the absence of one-cycles ($b_1 = 0$)
they will not have any observable consequences.

What happens for the $D5/{\bar D5}$ pairs? It turns out $-$ and
this is our third case discussed above $-$ that this situation is
a little more involved than the $D3/{\bar D3}$ pairs. To analyze
this case let us consider carefully all the possible orientations
for the $D5/{\bar D5}$ pairs. First, let the pairs be along
directions ($x, \theta_1$) and ($y, \theta_2, z$). The coupling on
a D5 brane and an anti D5 brane can be written as \dmukhi
\eqn\couplin{\int_{D5} D^+ \wedge (B_{NS} - F_1) - \int_{\bar D5}
D^+ \wedge (B_{NS} - F_2)} where $F_1$ and $F_2$ are possible
gauge fluxes on the D5 and the anti-D5 brane respectively, and
$D^+$ is the   four-form   of type IIB theory whose field strength
is a five form. The relative sign in the above equation signals
the presence of the   antibrane. Taking only the component $B_{NS}
= B_{x\theta_1} ~dx \wedge d\theta_1$ we see that the above
coupling gives us D3-branes with charge $Q$, where \eqn\chrdth{Q ~
= ~ \int_{\IP^1} (F_2 - F_1)} and the integral is over a sphere
with coordinates ($x, \theta_1$). Observe that this charge is
measured by gauge fluxes induced on the D5 and anti-D5 branes.
This is an important difference, because even though both D5 and
anti-D5 branes see the same $B_{NS}$ field, they would see
different gauge fluxes and therefore would create a D3-brane. This
new D3-brane will now wrap the three cycle along ($y, \theta_2,
z$) direction and appear as       a black hole    of charge Q.

It is now easy to see the two apparent differences from the $D3/{\bar D3}$ pairs.   First,   the $D3/{\bar D3}$ will annihilate as soon as they are formed, unless
the inflationary forces hurl them apart. On the other hand,
the $D5/{\bar D5}$ pairs would annihilate to give rise to wrapped D3 of charge Q even before the pair could be separated.

At this point one might wonder about the behavior of the tachyon\foot{This is not the D3/D7 tachyon, but the inherent tachyon of a brane-antibrane
pair.} in this system.
Due to the presence of $B_{NS}$ and two different gauge fluxes $F_1$ and
$F_2$, the mass   spectrum   is shifted. In fact the zero point energy $E$,
of the string modes between D5 and ${\bar D5}$ is worked out in \marina\ and can be written as
\eqn\masspect{E ~ = ~ \mp {1\o 2} \left( \vert \nu - {1\o 2} \vert \pm {1\o 2} \right)}
where $\nu$ is the shift in the mode number because of the fluxes. From this it is easy to study the behavior of the
tachyon profile for the system.
We can plot $E$ vs. $\nu$ and depict the profile as:
\vskip.1in

\centerline{\epsfbox{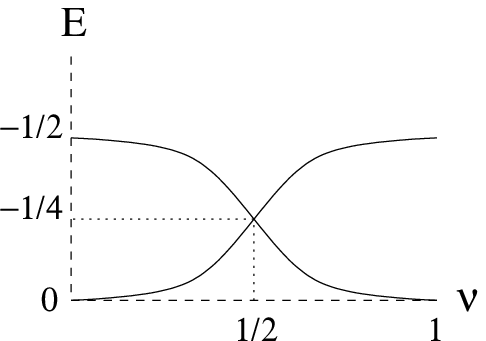}}\nobreak

\vskip.1in \noindent It is now clear that there will be {\it no}
tachyon in the system. In fact the tachyon becomes massless and
the net result is just  D3-branes with charge $Q$ (wrapping the
$S^3$ of the deformed conifold) from the annihilation of the
$D5/{\bar D5}$ pairs. All other orientation of the $D5/{\bar D5}$
will not have any observable effects on the D3-brane world volume.
Thus we see that other than black holes there is very little
possibilities of getting any other charged defects in this
scenario. Before ending this section, let us ask what happens for
the D3/D7 system in $K3 \times \IP^1$ background?

From the analysis  we did above, it is clear that we need at least
one three cycle to have black holes in four dimensions from
wrapped D3-branes. The third Betti number for $K3 \times \IP^1$ is
zero and therefore will not allow any three cycle. Naively this
means that there would be no possibly of getting any charged black
holes in four dimensions. What about other wrapped branes? From
the Betti numbers of our manifold: \eqn\betty{\{ b_i \} ~ = ~ 1, ~
0, ~ 23, ~ 0, ~ 23, ~0, ~1} we might expect charged strings and
three-branes in four dimensions. These charged strings are
extremely rare as they get annihilated by anti strings as soon as
they are formed (furthermore, as discussed before, they will have
no observable consequences because of the absence of one-cycles).
The D3-brane moving towards D7-branes will only detect them at the
final stage of the inflationary scenario. On the other hand, due
to compactness of the internal manifolds, these charged
three-branes are not allowed to form in our case. Also since $b_1
= 0$ there would be no domain walls (to get domain walls we need
D3 branes to wrap one-cycles and/or D5 branes to wrap three cycles
and/or D7 branes to warp five cycles. All of which are absent
here). Therefore it seems that D3/D7 inflationary model in $K3
\times \IP^1$ background has zero possibility of black holes
formation {from wrapped branes}. However before we come to a
definite conclusion let us contemplate another mechanism that
could in principle allow some possibilities of black holes
formation. This mechanism has to do with the choice of three form
fluxes in this background.

The fluxes on the K3 manifold $-$ on which we also have wrapped
D7-branes $-$ induce a configuration of intersecting D5 branes on
the D7-branes world volume. These D5 branes intersect along the $3 +
1$ space which forms our spacetime. This world volume configuration
is in fact U-dual to the deformed conifold setup that we studied
above! The wrapped D3-branes in the earlier configuration will be
U-dual to fundamental strings between the two D5 branes. However
there is one obvious problem with this map: we cannot separate the
two D5 branes, and therefore the strings between them will only
contribute massless hypermultiplets in four dimensions. One way out
of this is to allow different gauge fluxes on the two D5 branes.
This will shift the zero point energies of these strings.

In practice however it turns out more difficult to justify these
massive strings in the intersecting brane configurations. Therefore
the D5 brane configurations also will not be very useful to allow the
formations of black holes in the $K3 \times \IP^1$ background.

Before ending this section, let us contemplate another phenomenon
as the D3-D7 open string tachyon condenses: This is the possible
formation of non-BPS branes wrapped on appropriate cycles of the
compact geometry. We recall \senreview\ that the spectrum of
D-branes in Type IIA (B) theory included odd (even) dimensional
uncharged D-branes, besides the more familiar BPS D-branes that
carry RR-charges. These non-BPS branes have an open string tachyon
on their worldvolume and are unstable.  While their tensions also
scale as ${\cal O}({1\o g_s})$, they are dressed by the open
string tachyon potential which appears in the corresponding DBI
action.  Thus for instance the IIB theory on $K3 \times \IP^1$ in
principle will nucleate non-BPS 2-branes wrapped on 2-cycles
therein, since they are uncharged objects allowed simply on
energetic grounds as the D3-D7 tachyon condenses (on the other
hand, the approximately deformed conifold does not admit 2-cycles
so this process is not possible here). As possible defects, these
unstable branes would naively be expected to decay rapidly due to
the dominant open string tachyon. From the point of view of string
theory, the work of \llm\ (using Sen's boundary state
\senrolltach) shows that the dominant decay mode for these
unstable branes is into the full tower of massive closed string
states. On the other hand, it is {\it a priori} unclear what the
effective description of such decay processes is, via
supergravity. While one could naively imagine these wrapped
unstable branes to appear as point-like uncharged black holes in
the noncompact four dimensions, such objects would typically also
source the dilaton and other scalar fields. Thus it is not evident
that these systems describe black holes with smooth horizons as
opposed to naked singularities. Related work on possible
supergravity descriptions of brane-antibrane stacks appears in
e.g. \BraxMandalOz. It would be interesting to understand these
questions in greater detail.

\newsec{Classification and stability of cosmological solutions}

In Sec.~3 we studied various cosmological examples, by taking different
limits of the background fluxes and with some quantum corrections. We
also saw the necessity of higher derivative corrections to get any
reasonable answers. However a precise analysis of these higher
derivative terms has not been worked out in the literature, and
therefore the solutions that we presented in sec 3.1 to sec 3.3 are only
approximate. Furthermore these solutions are only valid in the regimes
discussed for various individual cases respectively.

In the absence of an exact solution, we would then like to
classify the possible background metrics that we can get using two
and three warp factors. The metric that we get in type IIB is of
the following generic form: \eqn\metwewant{ds^2 = {f_1\o
t^\alpha}(-dt^2 + dx_1^2 + dx_2^2) + {f_2\o t^\beta}~dx_3^2 +
{f_3\o t^\gamma}~g_{mn}~dy^m dy^n} where $f_i = f_i(y)$ are some
functions of the fourfold coordinates and $\alpha, \beta$ and
$\gamma$ could be positive or negative number. For arbitrary
$f_i(y)$ and arbitrary powers of $t$, the type IIB metric can in
general come from an M-theory metric of the form \netvix\ with
three different warp factors $A,B$ and $C$, given by:
\eqn\waabcc{\eqalign{& A = {1\o 2}~{\rm log}~{f_1f_2^{1\o 3}\o
t^{\alpha + {\beta\o 3}}} + {1\o 3}~{\rm log}~{\tau_2 \o \vert
\tau \vert^2}, ~~~~~~ B = {1\o 2}~{\rm log}~{f_3f_2^{1\o 3}\o
t^{\gamma + {\beta\o 3}}} + {1\o 3}~{\rm log}~{\tau_2 \o \vert
\tau \vert^2} \cr & ~~~~~~~~~~~~~~~~~~~~~~ C = -{1\o 3}\left[{\rm
log}~{f_2\o t^\beta} + {\rm log}~{\tau_2^2 \o \vert
\tau\vert}\right]}} where we have isolated the complex structure
dependences, and $\tau = \tau_1 + i \tau_2$ as before. The above
is therefore what we talked about in the beginning of Sec.~3, but
clearly there are many cases with only two warp factors. The two
warp factor cases would be when $B = C$, and $A$ different.  To
see what the possible choices are for such a background, we need
to find the difference $B-C$. This is given by: \eqn\bmc{B - C = {1\o
2}~{\rm log}~{f_2 f_3\o t^{\gamma + \beta}} + {\rm log}~{\tau_2 \o
\vert \tau \vert},} which could be simplified further if we
consider a background with trivial axion-dilaton i.e $\tau =i$
with the seven-branes charges cancelled locally\foot{This is when
the M-theory fiber is a $\IZ_2$ orbifold of $T^2$.}.  From here,
the two warp factor case would be when \bmc\ vanishes. Since the
space and time dependent parts of \bmc\ can be isolated, \bmc\ can
only vanish if \eqn\bmcvanish{f_2 = f_3^{-1}\cdot {\vert \tau
\vert \o \tau_2}, ~~~~~ \gamma + \beta =0} with $\alpha$ and
$f_1(y)$ remaining completely arbitrary. To simplify things a
little bit, let us also consider $\tau =i$ which means that we are keeping the seven branes
far away and also taking the limit when the dilaton is zero (see sec. 3 for the precise equations for which
such a situation could be realised). 
Then the condition
\bmcvanish\ would fall into the following cases: \vskip.1in

\noindent {\bf Case 1}: {$\gamma = \beta = \alpha = 0$, ~~$f_2 = f_3^{-1}$}

\vskip.1in

\noindent This is completely independent of time, and therefore the warp
factors are functions of the fourfold coordinates. We now expect that
the warp factors $A$ and $B$ would be related as $A = -2B$, as we have
seen in Sec.~2. Using the general forms of $A$ and $B$ in \waabcc, $A=
-2B$ would immediately imply $f_1 = f_2$.  This is therefore the
supersymmetric case studied earlier in various papers including
\sav,\gkp,\shiu. The M-theory lift of this will have the following
warp factors:
\eqn\mwar{A = {2\o 3}~{\rm log}~f_1, ~~~~~~~ B = C = -{1\o 3}~{\rm log}~f_1,}
which is of course the fourfold solution of \rBB.
\vskip.1in

\noindent {\bf Case 2}: {$\gamma = \beta = 0$, ~~$\alpha \ne 0$, ~~$f_2 = f_3^{-1}$}

\vskip.1in

\noindent There could be various choices here when $\alpha \ne 0$.
All of them are time dependent cases and therefore break
supersymmetry. The case where $\gamma = \beta = 0, \alpha = \pm 2, f_1
= f_2 = f_3^{-1}$ is our Examples 1 and 2 earlier (in Sec.~(3.1) and
Sec.~(3.2)). As discussed there for the case when $\alpha = 2$, this
is an approximate background when we take the background $G_{mnpq}$
fluxes to be very small on an almost non-compact fourfold. For the
case when $\alpha = -2$, we were not required to take fluxes very
small but solutions there only existed {\it iff}\ higher derivative
corrections are put in. In both cases the membrane is moving very
slowly and the warp factors in M-theory are (henceforth we will keep
$t_o=0$ for simplicity)
\eqn\watwo{A = {2\o 3}~{\rm log}~f_1\pm {\rm log}~t, ~~~~~~~ B = C = -{1\o 3}~{\rm log}~f_1.} 
The next
corresponding case would be when $\alpha \ne \pm 2$ with $f_1$ may
or may not be equal to $f_2$. With the analysis done in Sec.~3,
this condition seem not to be allowed to the order that we took
the quantum corrections. It could be that these cases are realized
when we certain terms in the background equations of motion are
ignored. It will be interesting to study these possibilities.
\vskip.1in

\noindent {\bf Case 3}: {$\gamma = -\beta$, ~~~~$f_2 = f_3^{-1}$}

\vskip.1in

\noindent Again there are a few choices here. The case studied in Example 3
earlier (Sec.~(3.3)) is when 
$\alpha = \beta = -\gamma = 1$, ~$f_1 = f_2= f_3^{-1}$.
For this case unfortunately the internal G-fluxes $G_{mnpq}$ become
time dependent. Therefore this case seems to be a viable solution only
when we remove the branes from our picture. Thus this case is
basically outside the D3/D7 inflationary setup, unless we incorporate
corrections to the anomaly equation \warpff\ which would allow such
kind of fluxes.  One interesting case to consider would be to find a
de-Sitter like solution here. From our earlier analysis \reasf, we
know that type IIB de-Sitter can be gotten from dimensional reduction
over a six dimensional space, if the warp factors satisfy:
\eqn\wasdfg{ \beta + 3\gamma = 2~~~~~~~ \Longrightarrow ~~~~~~~ \alpha = \beta = -1, ~~ \gamma = 1}
which is related to what we got in \metabvx\ by the replacement 
$t \to {1\o t}$ in \metabvx.
Therefore for the two cases we can consider fourfolds
in M-theory with warp factors given by:
\eqn\wathree{A = {2\o 3}~\left[{\rm log}~f_1 \pm {\rm log}~t\right], ~~~~~~~ B = C = -{1\o 3}~\left[{\rm log}~f_1 \pm {\rm log}~t\right]}
with the $+$ sign giving a de-Sitter solution  
and the $-$ sign giving the cosmology studied 
in Sec.~(3.3). In the latter case, the fourfold is effectively
non-compact. Similarly there are other interesting cases which could
possibly be seen under some restricted choices of flux backgrounds.

These are therefore all the possible allowed choices with two warp
factors. We now would like to come to the case with three warp
factors. For generic metric the warp factor choices are already given
in \waabcc. Of course the three warp factor cases are much more
flexible, but on the other hand the background equations of motion are
pretty involved now. We will perform a more systematic study, along
with possible derivations, of this in a sequel to this paper\foot{Of
course we should also consider the cases with four (or maybe more) warp
factors too. Having four warp factors would mean that the type
IIB six manifold is by itself a fibered product $-$ with the fiber (of
some dimension) warped differently than the base. Since not all
manifolds can have this inherent property, it will be interesting to
analyse this case in details to see what possible cosmology comes out
from here.}.  For the time being let us analyse all the possible
choices within this scenario.

\vskip.1in

\noindent {\bf Case 4}: {$\alpha = \beta = 2$, ~~$\gamma = 0$ ~~$f_1 = f_2$}

\vskip.1in

\noindent This is by far the most interesting case. The internal six manifold
is time independent. From our earlier analysis \reasf, this example
would correspond to an exact de-Sitter background.  And therefore this
would be an accelerating universe with the three warp factors given by:
\eqn\thwafaar{A = {2\o 3}~{\rm log}~{f_1\o t^2}, ~~~~~~ B = {1\o 2} \left[{\rm log}~f_3 + {1\o 3}~{\rm log}~{f_1\o t^2}\right], ~~~~~~
C = -{1\o 3}~{\rm log}~{f_1\o t^2}.}
We see that the internal fourfold has time dependent warp factors
although the type IIB six dimensional space is completely time
independent.  Such a background has the advantage that the four
dimensional dynamics that would depend on the internal space will now
become time independent.
\vskip.1in

\noindent {\bf Case 5}: {$\alpha = \beta$, ~~$\gamma \ne 0$ ~~$f_1 = f_2$}

\vskip.1in

\noindent Although interesting, case 4 is of course not the most
generic way to get a 3+1 dimensional de Sitter space. If we remove
the restriction that the internal space should be time
independent, then there are many other choices. From \reasf\ these
cases can be classified as: \eqn\desitc{ \alpha = \beta = \kappa,
~~~~~ \gamma = {2-\kappa \o 3}, ~~~~~ f_1 = f_2,} where $\kappa$
could be any number. We have also put no restriction on $f_3$,
although one might want to keep $f_3 = f_2^{-1}$ to comply with
the supersymmetric case. All these cases would give rise to
accelerating universes of course, with warp factors \eqn\laur{A =
{2\o 3}\left[{\rm log}~f_1 - \kappa ~{\rm log}~t\right], ~~~ B =
{1\o 6}~\left[{\rm log}~(f_1f^3_3) -2~{\rm log}~t\right], ~~~ C =
-{1\o 3}~\left[{\rm log}~f_1- \kappa~{\rm log}~t\right].} Other
cases with $\gamma$ not equal to the one above \desitc, therefore
will not be de-Sitter, but could be either accelerating or
decelerating universes. Furthermore, many of the earlier cases we
studied with two warp factors may also be studied with three warp
factors if we do not put any restrictions of $f_1, f_2$ and $f_3$.
This completes the classification.

Let us now briefly turn to the issue of moduli stabilization. This is
a bit subtle when the internal manifold becomes time dependent, as
sometimes the internal fourfold could have time-dependent warp factors
but the internal threefold in type IIB has time independent warp
factors.  The various cases can be classified as follows:

\noindent $\bullet$   The internal   fourfold in M-theory and internal
threefold in type IIB both have time-independent warp factors.

\noindent $\bullet$   The internal   fourfold has time-dependent warp
factors whereas internal threefold in type IIB has time-independent
warp factors.

\noindent $\bullet$ Both the internal fourfold in M-theory and the internal
threefold in type IIB have time-dependent warp factors.

\noindent The first case of course incorporates the supersymmetric
background where, unfortunately, there are still no models with all
moduli stabilized. Therefore the non-supersymmetric cases will
  definitely    be much more complicated to handle. This problem becomes
even more severe if the internal G-fluxes $G_{mnpq}$ in M-theory
are time dependent. Fortunately, some of the toy cosmologies that
we studied in sec 3.3 have G-fluxes that are time independent.
These solutions are not de-Sitter, but in case 4 above we gave an
example where, with three warp factors, it { might be}    possible
to realize a de-Sitter background.    Although a full analysis of
this background is beyond the scope of this paper, we will assume
that such a background may also allow a time independent
$G_{mnpq}$ flux on the fourfold.

Now the situation is much more tractable. The $G_{mnpq}$ fluxes
will induce a superpotential \gvw\ that will fix all the complex
structure moduli of the fourfold. Other K\"ahler structure moduli
may be fixed by non-perturbative effects. Some discussion of this
has appeared in \berghaack,\burt. From type IIB point of view, these
non-perturbative effects are related to the gaugino-condensate on
the D7-branes\foot{It is not too difficult to identify the gaugino
condensate terms from M-theory. In M-theory the singularities of
the fiber are related to the positions of the seven-branes (not
all are D7 though). On the other hand, G-fluxes can be decomposed
into   a {\it localized} and a {\it de-localized} part. The
de-localised part eventually becomes the $H_{NS}$ and $H_{RR}$ in
type IIB theory \sav, \bbdg, whereas   the localized part becomes
the seven-brane gauge fields \sav,\bbdg. In M-theory, the
localized G-fluxes couple with the supergravity fermions.
Expectation values of these fermion terms give rise to gaugino
condensates on type IIB D7-branes creating a superpotential. More
details on this will be presented elsewhere.} {or Euclidean
D3-brane instantons}.
 The original KKLT construction \kklt\ of de-Sitter vacua uses this
mechanism to fix the K\"ahler moduli, and an anti-D3-brane to break
supersymmetry. For our case we broke supersymmetry {\it
spontaneously} using non-primitive fluxes.
The second case could probably be dealt with  from the  type IIB point of view,
where the six manifold has a  time independent warp factor. However it
is not clear whether the fluxes could always be made time
independent. This then leads us to the question of moduli
stabilization when the fluxes themselves are varying wrt time. One
possible explanation could be that at a given time $t = t_0$, one  fixes
the fourfold moduli using a superpotential. Then, as the fluxes vary
 with respect to    time, the moduli are allowed to vary accordingly. Again, no
analysis has yet been performed to this effect that a concrete
statement could be made. We will therefore leave this discussion
here and continue  it in the sequel.

The third case is intractable so far, as the time dependences of
the internal manifolds in M-theory as well as in type IIB theory
make it hard to give a physical picture {regarding modulus
stabilization}. As mentioned in the Introduction, the transient
accelerations \`{a} la \town\ are too short to be used for
slow-roll inflation.

\newsec{Primordial black holes in cosmological backgrounds}

We now come to another interesting cosmological phenomenon: the
formation of primordial black holes in the   early universe.  As studied
in the literature, there could be various possibilities. Black
holes may form from the large density perturbations \carr,
\crawford\ induced near the end of the inflationary era.  This is
basically one of the key mechanisms of black hole formation. We
will soon show how the the $D3/D7$ model may realize some aspects
of this.  In a parallel scenario the formation of black holes
 starts   when the mass of the black hole is equal to the mass of the
horizon.

But there are other mechanisms. In some inflationary models that are
proposed in the literature, inflation ends by bubble nucleation. This
is a phase transition $-$ a strong first order phase transition that
proceeds explosively by the bubble nucleation. Due to this phase
transition primordial black holes can easily form.  Details of this
mechanism have appeared in \liddle.

Another mechanism is related to the thermal fluctuation during the
very early universe. The productivity rate of the black holes is
particularly efficient at very high temperature. A discussion of this
has appeared in \gross.

These black holes formed in the early universe (via any one of the
mechanisms listed above) are usually   light enough so that Hawking evaporation
can take place.  The lifetime $T$ of such a black hole of mass $m$ is
given by the celebrated formula of Hawking \hawrad
\eqn\lifetime{T ~ = ~ {m^3 \o M_p^4} \cdot {1 \o N}}
where $M_p = 1.22 \times 10^{19} ~{\rm GeV}$ is the Planck  mass,   
and $N$ denotes the number of particle degrees of freedom into
which the black hole can decay. Using the above relation, a simple
calculation tells us    that a black hole of mass $m \approx
10^{12} ~{\rm kg}$ will have a lifetime equal to the age of our
present universe (i.e $10^{10}$~years). These black holes
therefore will now be evaporating. Lighter black holes would have
evaporated long ago. What about heavier black holes?  These black
holes would still be present and therefore would contribute to the
energy density of the universe. Recall that if black holes form
early enough,   they can easily dominate the energy density of
the universe irrespective of whether they inhabited a very tiny
fraction or not at the beginning. Thus those heavier black holes
should not have energy density more than the critical density of
the universe. This is an important constraint. An interesting
possibility here is that, if the energy density of these black
holes is greater than the critical density, they might be a
candidate for cold dark matter. The black holes that are
evaporating today are also constrained by the critical energy
density. In fact studies of the $\gamma$ ray backgrounds \carr,
\carrtwo\ limits black holes in this mass range to contribute at
most {some}   orders of magnitude below the critical density. The
constraint can get a little weakened if one assumes the formation
of relics \liddletwo\foot{As a parallel comment, note that
classical black holes induced by metric perturbations, which were
in turn generated during the water-fall stage of the hybrid
inflation, can be sufficiently abundant $-$ since the inflation is
nearing its final stage $-$ to be dark matter candidate. An
analysis of this was done in \pisin\ following a {\it generalized
uncertainty principle} that can be inspired from string theory. It
was shown therein that these primordial black holes (which are
still extremely small, like $10^{10}$ times the Planck mass) can
stop their Hawking evaporation when they reach the Planck size.
These will then contribute to the dark matter. Based on hybrid
inflation, such primordial black holes remnants can indeed
saturate the dark matter $\Omega_m$. For more details the readers
may want to see \pisin.}.

From the discussion above one might wonder if there could be any
constraints on those black holes that have evaporated away
completely, i.e those black holes that have masses much below
$10^{12} ~{\rm kg}$. It turns out that these black holes are
strongly constrained \greenliddle, \glid. The constraint comes
from the fact that for black holes of masses $m \approx 10^7~{\rm
kg}$ would be evaporating during nucleosynthesis. Since
nucleosynthesis is an important part of the formation of the
present universe, an overproduction of these light black holes
would interfere with the process. Therefore these black holes are
also strongly constrained.  For further details the readers may
want to see the analysis given in table I and II of \greenliddle.

Before we end this section we should discuss the recent extension
to the standard cosmological scenario: the so called thermal
inflation \lyth. This is a second period of inflation which is
triggered by supersymmetric fields that have an almost flat
potential. In fact this second period has nothing to do with the
usual inflationary regime that solves the horizon, flatness and
the monopole problems. However, similar to the usual inflation,
the scalar fields that drive the thermal inflation also have a
false vacuum energy. Once the temperature drops to a certain
value, the false vacuum energy drives the new inflation till the
temperature drops further down  and   takes the scalar field out of
the false vacuum.

Identifying the scalar fields (that drive the new thermal inflation) as the
moduli fields of string theory (i.e the scalars that would come from the
non-trivial Betti numbers of the internal manifold) would in fact lead to
the famous moduli problem (also known as the Polonyi problem)
\moduli: namely, their expectation values of order $M_p$
would create many particles that remain till the nucleosynthesis and
therefore would be inconsistent with cosmological predictions. Many
solutions to this problem have been proposed \lisa. All of them are
more or less based on the fact that the thermal inflation, which is
basically a period of late inflation with Hubble expansion parameter
of the order of the weak scale, can effectively solve this problem
(see the second reference of \lisa\ for more details). The dilution
effect of the thermal inflation should also be small enough so that it
doesn't effect the density perturbations generated by the original
inflationary period.

The question now is to see the effect of the second stage of inflation
on the primordial black holes. It turns out that the main effect of
thermal inflation is to dilute the density of the primordial black
holes. Furthermore, during the two periods of inflation (i.e the
original one at the beginning and the later, thermal, one) a given
primordial black hole production corresponds to a {later} stage in the
original density perturbation generating epoch of inflation
\greenliddle. 
  In addition to that the thermal inflation introduces
missing mass ranges for the black holes by allowing those co-moving
scales to be pulled outside by thermal inflation that had originally
entered {\it before} thermal inflation had set in. 
Also since the
energy scale of the thermal inflation is much lower than the original
inflationary period, any new density perturbations are expected to be
small. This would mean that new black holes are not formed, after
re-entering the horizon again (see \carr,\greenliddle\ for more
details).

Having summarized some of the details related to the formation of
black holes in the early universe, it is now time to continue our
discussion of these issues in the hybrid inflationary model. This
will help us connect the picture with the wrapped $D3$ brane
intuition that we described earlier.

\subsec{Primordial black holes in D3/D7 system?}

To use the ideas  we developed earlier to study black holes in our
set-up, we have to give some reasons for the formation of
brane-antibrane pairs in this scenario. A
set of arguments that    motivates    to use the machinery that we
developed in sec 3 is as follows:

\noindent $\bullet$ As the D3-brane is moving towards the D7-branes,
there is a perturbation in the background due to the time dependent
evolution of background fields (recall that we have both $H_{NS}$ as
well as $H_{RR}$ fields that are evolving with time). If these
fields are fluctuating fast enough, then the time dependent
perturbations will be responsible in creating brane-antibrane pairs.

\noindent $\bullet$ As the D3-brane  approaches   the D7-branes,
the open strings stretched between them become tachyonic. These
tachyonic instabilities will signal a phase  transition.   Condensation of
these tachyons pumps energy into the system resulting in copious
production of brane-antibranes. From the dynamics of the system, one
would expect this to dominate over the first one.

From either of the two cases, once we assume  the existence of
brane-antibrane pairs,   the rest of the discussion should follow as given
earlier.  In fact we studied this kind of background in sec 3 and sec
4 (as case 4) above, where $\alpha = \beta =2, \gamma =0$ and $f_1 =
f_2$. Now if we replace $g_{mn}$ in \metwewant\ by the metric \finmet,
then this will be a precise background that allows us to study
primordial black holes in the inflationary scenario.   Thus combining,
\finmet, \metwewant\ and \thwafaar,   our background will take the
following simple form:
\eqn\priminfla{\eqalign{ & ~ ds^2 = ~ {f_1 \o t^2} (-dt^2 + dx_1^2 + dx_2^2 + dx_3^2) + f_3\Big[h_1~(dz + a_1 dx + a_2 dy)^2 + h_2~(dy^2 + d\theta_2^2)~ + \cr
&  h_6~dr^2 + h_4~(dx^2 + h_3~d\theta_1^2)
+ h_5~{\rm sin}~\psi~(dx~d\theta_2 + dy~d\theta_1) + h_5~{\rm cos}~\psi~(d\theta_1d\theta_2 - dx~dy)\Big]}}
where $h_i$ are given in \hideterm\ and the background threeform by \threeform. As we see
both the internal space and the background fluxes are time
independent.   The fluxes are non-primitive and they break supersymmetry.
The above
metric therefore captures the complete scenario wherein primordial
black holes are created in the D3/D7 inflationary  model.   This is
  expressed by the following figure:
\vskip.1in

\centerline{\epsfbox{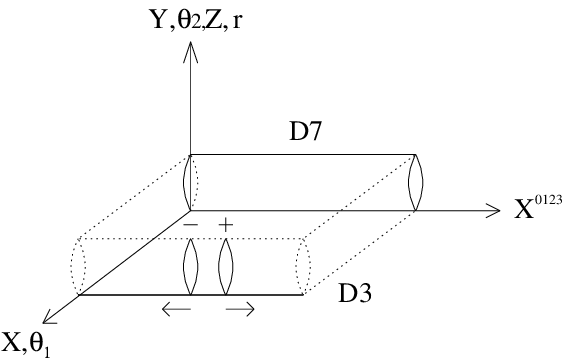}}\nobreak

\vskip.1in \noindent In this figure, the compact space is along ($x,
\theta_1, y, \theta_2, z, r$), with three-cycles denoted by
coordinates ($y, \theta_2, z$). The seven-branes therefore wrap the
directions ($y, \theta_2, z, r$) and are stretched along de-Sitter
spacetime $x^{0,1,2,3}$. We denote the surface of the D7-branes by a
cylinder, whose compact directions are along ($y, \theta_2, z$) and
also $r$. The world volume of D3 is along $x^{0,1,2,3}$. Both the
D7-branes and the D3-brane are points on ($x, \theta_1$) space, and
the D3-brane is slowly moving towards D7-branes with a velocity
$\dot y^m$ (that we determined earlier for various cases). As the D3
moves, there would be creation of small number of $D3/{\bar D}3$
pairs. These pairs are quickly separated by the   inflationary expansion,   as
depicted in the figure above. Finally,   when the D3-branes comes very
close to D7-branes, condensation of D3-D7 tachyons pumps energy
into the system creating more  $D3/{\bar D}3$
pairs. 

Of course,   this is not the only mechanism working here. Instead of the 
inflationary  expansion   separating D3-branes from antibranes, we could
have D3-branes created directly from D5 brane-antibrane pairs  by the
background $B_{NS}$ fields.  The production   rates for this are expected to
be bigger than for 
 the D3 pair productions, as   we discussed before\foot{The
$D5/{\bar D5}$ pairs do see some $B_{NS}$-fluxes, so that pair
production for them is not tied to tachyonic modes near the D7. And
there will be some probability for the D3 defect to be produced when
the D5s annihilate $-$ note that the dominant decay mode (i.e the
case when $F_2 \approx F_1$) for the D5s is still into just
radiation with no lower dimensional brane remnants. Thus this rate
will be smaller although nonzero. Therefore the production rates   of black
holes from the $D5/{\bar D5}$ annihilations are  expected to be higher
  than the ones
from separating the $D3/{\bar D3}$ pairs by  the inflationary expansion.   }.
From either of these two mechanisms, black holes in 3+1 dimensional
de-Sitter space will appear as D3-branes wrapping a three cycle ($y,
\theta_2, z$) with the intrinsic metric \eqn\volume{g =
\pmatrix{h_1+h_2& 0 & a_2 h_1\cr 0&h_2 & 0 \cr h_1a_2 & 0 & h_1}}
and with masses determined by the volume of the wrapped cycle. One
can now estimate the productivity rates of     black hole formation 
using two different techniques: (1) Evaluate the productivity rates
from an effective 3+1 dimensional theory by dimensional reduction
over the threefold that we discussed above, or (2) Evaluate the
productivity rates of brane-antibrane in a fluctuating background
directly from ten dimensional type IIB theory. Both of these
techniques should yield the same answer. However here we will
briefly sketch the first mechanism which is by now a standard way to
derive the productivity rates of charged black holes in any
dimensions. For more details the reader          is referred to        
\hawking,\dowker.

The analysis of \hawking,\dowker\  starts   by identifying {\it two}
different instanton actions $I_{\rm dS}$ and $I_{\rm BH}$ that are
responsible for the  production   rates:

\noindent $\bullet$ $I_{\rm dS}$ is the Euclidean action of four-dimensional
gravitational instanton that mediates the nucleation of a de-Sitter space
from nothing. In terms of our language, this would be the creation of the
four-dimensional part of the D3/D7 metric \priminfla\ via an instanton
from nothing. This four-dimensional part is of course an effective
theory when we reduce type IIB over the  non-Calabi-Yau                  background.

\noindent $\bullet$ $I_{\rm BH}$ is the Euclidean action of the gravitational
instanton that mediates the pair creation of charged black holes in
four-dimensions from nothing. Again, in terms of the analysis that we did
earlier, this would now be the dimensional reduction of the wrapped
$Dp/{\bar D}p$ case with $p=3,5$ in the D3/D7 setup.

\noindent Thus, in the language of \hawking, we have two different
instantons: one for the background and the other for the objects in this
background. The productivity rate $\Gamma$ is now given by
\eqn\prodrate{\Gamma = \eta ~e^{-2(I_{\rm BH} - I_{\rm dS})}}
where $\eta$ is the one-loop contribution from the quantum quadratic
fluctuations in the background fields \gins, \gross. A precise value of
$\eta$ is not very relevant for us. What we require are the values of the
instanton actions.

The instanton actions have been calculated in many papers (see for
example \hawking\ and citations therein) and for black holes of different
charges. Our
effective theory in four dimension will be the same except that we
now expect many scalars from the moduli of the theory. The potential
for these scalars will come from either \gvw\ (in M-theory) or
\tv,\gkp\ (in type IIB theory) or \bbdg,\bbdp\ (in heterotic
theory), including non-perturbative corrections. Assuming that the
potentials fix these scalars to some expectation values, we can
integrate them out to get our instanton actions. The final
productivity rates from all these     analyses   turn out to have a
universal form, and is given by: \eqn\finpro{\Gamma ~ \sim ~ e^{-{m
\o \sqrt{\Lambda}}}, ~~~~~~ m = {T\o g_s}\int dz dy
d\theta_2~\sqrt{h_1h_2(h_1 + h_2-a_2^2 h_1)}~f_3^3} where $m$ is the
mass of the black holes coming from the wrapped D3-branes with
tension $T$, $h_i$ are given in \hideterm\ and $g_s$ is the type IIB
coupling constant (i.e the value of the dilaton that comes from the
non-trivial complex structure $\tau$ of the fiber torus \axdil\ in
M-theory). Finally, $\Lambda$ is the cosmological constant that
would appear from the inherent non-commutativity on the D7-branes
\dtds\ (see also \kalth\ for a more updated discussion on this).

Although a full numerical analysis of productivity rates will now require
the values of the warp factors $f_1$ and $f_3$ $-$ which can only be
determined once we know the metric using three warp factors $-$ a
qualitative analysis (by evaluating \finpro\ with known values of $h_i$ \hideterm\ and 
$f_1, f_3$ satisfying $f^{-1}_1 = f_3 \approx \sqrt{j}$ as in \restonhf) 
  easily tells us that $\Gamma$ is very small here.
This is of course consistent with the inflationary scenario.

But this is not the full story yet. As the D3-brane moves towards
D7-branes the productivity rates $-$ which are initially given by
given by $\Gamma$ above $-$ will get further enhanced. This is
because, when the D3 finally reaches D7 the open string tachyon
between D3 and D7 condenses pumping energy to the system,    thereby 
enhancing the productivity rates of brane-antibranes.
This will continue till the moving D3-brane finally dissolves into
the D7-branes as a non-commutative instanton. We will not perform a
detailed analysis of $\Gamma$ here, but a qualitative analysis will
tell us that the productivity rates are still very small, because the last stage 
of the inflationary process where the tachyon condenses and the D3 is finally absorbed in the 
D7 branes is very short\foot{In fact the interval $\delta t$ can be estimated for our case once we know the 
D3 brane velocity $\dot y$ at the last stage when it is near the D7 branes, as $\delta t \approx {d_c\o \dot y}$. Here $d_c$  
is the critical distance at which an open string tachyon is formed and is given by \dtds\  $d_c \equiv {1\o 2} \pi \alpha' B_{y\theta_2}$ 
where $B_{y\theta_2}$ is the $B_{NS}$ field from \threeform\ with other components vanishing. Since we expect 
$\dot y$ to be large (its not slow roll anymore) and $d_c$ of order string scale (from above); the absorption process is almost instantaneous
resulting in a very small productivity rate of the blackholes from this process also.}.  
On the other hand, we saw earlier that the D3/D7 system on
$K3 \times \IP^1$ has zero productivity rates. Therefore with a non
Calabi-Yau threefold as an internal space and
$b_3 \ge 1$, there is a finite (albeit very small) probability of
detecting charged black holes in four-dimensional spacetime.

\newsec{Semilocal defects with higher global symmetries in $D3/D7$ model}

Having studied various cosmological solutions in the D3/D7 system, let us now
turn our attention to something a little different from our earlier
constructions, namely   the semilocal defects that can 
 arise in some regimes of the 
D3/D7 model.
In our earlier paper \semil\ we addressed the issue of
semilocal strings. Semilocal strings were discovered by     Ach\`{u}carro      and
Vachaspati in the   early nineties \anaV. These strings defy all the standard lore
of the stability and existence of solitons. In fact they are {\it
non-topological} solitons and their existence can occur   in theories that
have
both global and local symmetries.

The   standard analysis of the classification of solitons is   based on the
topology of the moduli space and the corresponding homotopy groups. In the
presence of global symmetries (shared by, say, only the moduli),  the above
naive 
classification  can fail.   This is where certain 
  non-topological objects, so called
semi-local defects, can surface. These defects could be made stable by
restricting the Higgs masses to lie between some range of energies. For one
dimensional defects $-$ the semi-local strings $-$ it was shown by Hindmarsh
\hindmarsh\ that when the Higgs mass is smaller than that of 
  the gauge bosons, the
semi-local
strings are absolutely stable despite the fact that the naive 
  vacuum manifold has a
trivial first homotopy group. In fact for the theory having a symmetry group
$G$ at {\it vanishing} gauge coupling,
the homotopy classification for semi-local defects can sometime be done if we
consider a much smaller subgroup $G_s$ that is in principle gauged when we
switch on
a non-trivial coupling (or alternatively switch on the gauge fields).
The classification then takes a somewhat familiar form:
\eqn\homclass{ \pi_n \left({G_s\o H_s}\right) ~ = ~ \IZ, ~~~~~~~
\pi_n\left({G\o H}\right) ~ = ~ 1}
for a given $n$ where $n$ determines the kind of defects one would see in a
given model and
$H_s$ is the unbroken subgroup of $G_s$.
Using this one can show that, although the
vacuum manifold may have trivial homotopy, the theory could still have stable
defects. A more detailed analysis has been  carried out 
in \hindmarsh, \preskill\  to   which the
reader is referred for further details. 

It turns out that the D3/D7 model that we studied here is a   good
place to realize these defects as non-topological solitons of the
theory\foot{Semilocal    vortices    
based on unitary symmetry groups have been constructed  in \tong\ using Hanany-Witten type brane configurations \hw\  in
 type IIA string theory.}, 
although   the regime    we shall use to study these
solitons may not necessarily be the inflationary one. This is
because of the existence of global symmetries, which  generically require 
many scalars that would typically ruin the inflationary nature of
the model. Furthermore there are few more points that we have to
take care of before we can   embed non-topological defects in our
theory:

\noindent $\bullet$ The dynamics of the seven branes should be isolated from
the dynamics of  the   D3 branes. As discussed in \semil\ for the case when the
internal six manifold is of the form $K3 \times \IP^1$, the seven branes wrap
the K3 manifold which can be fixed to a large size. For our case the
seven manifold wrap a four-cycle of the metric \finmet\ oriented along
($z,y, \theta_2, r$),   whose unwarped metric can be written as
\eqn\metunw{ g_{mn} = \pmatrix{h_1&a_2 h_1& 0 & 0\cr \noalign{\vskip -0.20 cm}
\cr a_2h_1&h_2 + a_2^2 h_1& 0 & 0\cr
\noalign{\vskip -0.20 cm}  \cr   0&0 & h_2 & 0\cr \noalign{\vskip -0.20 cm}
\cr  0 &0& 0 & h_6}}
where $h_i$ and $a_2$ are defined   in \hideterm. Thus starting with
the seven brane coupling $g_7^2$, the effective three brane coupling can be
easily deduced to be
\eqn\thbrco{{1\o g_3^2} \equiv {1\o g_7^2} \int_{\Sigma_4}~e^{4B}h_2 \sqrt{h_1
h_6}}
where $\Sigma_4$ would be the four-cycle with the metric $g_{mn}$ \metunw\ on
which  the  seven-branes are wrapped.  Using arguments of moduli stabilisation,
(as we would require to fix the size of the internal manifold against any runaway) this
effective coupling can be made very small and therefore the seven branes would
effective decouple from the three brane dynamics.

\noindent $\bullet$ The fourfold   we use to describe our new     six-manifold 
should have a representation in terms of a Weierstrass equation.
In fact,   this is exactly what we achieved when we derived our background in
example 5 of sec 3. As mentioned therein, the Weierstrass equation is
written as $y^2  = x^3 +  x f + g$ with $f \oplus g$ being sections of
  ${\cal L}^4 \oplus {\cal L}^6$ with ${\cal L}$ being
${\cal K}_{{\cal B}_\mu}^{-1}$, and   ${\cal B}_\mu$ as  in   \defconi. Thus
the fiber would have to degenerate over a sub-space of our six manifold (it
could
be the full six manifold, but then the interpretation of the singularities in
terms of type IIB branes becomes complicated).

The next question therefore would be to specify  this   subspace,
along with its possible metric. This is not difficult. We want the
seven branes to be points     in   the ($x, \theta_1$) directions. The
metric along these directions can be worked out from \finmet. The
result is a warped torus with the following metric:
\eqn\wartor{ds^2 = {e^{2B}\o C}~\vert dz \vert^2 ~ = ~ {e^{2B} \o
C}~\left[dx^2 + (C-\beta_1^2 E^2)C~d\theta_1^2\right],} where all
the quantities appearing above are defined in \hideterm, and the
complex structure of the torus can be easily inferred to be $\tau
= i~\sqrt{(C-\beta_1^2 E^2)C}$.

{}From the above metric \wartor\ it is now possible to see how the seven branes
are placed. Observe that the warp factor $e^{2B}$ will contain the information
of the
   seven-branes,   as the other warp factors are known in terms of the variables ($E$
and $F$) before the   geometric transition. From  the   M-theory view point,
  the
warp factors satisfy an equation of the form \warpff\ with the condition \abrela\
(for the two warp factor case). Of course the kind of metric that we want can
only be derived from three warp factors by choosing case 4 of sec 4, and
therefore the warp factors would change suitably. On the other hand we could
also study this
from type IIB point of view. An analysis of warp factors (for the
supersymmetric case)   has   been done in \shiu. The result is presented in terms
of          Green's     functions that pick up singularities at the points where we expect
seven branes and three branes. We will not go into the details of those
calculations
because our system is a little more complicated due to  the 
   non-primitivity of      the 
background fluxes, instead we will try to estimate the possible values of
$e^{2B}$
by going to   a regime where we could apply the exact result of \greeneyau.

The regime that we are interested in will be when the complex
structure of the base is $\tau \approx i$, and the fluxes  are 
small there (so that the warp factor equation is solely affected
by the brane singularities). In terms of the variables used to
define the metric \wartor, this is possible if we restrict
ourselves to the region: 
\eqn\region{{1\o \alpha_0}~\left(C - {1\o
C}\right)\left(D - {1\o D}\right) + C^2 - \alpha_0 D E^2 e^{2\phi}
~ = ~ 1.} This regime, as mentioned earlier, is non-inflationary
because the fluxes are effectively vanishing and therefore the
membrane (or the D3 brane) is almost stationary \tempom\foot{It is also interesting to note that
in this region \region\ the two backgrounds that we studied i.e the {\it deformed} conifold of sec. 3.5 and the 
$K3 \times \IP^1$ background of \dtds\ would look effectively similar, and therefore would have an approximate 
${\cal N} = 2$ supersymmetry. Globally the two backgrounds are quite different of course as we saw earlier.}. 
We need not         
restrict ourselves to this regime, and can study the issue of
semi-local defects in the full inflationary set-up as was done
earlier in \semil. The point however is that, as we increase the
global symmetries, the number of scalars increases and therefore
usual inflationary dynamics may not be easy to realise here.
For simplicity, we will
restrict ourselves to a region on or near \region. The warp factor
$B(y)$ now has the following representation: \eqn\wabba{ B(y) ~ =
~ b(y) -{1\o 12}~ \sum_{i=1}^n~{\rm log}~(z - z_i) + {\rm c.c} +
{\rm modular~ functions}} where $b(y)$ would be some as yet
undetermined function of the internal space, and we have put in
modular functions to have a consistent description of warp factors
on a torus. Finally,  the integer $n$ determines the number of seven
branes,   with $z_i$    being         the position of seven branes on the ($x,
\theta_1$) torus.

The $z_i$ behavior of the warp factor $B$ will not change even
when we switch on non-primitive fluxes, although the function
$b(y)$ might get more complicated. Thus going to a slightly
different regime of our model, we are able to infer the $y^m$
dependences of the warp factor $B$. The other warp factor $A(y)$
still remains undetermined (at least using this technique),
although if there is some special relation between $A$ and $B$,   we
might expect to get some information about $A$ also.

\subsec{Examples of semilocal defects}

\noindent Having set-up the stage to study semilocal defects, we
should now venture towards concrete examples. The semilocal
defects are determined   with respect to   global as well as local symmetries.
The global symmetries are related to the singularities of our
manifold \tate,\bikmsv,\vwc. Various singularity types would
determine what global symmetries we should expect in this set-up.

The local symmetries are determined by the number of D3 branes (or M2 branes in
M-theory) that we would put in. Once we determine both the local and
global symmetries, we look for energetics of the system to see whether we
should expect semilocal defects or not. The energetics can sometime be
specified
by the ratio of Higgs mass $m$, and vector Boson mass $M$ as $\beta = {m^2\o
M^2}$ from the D3 brane point of view. For example, consider a case where the
Higgs boson is much heavier than the vector boson. Having a massive vector
means that we expect magnetic fluxes to confine to a finite region. However as
shown in \hindmarsh\ and \preskill\ this doesn't quite happen because there are
no stable vortices. Thus energetics plays   an important role here,
 and  that is   one   of the key 
 reasons   why these defects are termed   non-topological.

Coming back to the issue of global symmetries, it is by now well
known that ADE type of singularities can easily arise here once we
have the following situation: \eqn\ades{\eqalign{& {\bf
A_n}:~~z^{n+1}, ~~ n \ge 1, ~~~~~~~~ {\bf D_n}: ~~ x^{n-1} + x
z^2, ~~ n \ge 4 \cr & {\bf E_6}:~~x^3 + z^4, ~~~~ {\bf E_7}: ~~x^3
+ x z^3, ~~~~ {\bf E_8}: ~~ x^3 + z^5}} where $x, z$ would be
related to the  coordinates of the $T^2$ fiber or/and              to the 
($x,\theta_1$) space in a suitable way.

To determine the relation between the Weierstrass equation that  we presented
earlier   and   the singularity types above,  \ades, we have to write an
equation that is more generic than the Weierstrass equation \bikmsv, namely:
\eqn\weinow{y^2 + a_1 xy + a_3 y ~ = ~ x^3 + a_2 x^2 + a_4 x + a_6}
where $a_i$ are in general functions of the base coordinate $z$, and if we
replace $x$ in the above equation \weinow\ by $x - {b_2 \o 12}$, then \weinow\
can take the following form:
\eqn\weiwei{\eqalign{&~~ \left(y + {12 a_1 x - a_1 b_2 + 12 a_3 \o 24}\right)^2
= x^3 - {x\o 48}\left(a_1^4 + 16 a_2^2 + 8 a_1^2 a_2 -24 a_1 a_3 -48
a_4\right)~+ \cr
& ~~~~~~~~~ + {1\o 864}\big(a_1^8 + 12 a_1^4 a_2 + 48 a_1^2 a_2^2 + 64 a_2^3 +
216 a_3^2 - 36 a_1^3 a_3 -72 a_1^2 a_4 + \cr
& ~~~~~~~~~~~~~~~~~~~~~~~~~~~~~ - 144 a_1 a_2 a_3 -288a_2a_4 + 864a_6\big)}}
with $b_2 = 4a_2 + a_1^2$.
In this form it is now easy to identify   the Weierstrass equation that we
presented earlier; and once we have the Weierstrass equation it is a simple
exercise to extract the singularity types that we want for our case. In our
earlier paper \semil\ we gave a possible F-theory configuration
that is responsible for exceptional global symmetries in the $K3 \times \IP^1$
background.
These configurations have been discussed earlier in \dm.
The F-theory curves can be extracted from there. What we now require is an
analysis similar to the one done by
\hindvac\ for the existence of semilocal strings in our model with higher
global symmetries.

The $SU(n+1)$ global symmetry follows in a straightforward manner.
We have already given a concrete way to realize $SU(2)$ global
symmetry in \semil. What we need is simply to fix the discriminant
behavior as \eqn\discbe{\Delta~\sim~  z^{n+1} + {\cal O}(z^{n+2})}
upto an overall numerical factor. The $A_1, A_2, A_3..$ cases
would therefore go as $z^2, z^3, z^4..$ respectively. A more
detailed representation of the singularities can also be  obtained   when
the $T^2$ fiber degenerates over a four-dimensional subspace of
the base. This is when we would need ($z_1, z_2$) coordinates to
specify $f,g$ in the Weierstrass equation. For our case we have to
fix $f(z)$ and $g(z)$ in such a way   as to realize \discbe\
correctly. The local symmetry is just $U(1)$ coming from the D3
brane that is fixed far away. The breaking pattern here will be:
\eqn\grop{{SU(n+1)_g \times U(1)_l \o \IZ_{n+1}}~~~ (n > 1)~~
{}^{~\Phi}_{\longrightarrow} ~~~ U(n)_g,} where the subscript $g,l$
refer to global and local respectively. The vacuum manifold is
$S^{2n+1}$,   which basically comes from the coset space of our
groups; and since $\pi_1\left({U(n+1)\o U(n)}\right) = 1$, there
could only be semi-local strings. Once the energetics favor them,
the Higgs fields 
 wrap the $S^1$ directions of the Hopf
fibration \hindvac: \eqn\semwrap{S^{2n+1}~~
{}^{~S^1}_{\longrightarrow} ~~ \IC\IP^n.} These $S^1$ are related
to the local $U(1)$ gauge symmetry that the single D3 brane has on
its world volume. Thus there would be infinite set of strings {with Higgs fields   } 
wrapping various $S^1$ that cannot be deformed into one another
with {\it finite} energies.

The $U(1)$ local symmetry that we discussed for the D3 brane is
only residual. The actual local symmetry would be $Sp(1)$
broken to $U(1)$ at all points  {in moduli space   }     \seiwit. In fact for $m$ D3 branes
we expect to see an unbroken gauge group of $Sp(m)$ \ds\ (where
only a subgroup of this will be realized in the Coulomb branch of
the corresponding ${\cal N} = 2$ gauge theory\foot{Recall that we
are in the subspace \region\ where we may see an approximate
${\cal N} = 2$ theory on the D3 brane probes.}). For the
semi-local defects, then we would be interested in studying 
global symmetries that are of the form $Sp(n+1)$.

The $Sp(n+1)$ symmetries are not that easy to realise in a
conventional way from Tate's algorithm (see discussion in
\bikmsv). We will therefore give an approximate way to see these
from \weiwei, and then try to realize them directly from     the      
    D3 brane
world volume using the construction of \wolf, \alex, \galicki. The
Weierstrass equation from \weinow\ for this case can be written as
\eqn\wespk{y^2 - x^3 + a_1 xy +a_2 x^2 +  a_3 yz^{n+1} + a_4 x
z^{n+1} + a_6 z^{2n+2} = 0,} where $a_i$ are in general arbitrary,
and the behavior of the discriminant for the case when we convert
\wespk\ to the standard Weierstrass form \weiwei, will tell us
that \eqn\disspk{ \Delta ~ \sim ~ z^{2n+2} + {\cal O}(z^{2n+3})}
up to an overall numerical factor. One can verify that for
$Sp(1) \approx SU(2)$ the leading order behavior of the
discriminant is $z^2$, which is what we got earlier. The above
analysis therefore gives us global $Sp(n+1)$ symmetries,  up to 
possible subtleties mentioned in \bikmsv. Therefore for a D3 brane
near the seven branes we expect to see the following breaking:
\eqn\brspk{{Sp(n+1)_g \times Sp(1)_l \o \IZ_2} ~~
{}^{~\Phi}_{\longrightarrow} ~~{Sp(n)_g \times SU(2)_g \o \IZ_2},}
 which is exactly what was discussed in \hindvac. Our case can however be even
more general that the one presented in \hindvac\ and mentioned above, because
we can put $m$ D3 branes to get an additional local symmetries of $Sp(m)$.

The vacuum manifold is a $4n+3$ dimensional sphere $S^{4n+3}$. What about gauge orbits?
The local gauge symmetry is $Sp(m)$, and for D3 branes one would expect (at
least
classically) a gauge orbit for $SU(2)$ group, which is of course the three
sphere $S^3$. Thus the equivalent picture of \semwrap\ for this case would be
\hindvac
\eqn\spkwrap{S^{4n+3} ~~ {}^{~S^3}_{\longrightarrow}~~ \IH\IP^n}
where $\IH\IP^n$ is the {\it quaternionic} projective space (recall that
earlier we had complex projective space \semwrap).

{}From our D3 brane(s) picture, such a vacuum manifold is not difficult to find.
The ($n+1$)-plet complex scalar fields (i.e the strings to the different seven
branes)
can be theoretically replaced by ($n+1$)    quaternions.    The space that we get
from doing this is almost equivalent to a K\"ahler manifold in the sense that
there may sometime exist
a globally defined complex structure $J^{~i}_j$ which is written in terms of
three locally defined (1,1) tensors as
\eqn\tenfu{J^{~i}_j = \alpha_{\rm x}J^{{\rm x}~i}_j + \alpha_{\rm y}J^{{\rm
y}~i}_j + \alpha_{\rm z}J^{{\rm z}~i}_j}
where $\alpha_{\rm x, y, z}$ are constants. The complex structure $J^{~i}_j$ may or may not be
covariantly constant. When it is covariantly constant then it is K\"ahler of course. On the other hand, the
tensors $J^{{\rm x}~i}_j,J^{{\rm y}~i}_j,J^{{\rm z}~i}_j$ satisfy a
quaternion algebra \eqn\quatro{J^{{\rm x}~j}_i J^{{\rm y}~k}_j = - \delta^{\rm xy}\delta^{~k}_i +
\epsilon^{\rm xyz} J^{{\rm z}~k}_i.}
A quick consistency check of the above analysis would be to ask whether restoration of
supersymmetry would allow such vacuum manifold. From the analysis that we presented above,
vanishing flux condition would lead to an approximate ${\cal N} = 2$ 
             supergravity                   This
is perfectly consistent with our picture as pointed out long ago by Bagger-Witten \bagger\
that local ${\cal N} = 2$
supersymmetry requires the vacuum manifold to be a quaternionic K\"ahler manifold with
negative scalar curvature that is proportional to the Newton's constant\foot{On the other hand, ${\cal N} =1$
supersymmetry requires the vacuum manifold to be a K\"ahler one \bagger.}. In the limit of vanishing scalar
curvature we expect to get a hyper-K\"ahler manifold as our vacuum manifold. This is of course    a     well known story \rocek.

For the example that we presented here  in       \spkwrap, the simplest case
would be when the base manifold is $\IH\IP^1$. This is a
four-sphere $S^4$ which is a quaternionic K\"ahler manifold {\it
without} a complex structure.           There      are also other
quaternionic spaces that are not of the form $\IH \IP^n$. In fact
as shown by \wolf\ and \alex, there are two more varieties given
by \eqn\xpnypn{X\IP^n ~=~ {SU(n+2)\o U(n) \otimes U(2)}, ~~~~~
Y\IP^n ~ = ~ {SO(n+4) \o SO(n) \times SO(4)},} where $\otimes$ is
used to denote the product $U(n) \times U(2)$ with unit
determinant.

Comparing $X\IP^n$ with our earlier analysis, we see
that once we know the singularity equation for     the    $A_n$ algebra it is
not difficult to realise this coset. When $n =1$, $X\IP^1 = \IC\IP^2$,
which we have already determined. Therefore it only remains to
determine the Weierstrass equation for $SO(n+4)$, then the second
coset can also be realised. This equation is rather easy to
extract from Tate's algorithm so we will not do it here.
Interested readers may want to work it out    by themselves.         We simply
point out an obvious group equality \eqn\grequa{Y\IP^1~ = ~ {SO(5)
\o SO(4)}~ =~ {Sp(2) \o Sp(1)\times Sp(1)} ~ = ~ S^4 ~ = ~
\IH\IP^1,} which is again consistent from our vacuum manifold
construction. Thus this way all the defects associated with $A_n,
B_n, C_n$ and $D_n$ global symmetries can be realised in our
model. 
 What remains now are the exceptional ones.

To get the semilocal defects associated with exceptional global
symmetries, we have to back up a little for more generality. In
the process we will also be able to shed some light on the
classification of {\it homogeneous} quaternionic K\"ahler manifold
done many years ago by Wolf \wolf\ and Alekseevski \alex\     and    generalized by  
   \dWVP.    The key
idea governing the formation of semilocal defects can be rephrased
in the following way:

\noindent $\bullet$ Given any global group with the corresponding algebra ${\cal G}$, first find the {\it maximal} regular subalgebra ${\cal H}$. Having a
maximal subalgebra would mean that we can ignore the $u(1)$ groups.

\noindent $\bullet$ The subalgebra should be expressible in terms of product of two smaller subalgebras, namely ${\cal H} = {\cal H}_1 \times {\cal H}_2$.

\noindent $\bullet$ One of ${\cal H}_i$ should form the {\it
local} gauge symmetry on the D3 brane(s). For example let us
assume ${\cal H}_1$ to be the allowed local algebra (or the local
group) on the D3 brane(s).

\noindent $\bullet$ The local group\foot{We use the same notation for the group and its algebra. The distinction between them doesn't affect the analysis below.}
${\cal H}_1$ should have the required homotopy classification $\pi_n({\cal H}^c_1) = \IZ$, where ${\cal H}^c_1$ is the coset       
for the group ${\cal H}_1$.

Then semilocal defects can form on the D3 brane(s) world volume with a global symmetry ${\cal G}$, provided the energetics also   allows.   The generic breaking
pattern of the groups in this case will be:
\eqn\grbrpat{ {\cal G}_g \times ({\cal H}_1)_l ~~ {}^{~\Phi}_{\longrightarrow} ~~ ({\cal H}_2)_g \times ({\cal H}_1)_g}
where the subscript $g$ and $l$ refer to global and local symmetries respectively, as before. The corresponding coset manifold ${\cal M}_G$,   for which
  every point
$p$   corresponds to   a semilocal defect that cannot be deformed into 
 another    one   with finite energy, is given by
\eqn\semmdef{{\cal M}_G ~ = ~ {{\cal G} \o {\cal H}_1 \times {\cal H}_2}.}
Many of the cases that we studied so far (or have been addressed in the literature) can be seen to follow from the above framework. Therefore let us first use this
to study the case when ${\cal G} = E_6$. The maximal regular subalgebra of $E_6$ 
can be extracted from the extended Dynkin diagram 
\vskip.1in

\centerline{\epsfbox{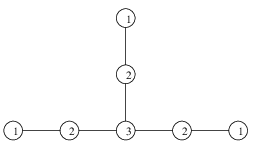}}\nobreak

\vskip.1in
\noindent and is given by ${\cal H} = su(6) \times su(2)$. This immediately tells
us two things: One, the world volume gauge group ${\cal H}_1 = SU(2) = Sp(1)$, and two, the manifold ${\cal M}_{E_6}$ is
\eqn\maniesix{{\cal M}_{E_6} ~ = ~ {E_6 \o SU(6) \times Sp(1)}.}
Since $SU(2) \sim S^3$, the homotopy classification will tell us that $\pi_3(S^3) = \IZ$. These are the instantons, and therefore should have a construction
via the quaternion as we discussed before. Having a quarternionic framework also means that ${\cal M}_{E_6}$ should be a quaternionic K\"ahler manifold if
whatever we said is consistent. One can easily check from \wolf, \alex\ that this is exactly the case!

Therefore we seem to get our first exceptional semilocal defect in
this model. However in the process of deriving this we have
ignored a subtlety. This subtlety cannot be seen at the level of
group structure, but is visible when we look at the F-theory curve
associated with our manifold. Therefore let us construct the
corresponding F-theory curve. From Tate's algorithm this is given
by \eqn\esix{y^2 - x^3 - a_1z^5 + a_2 xyz - a_3 x^2z^2 -a_4 xz^3 +
a_5 yz^2 ~ = ~ 0} with $a_i$ arbitrary. Once we convert this
equation to the familiar form of \weiwei\ the discriminant locus
can be easily worked out. For us this will be given by
\eqn\disesix{\Delta ~ \sim~ z^8 + {\cal O}(z^9)} up to an overall
numerical factor. Knowing the discriminant we can in
principle extract the corresponding subalgebra associated with the
global group ${\cal G} = E_6$ provided the background space is
specified. We have already chosen our subspace that allowed us
earlier to realise an approximate ${\cal N} =2$ supersymmetry.
This is of course the space given by the equation \region. The
F-theory curve \esix\ with discriminant \disesix\ will then give
us the following subalgebra on this space: \eqn\subalesix{ su(5)
\times su(2) \times u(1)} which is almost the maximal subalgebra
that we wanted, but not quite. In fact $su(6)$ is broken
to $su(5) \times u(1)$. Thus this is the closest we come to
getting the full structure of the coset space directly from type
IIB string theory (or F-theory). How do we then realize the full
configuration? First we decompose the $E_6$ adjoint in terms of
the subalgebra \subalesix\ as \eqn\seveig{{\bf 78}~ = ~ ({\bf 24},
{\bf 1})_0 + ({\bf 1},{\bf 1})_0 + ({\bf 1}, {\bf 3})_0 + ({\bf
10}, {\bf 2})_{-3} + ({\bf 5}, {\bf 1})_6 + {\rm c.c}} where the
subscripts refer to the $U(1)$ charges and the c.c are associated
with $\bar{10}$ and $\bar 5$ with $U(1)$ charges 3 and $-6$
respectively.

Secondly, having given the decomposition,   the   rest of the discussion now should follow the familiar line developed in the series of papers \dm, \barton. We will not
elaborate on this aspect as the readers can look up the details in those papers. It'll simply suffice to mention that the non-trivial configuration
required to get the full group structure lies in the process of brane creation via the Hanany-Witten effect \hw\ leading to strings with multiple
prongs \schkol, \dmbps, \senlater\ that fill out the rest of the group generators \barton.

The above construction therefore gives us the semilocal instanton
configurations associated with global symmetry $E_6$. Let us now
turn towards the next group ${\cal G} = E_7$. The extended Dynkin 
diagram of $E_7$ 
\vskip.1in

\centerline{\epsfbox{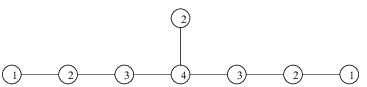}}\nobreak

\vskip.1in
\noindent can be {\it cut} in different ways to give rise to various 
maximal regular subalgebras of $E_7$. They are
given by \eqn\mare{su(8), ~~~~
{\rm spin}(12) \times su(2), ~~~~ su(6) \times su(3),~~~~su(4) \times su(4) \times su(2)} where
spin(12) actually comes from $so(12)$ with some identification
between the generators. From the set of steps that we mentioned
earlier, we can immediately ignore $su(8)$ and $su(4) \times su(4) \times su(2)$ as they are not 
product of two subalgebras. What about the other two cases? To see which one we
could keep, we have to look for the {\it homotopy} classification
of the local gauge group on the D3 brane(s). Taking ${\cal H}_1$
as either $su(2)$ or $su(3)$ we see that\foot{In fact generically 
$\pi_3\left(su(n)\right)\vert_{n\ge 2} =  \IZ$. Similarly  
$\pi_3\left(so(n)\right)\vert_{n\ge 3, n \ne 4} =  \IZ$ and 
$\pi_3\left(so(4)\right) = \pi_3\left(su(2) \times su(2)\right)  = \IZ \oplus \IZ$.
For Exceptional groups $\pi_m \left(E_n\right) ~ = ~ \IZ \delta_{m3}$. For more details see \nakahara.}
\eqn\homclass{\pi_3\left(su(2)\right)~ = ~ \pi_3\left(su(3)\right) ~ = ~ \IZ.}
Thus homotopically both the groups can exist. But since on a single D3 brane we can 
either have $U(1)$ (quantum mechanically) or $Sp(1)$ (classically), the case with ${\cal H}_1 = SU(3)$ 
cannot arise here.
The above consideration immediately gives
us the corresponding unique coset manifold for the global symmetry
$E_7$ as \eqn\coseseven{{\cal M}_{E_7} ~ = ~ {E_7 \o {\rm
Spin}(12) \times Sp(1)}.} Our previous consideration will require
us to view this as a homogeneous quaternionic K\"ahler manifold.
From the classification of \wolf, \alex\ we see that this is indeed
the case.

Here again we face the same kind of subtlety, as for the $E_6$
case, when we look into the F-theory curve. We will not go into
the details of this as the curve and its behavior in the sub-space
\region\ can be easily worked out. The discriminant locus and the
corresponding subalgebra that we can realise here will
  be, respectively,         \eqn\dissub{\Delta ~ \sim ~ z^9 + {\cal
O}(z^{10}), ~~~~~~~~ su(6) \times su(2) \times u(1)} and therefore
once we decompose ${\bf 133}$ over this subalgebra we can use
multi-prong strings to realize the full group structure \barton.

The final exceptional global symmetry that we want to study here
is $E_8$. This is straightforward. 
The extended Dynkin diagram is now given by
\vskip.1in

\centerline{\epsfbox{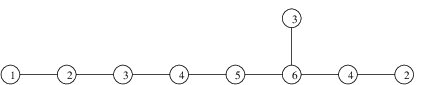}}\nobreak

\vskip.1in
\noindent From here the relevant allowed maximal 
subalgebra is $E_7 \times su(2)$. Other choices can be discarded because of the absence of $su(2)$ factor or 
non-decomposability as products of two smaller subalgebras.
Thus this will give us the next coset manifold:
\eqn\eeight{{\cal M}_{E_8} ~ = ~ {E_8 \o E_7 \times Sp(1)}.} From
the construction of \wolf, \alex\ the readers can check
that we have realized the final homogeneous quaternionic K\"ahler
manifold. Next looking at the discriminant on the sub-space
\region\ we get the manifest subalgebra: $su(7) \times su(2)
\times u(1)$, on which we could decompose the adjoint ${\bf 248}$
and recover the full structure via \barton. This way we get a ``physical''
reason for the existence of the homogeneous quaternionic K\"ahler manifolds.
Namely: 
they are related to the existence of semilocal defects on a D3 brane!

\noindent Before finishing, we
should point out two things:

\noindent $\bullet$ First, instead of the choice of {\it maximal} subalgebras, we could have also asked for {\it symmetric} 
subalgebras of the groups. The symmetric subalgebras for various groups have been listed in \slanskie. For the $A_n, B_n, C_n, D_n$ cases, they are
\eqn\abcdcase{\eqalign{& su(p+q)~ \to~  su(p) \times su(q) \times u(1), ~~~~~~  so(p+q) ~ \to~  so(p) \times so(q)\cr
&~~~~~~~~~~~~~~~~~~  sp(2p+2q) ~ \to ~ sp(2p) \times sp(2q)}}
where $p$ and $q$ form the various distribution (as even or odd integers). For the $E_n$ cases one would have
\eqn\encases{\eqalign{& E_8 ~ \to ~ so(16), ~~~ su(2) \times E_7 \cr
& E_7 ~ \to ~ su(8), ~~~ su(2) \times so(12), ~~~ E_6 \times u(1) \cr
& E_6 ~ \to ~ sp(8), ~~~ su(2) \times su(6), ~~~ so(10) \times u(1), ~~~ F_4.}}
From the list one has to 
extract out the relevant algebras that we would require for our case.

\noindent $\bullet$ Second, there are two more coset spaces given in
terms of $G_2$ and $F_4$ algebra as \eqn\gtwoffour{{G_2 \o SU(2)
\times Sp(1)}, ~~~~~~~ {\rm and}~~~~~~ {F_4\o Sp(3) \times
Sp(1)}.} The structure of these follow exactly the criteria that
we laid above (notice the $Sp(1)$ factor)\foot{The symmetric subalgebras for 
the group $F_4$ are $sp(1) \times sp(3)$ and $so(9)$; and for $G_2$ is $su(2) \times sp(1)$.}. 
The F-theory curves for
these cases can be worked out from Tate's algorithm. But a full
analysis using multi-prong strings \schkol, \dmbps, \senlater\ {\it \`{a} la}
\barton\ has not been done. So we will leave this for the sequel
to this paper.

\vskip.2in

\centerline{\bf Acknowledgements}

\vskip.1in

\noindent We would like to thank Raphael Bousso, Rich Corrado,
Atish Dabholkar,
Carlos Herdeiro, Shinji Hirano, Nori Iizuka, Shamit Kachru, Renata
Kallosh, Sheldon Katz, Andrei Linde, T.~Padmanabhan, Ashoke Sen, David Tong, 
Prasanta Tripathy, and especially Sandip Trivedi for many useful
discussions and valuable comments; and Yonatan Zunger for providing us
his GRONK program to do some of the computations.
K.D would also like to thank
Carlos Herdeiro and Shinji Hirano for an earlier collaboration in
2002 which influenced some of the ideas presented here. K.N.
thanks the organizers of the String Cosmology Workshop, IUCAA,
Pune, and the Indian Strings Workshop, Khajuraho, for stimulating
workshops where some of this work was done. The work of P.C.  and
M.S. are supported by DOE grant DE-AC03-76SF00515. The work of K.D
is partially supported by NSF grant number DMS-02-44412, and the
dept. of Maths and Physics at the University of Illinois at
Urbana-Champaign. The work
of M.Z is supported by an Emmy-Noether-Fellowship of the German
Research Foundation (DFG) ZA 279/1-1.


\listrefs

\bye